\documentclass[apjs,square,numbers,iop]{myemulateapj}
\usepackage{graphicx,url}
\usepackage{amsmath,amssymb}
\usepackage{hyperref}
\hypersetup{colorlinks,
	linkcolor=blue, 
	urlcolor=magenta, 
	anchorcolor=blue,
	citecolor=blue
}
\usepackage{bm}
\usepackage{xcolor}
\usepackage{mathrsfs}
\usepackage{natbib}
\begin{document}

\title{Recovering the CMB Signal with Machine Learning}

\author{Guo-Jian Wang$^{2,3}$, Hong-Liang Shi$^{4}$, Ye-Peng Yan$^{1}$, Jun-Qing Xia$^{1}$, Yan-Yun Zhao$^{4}$, Si-Yu Li$^{5}$, Jun-Feng Li$^{6}$}

\affil{$^{1}$Department of Astronomy, Beijing Normal University, Beijing 100875, People's Republic of China; \textcolor{blue}{xiajq@bnu.edu.cn}}

\affil{$^{2}$School of Chemistry and Physics, University of KwaZulu-Natal, Westville Campus, Private Bag X54001, Durban, 4000, South Africa}

\affil{$^{3}$NAOC-UKZN Computational Astrophysics Centre (NUCAC), University of KwaZulu-Natal, Durban, 4000, South Africa}

\affil{$^{4}$School of Information and Communication Engineering, Beijing University of Posts and Telecommunications, Beijing 100876, People's Republic of China}

\affil{$^{5}$Key Laboratory of Particle Astrophysics, Institute of High Energy Physics, Chinese Academy of Science, P. O. Box 918-3, Beijing 100049, People's Republic of China}

\affil{$^{6}$Institute of Acoustics, Chinese Academy of Sciences, No. 21 North 4th Ring Road, Beijing 100190, People's Republic of China}

\received{2020 November 29}
\revised{2022 March 14}
\accepted{2022 March 17}
\published{2022 MM DD}

\begin{abstract}
The cosmic microwave background (CMB), carrying the inhomogeneous information of the very early universe, is of great significance for understanding the origin and evolution of our universe. However, observational CMB maps contain serious foreground contaminations from several sources, such as galactic synchrotron and thermal dust emissions. Here, we build a deep convolutional neural network (CNN) to recover the tiny CMB signal from various huge foreground contaminations. Focusing on the CMB temperature fluctuations, we find that the CNN model can successfully recover the CMB temperature maps with high accuracy, and that the deviation of the recovered power spectrum $C_\ell$ is smaller than the cosmic variance at $\ell>10$. We then apply this method to the current Planck observation, and find that the recovered CMB is quite consistent with that disclosed by the Planck collaboration, which indicates that the CNN method can provide a promising approach to the component separation of CMB observations. Furthermore, we test the CNN method with simulated CMB polarization maps based on the CMB-S4 experiment. The result shows that both the EE and BB power spectra can be recovered with high accuracy. Therefore, this method will be helpful for the detection of primordial gravitational waves in current and future CMB experiments. The CNN is designed to analyze two-dimensional images, thus this method is not only able to process full-sky maps, but also partial-sky maps. Therefore, it can also be used for other similar experiments, such as radio surveys like the Square Kilometer Array.
\end{abstract}
\maketitle

\section{Introduction}\label{sec:introduction}

Over the past two decades, the temperature anisotropies of the cosmic microwave background (CMB) have been increasingly measured. These measurements have ushered research into cosmology into the era of precision \citep{Planck2018:VI}. Additional information about the history of the universe is contained in the polarization anisotropy of the CMB. Thus, many experiments, such as the Simons Array \citep{Arnold:2014}, Atacama Cosmology Telescope \citep{Thornton:2016}, SPT-3G \citep{Benson:2014}, BICEP3 \citep{Kang:2018}, the Simons Observatory \citep{Ade:2019}, CMB-S4 \citep{CMBS4}, LiteBIRD \citep{Sugai:2020} and POLARBEAR \citep{POLARBEAR} , are measuring or planning to measure the polarized CMB with improved sensitivity. For these experiments, the component separation of CMB the raw CMB maps during the data analysis is a key step in obtaining the foreground-cleaned information for the cosmological studies to follow.

The observed CMB maps are contaminated by various foreground components, in our galaxy or beyond. Therefore, the purpose of component separation is to separate the CMB signal from the contaminated data to obtain a foreground-cleaned CMB map for cosmological studies. To achieve this, two kinds of method are usually adopted in the literature: (1) the nonparametric method, which only makes minimal assumptions concerning the foregrounds and seeks only to minimize the variance of the CMB signal---such as the traditional Internal Linear Combination algorithm (ILC, \citet{ilc}), which can quickly provide a foreground-cleaned CMB map, but not detailed information about the various foreground emissions; or (2) the parametric method, which uses parametric modeling of the foregrounds and explores model parameters through Bayesian parameter estimation techniques, by fitting a parametric signal model per pixel---such as the Commander method \citep{commander}, which allows us to exploit the features of the foreground-cleaned CMB map in detail, as well as those of the foreground emissions, but is typically a time- and resource-consuming procedure, and only sufficient for low-resolution CMB maps. Apparently, all these algorithms have their pros and cons.

With the remarkable progress of computer science over recent years, methods based on machine learning (ML) have shown outstanding performance in solving cosmological problems, in terms of both accuracy and efficiency. For example, ML performs excellently in analyzing gravitational waves \citep{George:2018a,George:2018b,George:2018c,Shen:2019,LiXiangru:2020}, estimating the parameters of a 21 cm signal \citep{Shimabukuro:2017,Schmit:2018}, discriminating cosmological and reionization models \citep{Schmelzle:2017,Hassan:2018}, searching for and estimating the parameters of strong gravitational lenses \citep{Jacobs:2017,Hezaveh:2017,Petrillo:2017,Pourrahmani:2018,Schaefer:2018}, classifying the large-scale structure of the universe \citep{Aragon-Calvo:2019}, estimating cosmological parameters \citep{Fluri:2018,Fluri:2019,Ribli:2019,Ntampaka:2020,Wanggj:2020a}, studying the evolution of dark energy models \citep{Escamilla-Rivera:2020}, reconstructing functions from cosmological observational data \citep{Wanggj:2020b,Wanggj:2021}, reconstructing the lensing potential of the CMB \citep{Caldeira:2019}, simulating and separating the dust foregrounds \citep{Aylor:2020,Krachmalnicoff:2021}, recognizing foreground models \citep{Farsian:2020} and inpainting foreground maps \citep{Puglisi:2020}, and even recovering the CMB temperature anisotropy signal from power spectra using multilayer perceptron \citep{Norgaard-Nielsen:2010}.

As a method of ML, deep convolutional neural networks (CNNs) have shown excellent capabilities in image recognition and classification tasks. Recent research shows that CNNs are capable of accurately extracting full-sky temperature maps of the CMB from observational data \citep{Petroff:2020}. In their analysis, a {\it DeepSphere} graph-based convolution \citep{Perraudin:2019} is adopted to the HEALPix sphere to extend traditional CNNs to the sphere, making this method suitable for full-sky CMB maps. As an improvement, we propose that traditional CNNs are also capable of separating various foreground emission sources from CMB observations, to provide a foreground-cleaned CMB map from both full-sky maps and partial-sky maps. For the recovery of the CMB temperature signal, we simulate components of CMB observations to train the CNN model, then apply it to the Planck mission; while for the recovery of the CMB polarization signal, based on the CMB-S4 experiment, we simulate CMB Q and U maps to train the network model and to test its capability of recovering the polarization signal. The main difference between our analysis and that of \citet{Petroff:2020} is that we use traditional CNNs to process the spherical sky map and generate foreground components using random spectral parameters, which may be important for the recovery of the CMB signal. In addition, we also use different frequency bands to train the network. The code and examples are available online.\footnote{\url{https://github.com/Guo-Jian-Wang/cmbNNCS}}

This paper is organized as follows. In section \ref{sec:method}, we illustrate the methodology of using a neural network to recover the CMB signal, which includes the architecture of the neural network, data simulations, the generation of the training data, data preprocessing, and the training process for the network model. In section \ref{sec:application_to_cmb}, we apply the neural network to the CMB experiments---specifically, we first test it using simulated data, then apply it to Planck full mission data sets. In section \ref{sec:discussion}, discussions about using the CNN method for the recovery of the CMB signal are presented. Finally, conclusions are shown in section \ref{sec:conclusions}.

\section{Methodology}\label{sec:method}

\subsection{Network Architecture}

A feedforward neural network is a mathematical model that consists of multiple layers. Each layer accepts a vector from the former layer as an input, applies a linear transformation to the input, and gives an output to the next layer after a nonlinear activation. Formally, the output of an individual neuron can be written as $f(w{\bm x} + b)$, where $\bm x$ is the input vector, $w$ and $b$ are the linear weights and biases to be learned, and $f$ is the corresponding nonlinear activation function. The weights $w$ and biases $b$ are optimized through a backpropagation algorithm when training the network on a data set $\{\bm X_{\rm train}, Y_{\rm train}\}$, by minimizing the deviation, which is expressed quantitatively by a loss function, of the predicted result $\hat{\bm Y}=f_{w,b}(\bm X_{\rm train})$ and the ground truth $\bm Y_{\rm train}$. A CNN, which is designed for image recognition, is a class of feedforward artificial neural network that consists of at least one convolutional layer. A convolutional layer accepts inputs of images and convolves them with small filters (or kernels), whose values are weights to be learned. A filter is applied across an entire image with a bias, and a feature map is generated after a nonlinear activation function. 

\begin{figure}
	\centering
	\includegraphics[width=0.48\textwidth,angle=0]{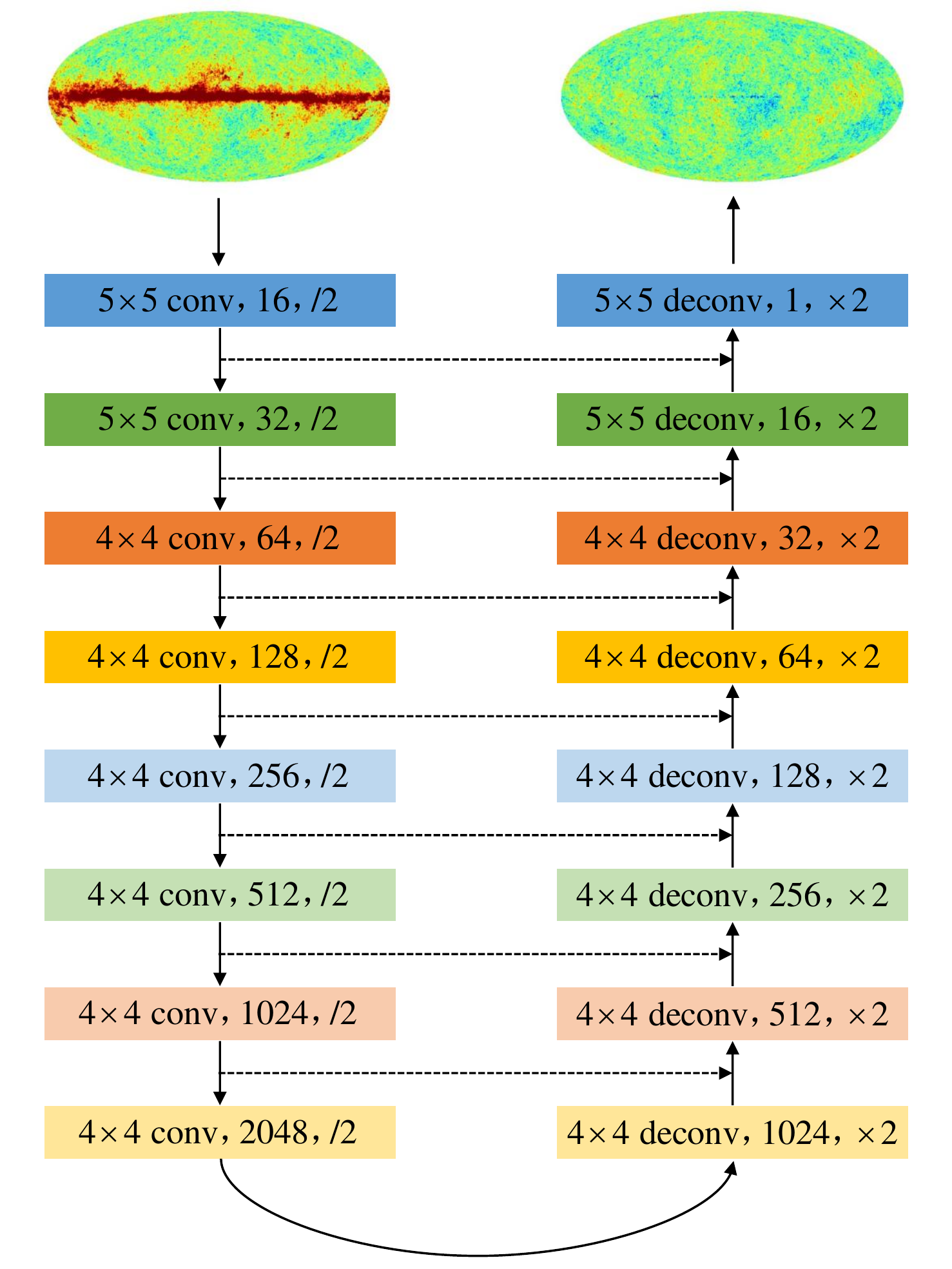}
	\caption{The general structure of the network model used for the component separation of a CMB map. The network contains symmetric convolutional layers (``conv'') and deconvolution layers (``deconv''). Skip connections (the dashed lines) are connected symmetrically from each convolutional layer to its corresponding mirrored deconvolutional layer.}\label{fig:nn_model}
\end{figure}

Previous work has shown that a deep network based on a CNN can be used for image denoising \citep{Vincent:2008}. Therefore, a CNN-based network can be an appropriate choice for the component separation of CMB observations. Here, we adopt the widely used U-Net architecture \citep{Ronneberger:2015} to achieve our analysis. The model consists of two parts: the encoder part (the left part of Figure \ref{fig:nn_model}) and the decoder part (the right part of Figure \ref{fig:nn_model}). The encoder part aims to learn a representation of the input data, while the decoder part reads the representation to generate an output sequence. In this structure, skip connections (the dashed lines in Figure \ref{fig:nn_model}) are added symmetrically from each convolutional layer to its corresponding mirrored deconvolutional layer, to ensure that more information about the convolutional feature maps is passed to the corresponding deconvolutional layers \citep{skip}. The passed convolutional feature maps are copied and concatenated with the deconvolutional feature maps.

In practice, we design a deep convolutional neural network with eight convolutional layers and eight deconvolutional layers (see Figure \ref{fig:nn_model}) for the component separation of raw CMB maps. Each convolutional or deconvolutional layer is followed by a nonlinear layer. Here, we take the parametric rectified linear unit \citep{prelu} as the nonlinear activation function. In addition to the last layer, the batch normalization \citep{batchnorm} is implemented before each nonlinear layer. We adopt Adam \citep{adam} as the optimizer, and the least absolute deviation is used as the loss function, averaged over all elements of the target maps. The first two convolutional layers and the last two deconvolutional layers have a kernel size of $5\times5$, while the other parts have a kernel size of $4\times4$. The output channels of each layer are 16, 32, 64, 128, 256, 512, 1024, 2048, 1024, 512, 256, 128, 64, 32, 16 and 1, respectively. The stride of all convolutional and deconvolutional layers is set to 2, which makes the size of the feature maps decrease in the encoder part and increase in the decoder part. Padding is applied to ensure that the sizes of the feature maps that are output are reduced by half for the convolutional layers and doubled for the deconvolutional layers.

\subsection{Data Simulations}\label{sec:simulation}

The network is trained through simulated maps in several frequency bands, which are generated by using the public software Code for Anisotropies in the Microwave Background (CAMB; \citet{camb}), HEALPix \citep{healpix}, and PySM \citep{pysm}. The sources we consider here include the CMB as the signal, and thermal dust, synchrotron, free-free, and anomalous microwave emission (AME) as the contaminations. Besides, the Gaussian beam and instrument noise of the Planck mission are also considered in our analysis.

It should be noted that the network aims to learn a mapping between contaminated CMB maps and foreground-cleaned CMB maps. Therefore, the data space of the contaminated CMB maps fed to the network should be large enough to cover the true observational data (such as the Planck contaminated CMB maps). In order to expand the coverage, we manually add uncertainties at the simulation step to both the CMB signal and the foregrounds. Specifically, for the CMB part, we compute the lensed CMB power spectra with CAMB\footnote{\url{https://github.com/cmbant/CAMB}} in the $\Lambda$ cold dark matter framework ($H_0$, $\Omega_bh^2$, $\Omega_ch^2$, $\tau$, $A_s$, $n_s$). The value of each cosmological parameter is subject to a Gaussian distribution $\mathcal{N}(P, \sigma^2)$, where $P$ and $\sigma$ are the best-fit value and the $1\sigma$ standard deviation obtained from the Planck2015 results \citep{planck2015:I}, respectively. Then we use the function {\it synfast} of the public software Healpy\footnote{\url{https://github.com/healpy/healpy}} (a python wrapper of HEALPix) to compute the CMB maps, with a pixel resolution of $N_{\rm side}=512$. 

For the foregrounds, the maps are simulated by using PySM, which is a public software that generates full-sky simulations of galactic foregrounds in intensity and polarization for various CMB experiments. We consider four dominant components for the temperature fluctuation maps: thermal dust at high frequencies, synchrotron, free-free, and AME at low frequencies. The model for thermal dust that we use here is the {\it d1} model, referring to the single-component modified blackbody model, with templates for emission at 545 GHz in intensity from the Planck2015 analysis \citep{planck-dust}. The model for synchrotron is $\it s1$, referring to a power-law model with the temperature templates at 408 MHz \citep{syn-temp}. The free-free model is {\it f1}, referring to an analytic model with the temperature template at 30 GHz, which is derived from the Commander fit to the Planck2015 data\citep{planck-free}. The model for AME is {\it a2}, referring to the sum of two spinning dust populations, based on the Commander code \citep{planck-dust}. The equations for generating the foreground maps are listed below:
\begin{eqnarray}\label{equ:spectral_params}
T_{d, \nu} &=& A_{T, d}{(\frac{\nu}{\nu_{0, d, T}})}^{\beta_d}\frac{B_{\nu}(T_d)}{B_{\nu_{0}}(T_d)}~, \nonumber\\
T_{s, \nu} &=& A_{T, s}{(\frac{\nu}{\nu_{0, s, T}})}^{\beta_s}~, \nonumber \\
T_{f, \nu} &=& A_{T, f}{(\frac{\nu}{\nu_{0, f, T}})}^{\beta_f}~,  \\
T_{a, \nu} &=& A_{\nu_{0,1}}\epsilon(\nu, \nu_{0,1}, \nu_{p,1}, \nu_{p0}) + A_{\nu_{0,2}}\epsilon(\nu, \nu_{0,2}, \nu_{p,2}, \nu_{p0})  ~, \nonumber
\end{eqnarray}
where $T_d$ stands for the blackbody temperature of the thermal dust, and $A$ and $\beta$ for the template maps of amplitude and spectral index, respectively. 

In order to increase the data space of the foreground maps, we randomize the spectral parameters $A$ and $\beta$ for each component in each frequency band when generating the training data. For the sake of simplicity, here we only consider randomizing the spectral index $\beta$, and we simply randomize all the pixels together. More analysis of the randomness of the spectral parameters will be discussed in section \ref{sec:variability_of_foregrounds}. Specifically, we generate maps for each component in each frequency band, through randomizing the values of $A$, according to a Gaussian distribution with its mean value equal to the spectral index of the template and a 10\% standard deviation. We note that the method of generating foregrounds through random spectral parameters can also reduce the dependence of the results on the foreground template, which may improve the reliability of the results.

In our analysis, we test the CNN method with four frequency bands: 100 GHz, 143 GHz, 217 GHz, and 353 GHz. The beam file used here is {\it Bl\_T\_R3.01\_fullsky\_X$\times$X.fits}, which is taken from the Planck Legacy Archive,\footnote{\url{http://pla.esac.esa.int/pla}} where {\it X$\times$X} stands for 100$\times$100, 143$\times$143, 217$\times$217, or 353$\times$353. The Planck frequency maps (at 100 GHz, 143 GHz, 217 GHz, and 353 GHz) contain effective Gaussian beams of FWHM=(9$^\prime$.66, 7$^\prime$.27, 5$^\prime$.01, and 4$^\prime$.86); thus, these beam effects are considered in our simulation. The instrument noise adopted here is the 300 realizations of noise and systematic residual maps that are publicly available on the Planck Legacy Archive, and these noisy maps are added to the mock data randomly.

\subsection{Training Set}\label{sec:training_set}

We generate 1200 mock data sets at four frequency bands with the pixel resolution $N_{\rm side}=512$, and split them into three groups: 1000 are taken as the training set, 100 as the validation set, and the others as the test set. We train the networks on the training set, select the optimal on the validation set, and then test it on the test set. As we illustrated in section \ref{sec:simulation}, both the CMB and the foreground maps are generated randomly. Therefore, the 1200 mock data are also generated randomly, which means that the cosmological and spectral parameters of each set of maps in the 1200 mock data are different. The inputs of the network are the contaminated CMB maps in four frequency bands, which contain the foregrounds, beam, and noise information, while the output is a foreground-cleaned CMB map with beam FWHM=7$^\prime$.27.

\subsection{Data Preprocessing}\label{sec:data_preprocessing}

The simulated CMB and foreground maps are one-dimensional arrays in the RING numbering scheme of HEALPix. However, the filter of the convolutional layer used in this work is designed to process two-dimensional image data. Therefore, these one-dimensional arrays should be transformed into two-dimensional arrays before feeding them into the network.

Starting from the one-dimensional array, we first transform it into the NESTED scheme of HEALPix, then arrange it into 12 small pieces of maps, with the size of $N_{\rm side}\times N_{\rm side}$, according to the NESTED scheme. Finally, these small maps are combined into a large plane map (with the size of $4N_{\rm side}\times 3N_{\rm side}$) for network training. We note that the combined two-dimensional map is an approximation of the spherical sky map. After the training of the network, we can feed the test set or the observational data to the well-trained network model to obtain a foreground-cleaned CMB map. The recovered CMB map is a two-dimensional map with the size of $4N_{\rm side}\times 3N_{\rm side}$. Therefore, we can finally obtain a spherical CMB map using the output two-dimensional map via inverse operation.

Besides, in order to facilitate the training of the network, we scale the target to smaller values. Specifically, we divide the foreground-cleaned CMB maps (i.e. the target in the training set) by five to make the network easier to converge. Finally, we can obtain the final estimated CMB signal by inverse operation. Note that the scaling to the target is just a trick to ensure that the network can be better trained, and the number five here is based on experiments.

\subsection{Training Process}\label{sec:training_process}

In our pipeline, the batch size is 12, the initial learning rate is set to 0.1 and gradually decays to $10^{-6}$ with the number of iterations, and the network model will be trained through 10,000 iterations by minimizing the loss function. For each iteration, a subsample with the size of the batch size will be loaded from the disk and fed to the network, and then the loss between the predicted maps and the target maps will be calculated, and, finally, the gradient of the loss will be calculated and passed backward to update the network parameters. The network models are trained on four NVIDIA 1080 Ti graphics processing units, and it takes $\sim6$ hr for one network model.

\begin{figure}
	\centering
	\includegraphics[width=0.45\textwidth]{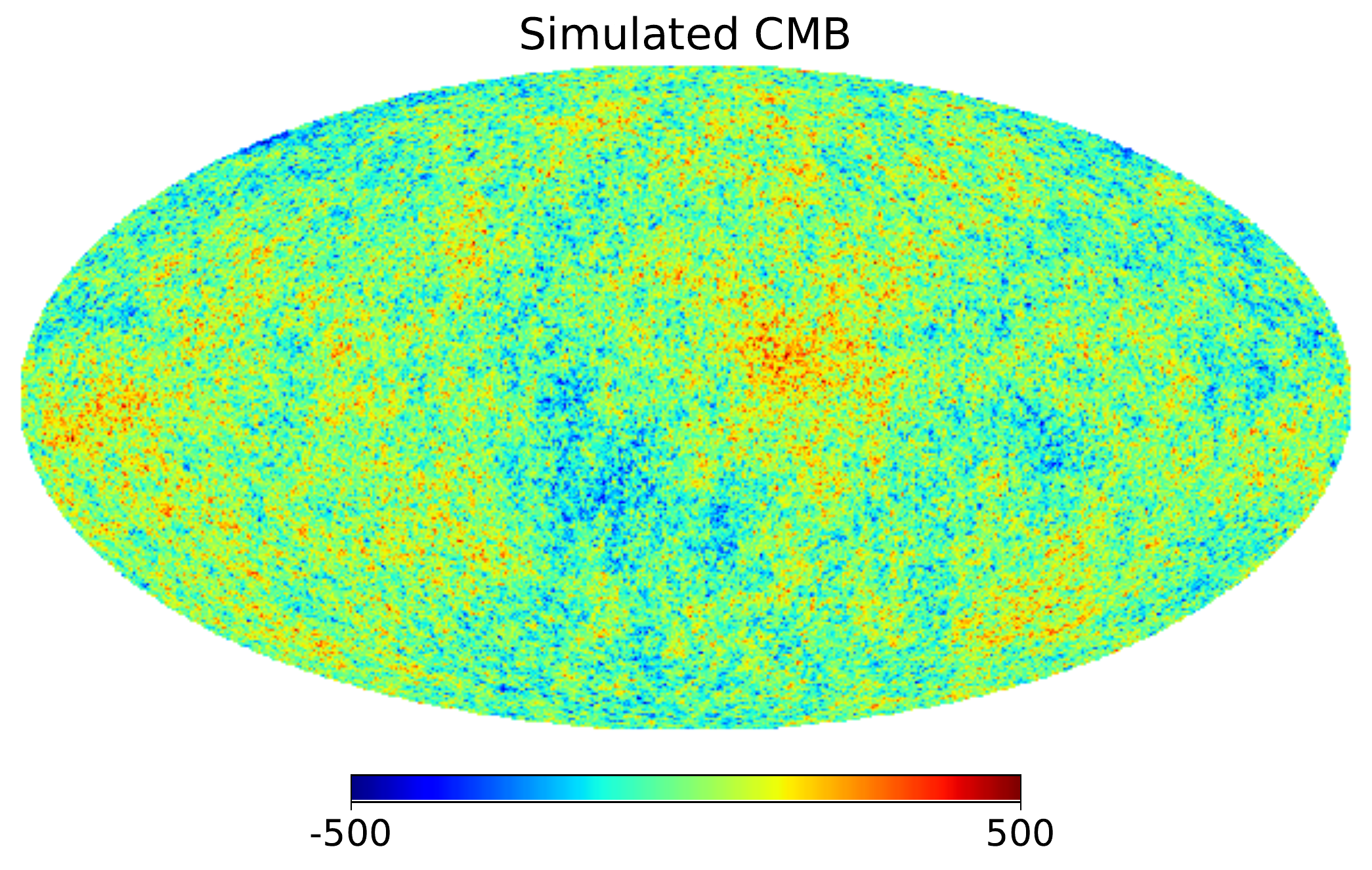}
	\includegraphics[width=0.45\textwidth]{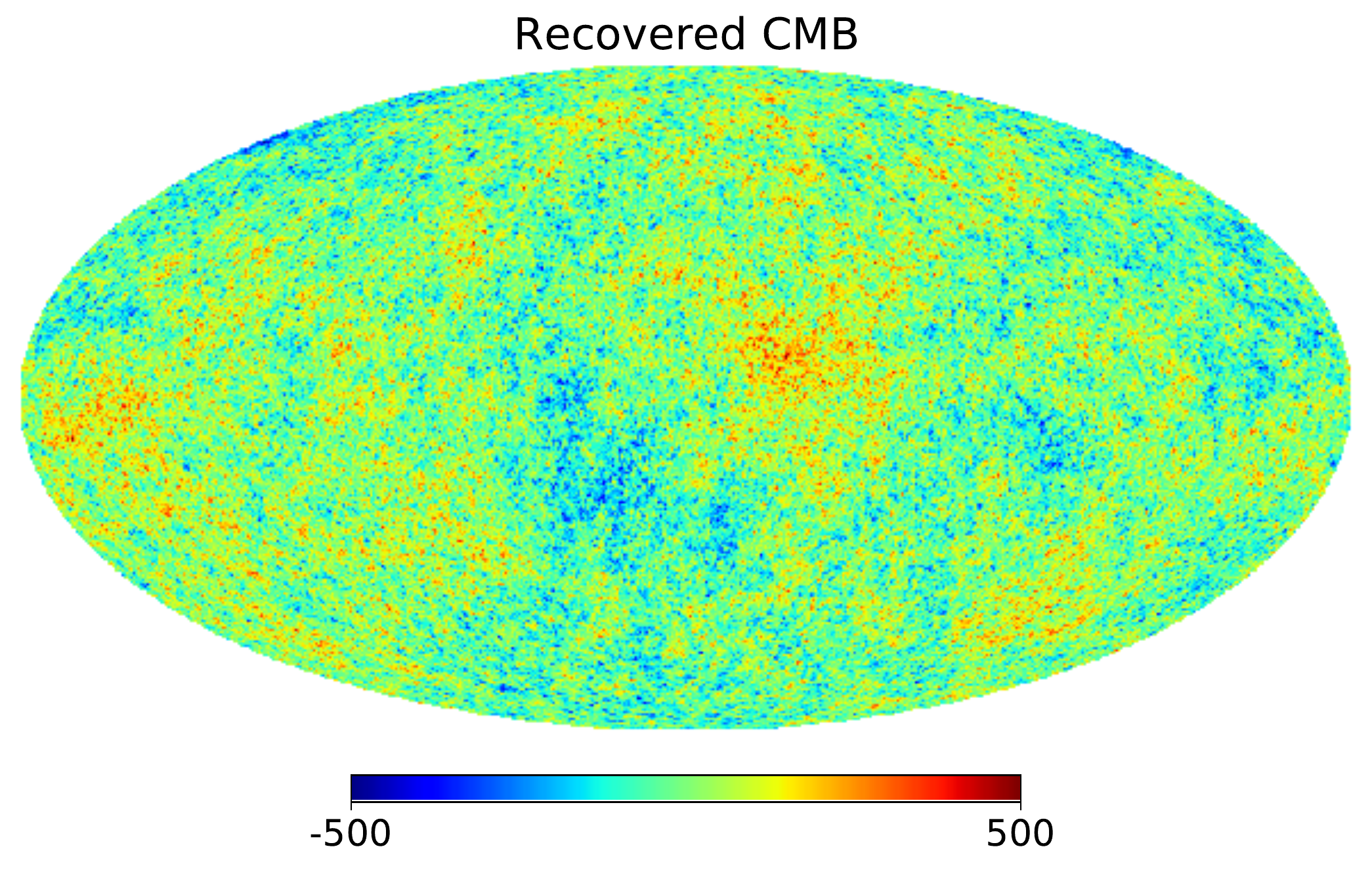}
	\includegraphics[width=0.45\textwidth]{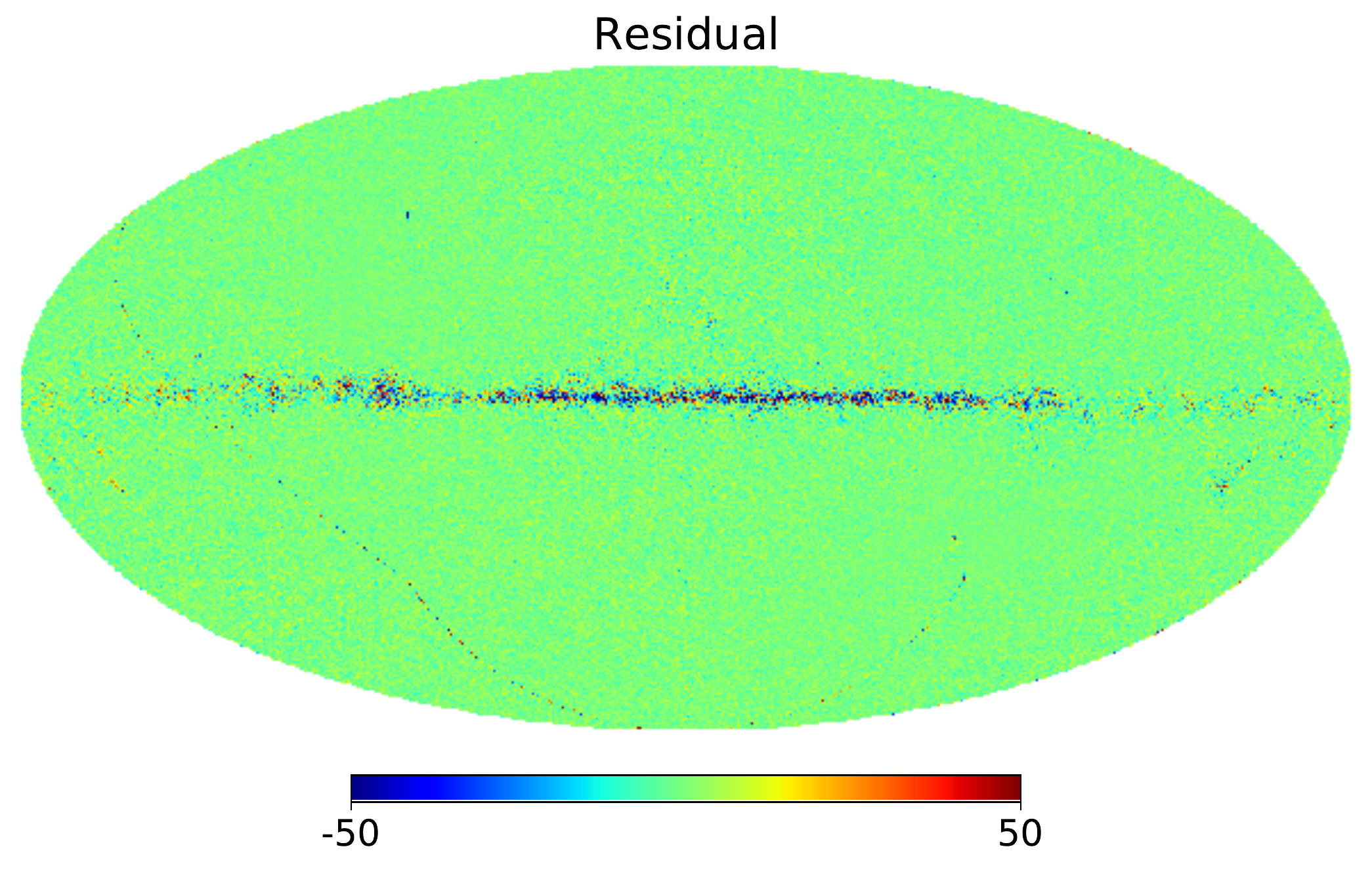}
	\caption{Recovered CMB temperature map using the neural network. {\it Upper panel:} the simulated pure CMB map. {\it Middle panel:} the CMB map recovered by the neural network. {\it Lower panel:} the residual map of the recovered CMB map.} \label{fig:sim_cmb_map}
\end{figure}

\begin{figure*}
	\centering
	\includegraphics[width=0.9\textwidth]{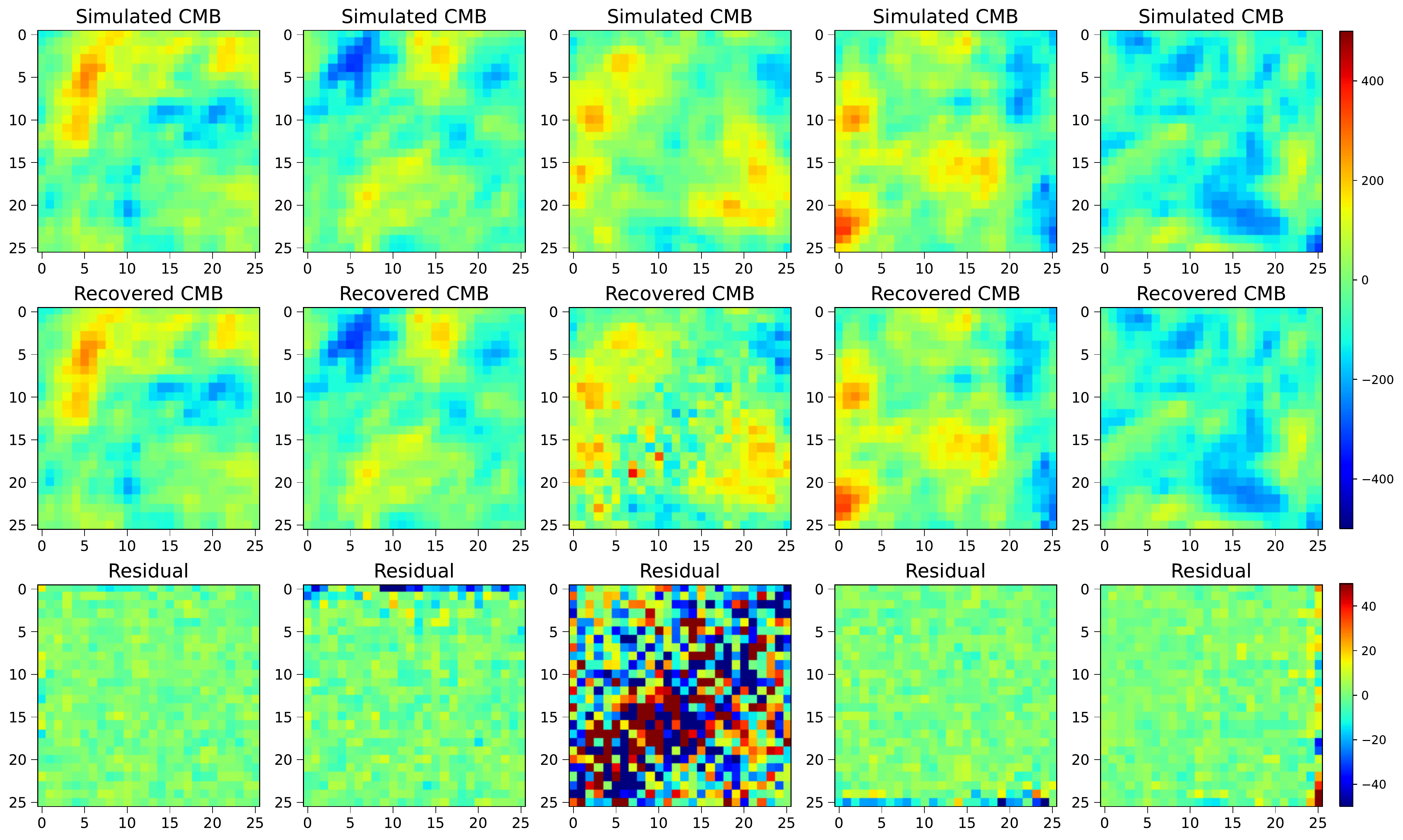}
	\caption{Five small patches with $3\times3$ deg$^2$, selected from Figure \ref{fig:sim_cmb_map}. These patches are selected from the north pole to the south pole, with different latitudes.} \label{fig:sim_cmb_map_miniPatch}
\end{figure*}

\section{Application to CMB Experiments}\label{sec:application_to_cmb}

The main procedure in using this method is to train a network model using simulated data and then apply the well-trained model to observational data; thus, in this section, we first test our method on simulated CMB maps, and then test it on the Planck CMB maps.

\begin{figure*}
	\centering
	\includegraphics[width=0.43\textwidth]{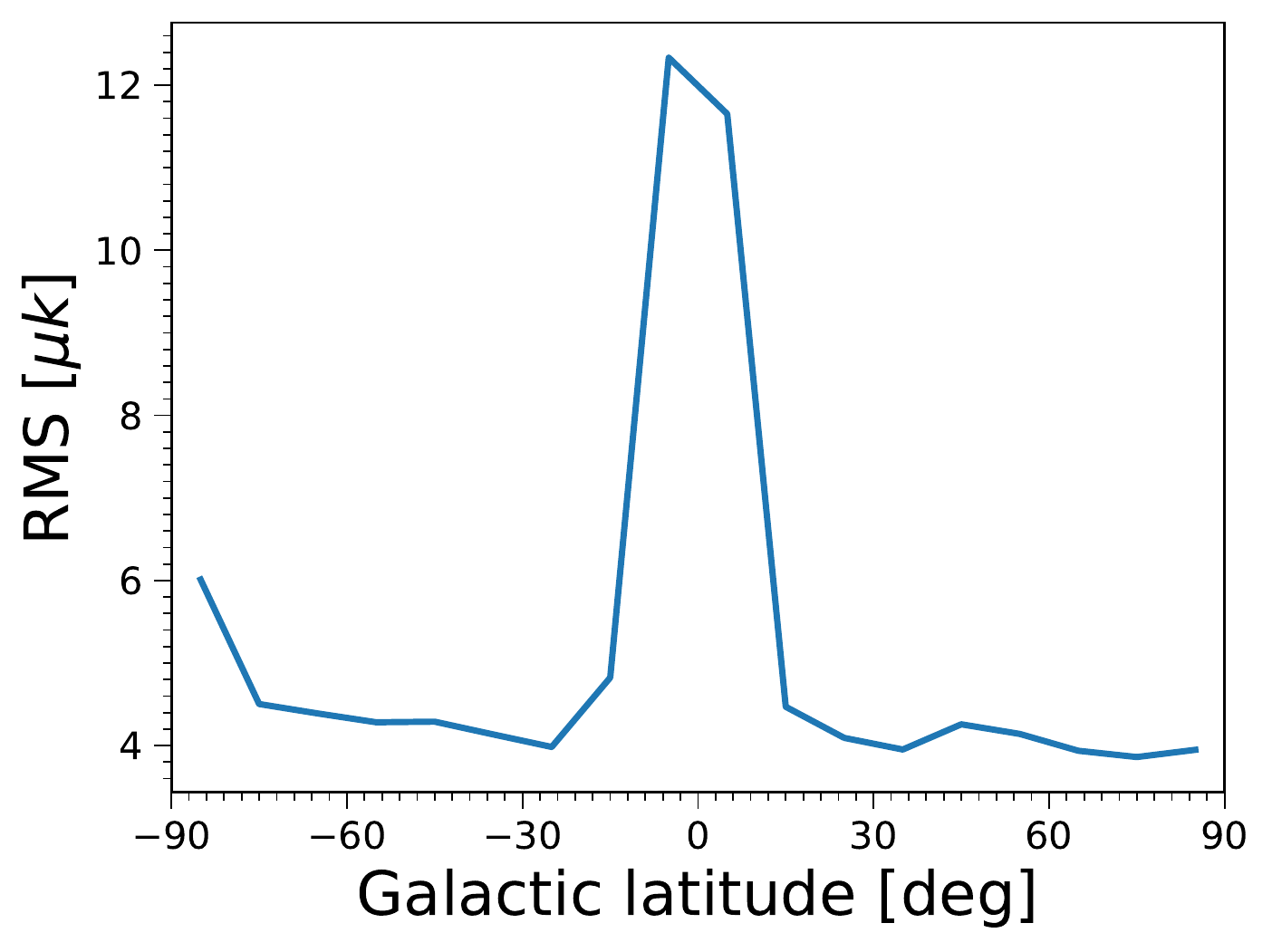}
	\includegraphics[width=0.45\textwidth]{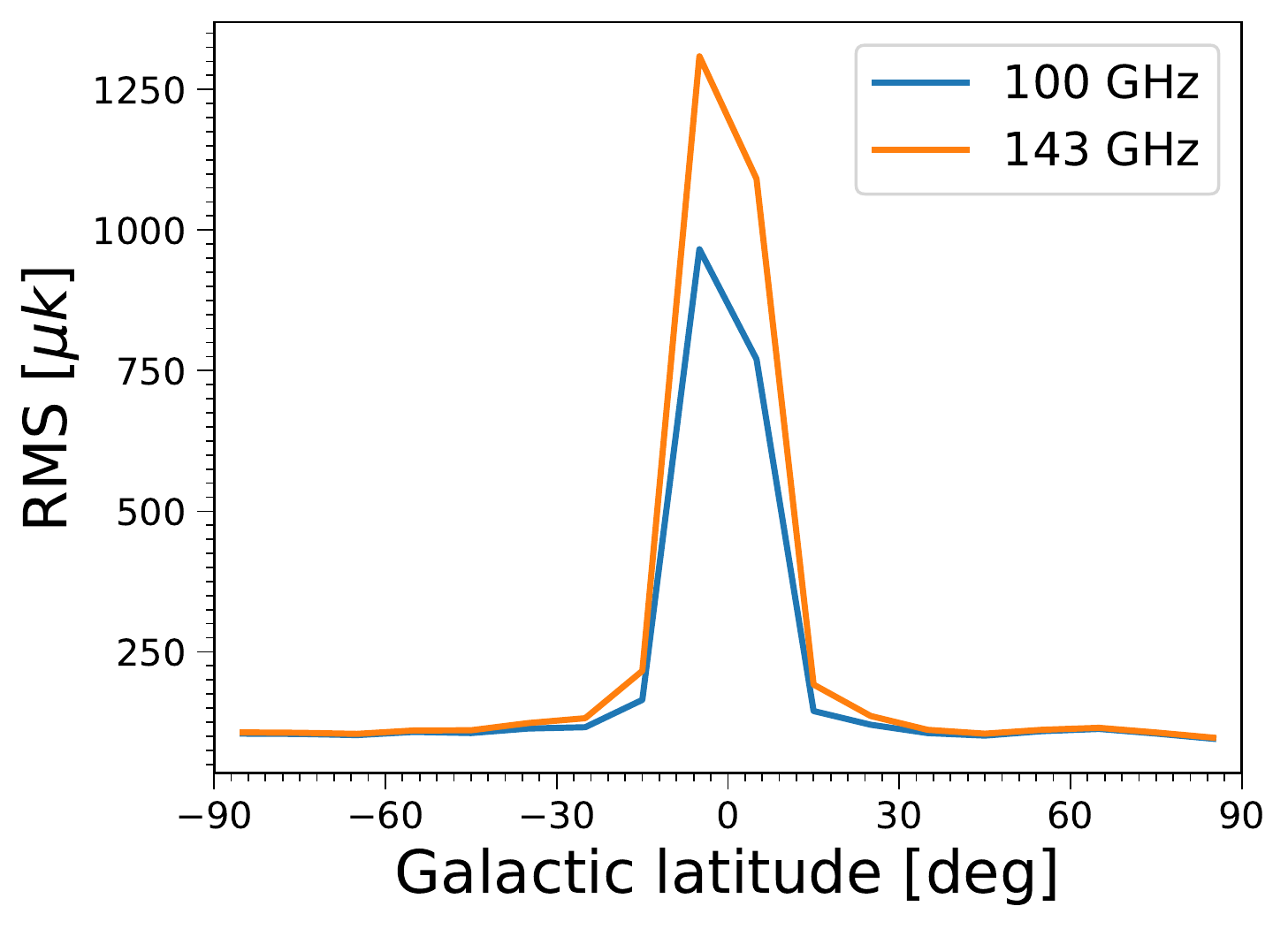}
	\caption{{\it Left panel}: rms for 18 zonal bands in the residual map, each being 10$^\circ$ wide in galactic latitude. {\it Right panel:} rms of the contaminated CMB maps at 100 GHz and 143 GHz.}
	\label{fig:sim_cmb_rms}
\end{figure*}

\begin{figure}
	\centering
	\includegraphics[width=0.45\textwidth]{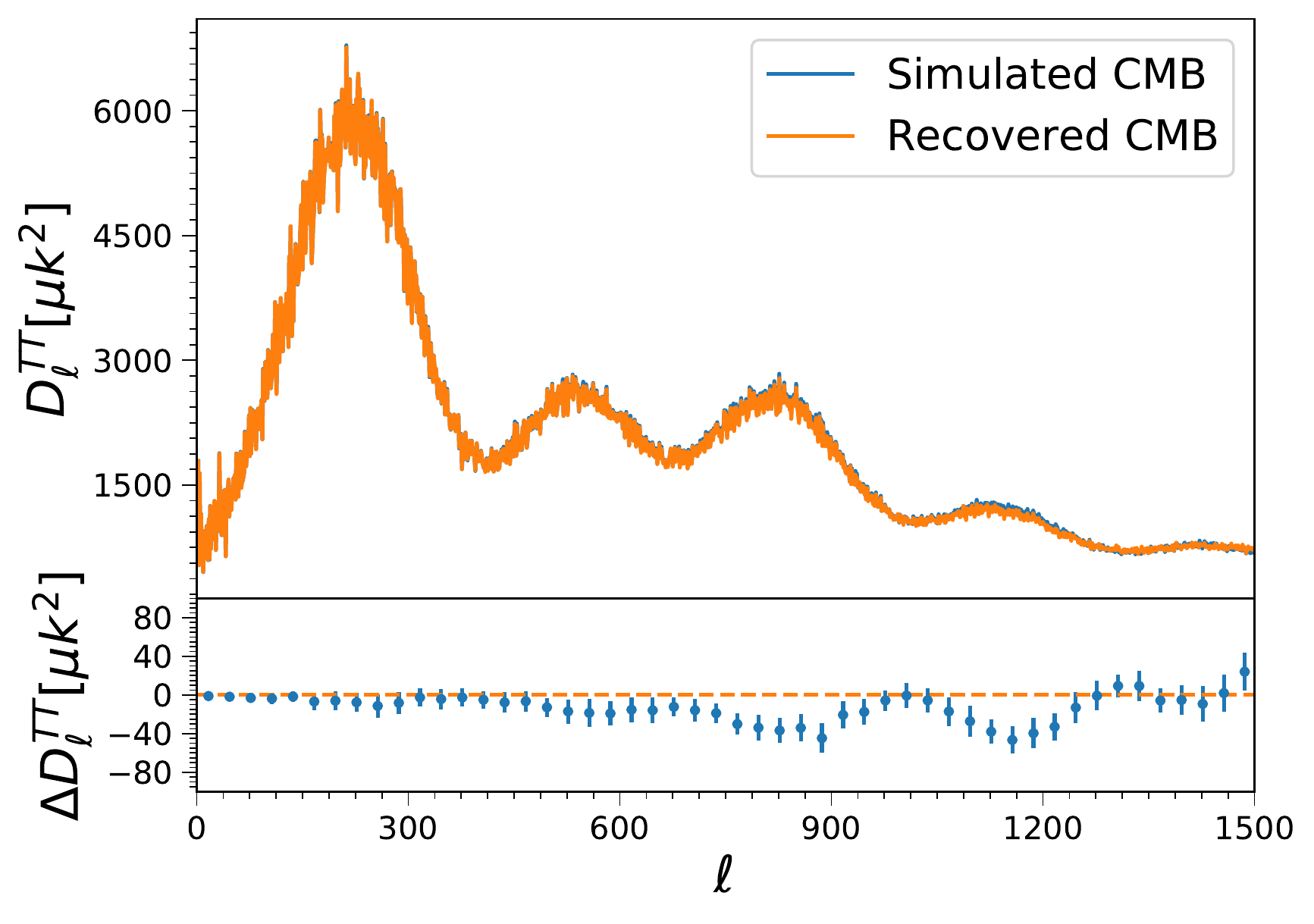}
	\caption{Power spectrum of the recovered CMB map (upper panel, using a bin size of 1), and the difference between this spectrum and that of the simulated CMB map (lower panel, using a bin size of 30).} \label{fig:sim_cmb_spectra}
\end{figure}

\begin{figure}
	\centering
	\includegraphics[width=0.45\textwidth]{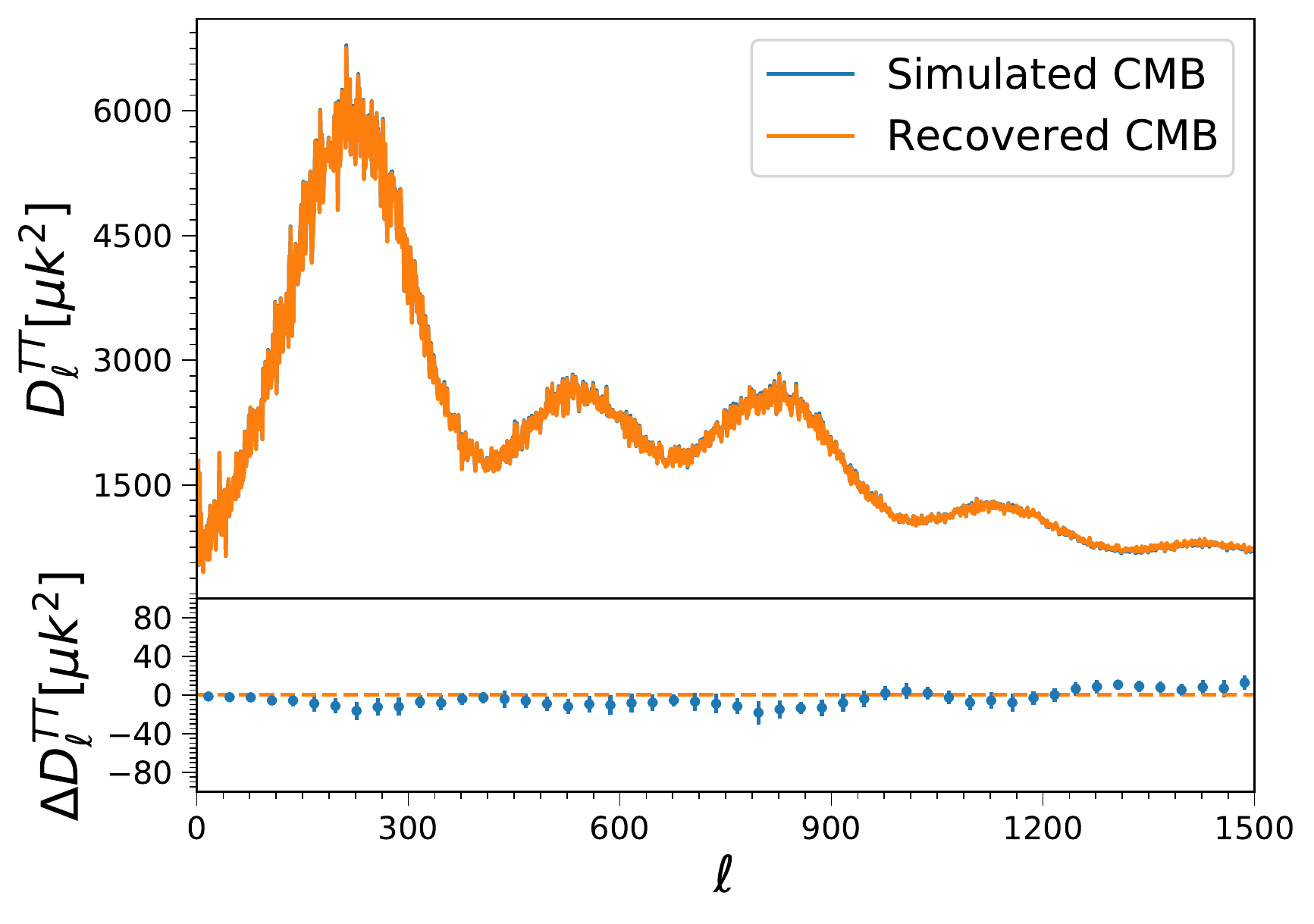}
	\caption{The same as Figure \ref{fig:sim_cmb_spectra}, but without instrument noise and instrument beam effects being included in the training data and the testing data.} \label{fig:sim_cmb_spectra_NoBeamNoise}
\end{figure}

\subsection{Recovering the CMB Signal}\label{sec:recover_sim_cmb}

In this section, we train the networks to recover the CMB temperature signal from simulated CMB observations. After selecting the optimal model with the validation set, we feed the maps of the test set to the network model to obtain a series of foreground-cleaned CMB maps. In Figure \ref{fig:sim_cmb_map}, we show one set of maps of the test set. The upper panel is the simulated pure CMB signal, the middle panel is the CMB map recovered by the neural network, and the lower panel is the corresponding residual map. Based on the residual map, we can see that the CMB information in the high galactic latitudes has been recovered very well, while in the low galactic latitudes, especially close to the galactic plane, the performance of the network seems a little unsatisfactory, due to the strong foreground contaminations. In order to have a deeper look at the recovery of the CMB map, we selected five small patches with $3\times3$ deg$^2$ from the north pole to the south pole with different latitudes, as shown in Figure \ref{fig:sim_cmb_map_miniPatch}. The first and last patches are located near the north and south poles, respectively. The second and fourth patches are located between the galactic plane and the north (south) pole, while the third patch is located in the galactic plane. For the first, second, fourth, and fifth patches, the recovered CMB maps are almost the same as the simulated ones, and there is little information left in the residual maps. On the contrary, the recovered CMB map for the third patch looks different to the simulated one, and there is a lot of information in the residual map. As we have shown in the residual map of Figure \ref{fig:sim_cmb_map}, this should be caused by the foreground contaminations in the galactic plane.

For a quantitative comparison, we calculate the mean squared error (MSE) for the recovered CMB map: ${\rm MSE}=40.055~\mu \text{K}^2$. Furthermore, we also calculate the root mean square (rms) for 18 zonal bands in the residual map, each being 10$^\circ$ wide in galactic latitude, shown in the left panel of Figure \ref{fig:sim_cmb_rms}. The rms range of the residual CMB map is from $3.861~\mu$K in the high galactic latitudes to $12.335~\mu$K in the galactic center, with a sudden increase close to the galactic plane $b < 20^\circ$. For comparison, we also calculate the rms of the contaminated CMB maps at 100 GHz and 143 GHz, shown in the right panel of Figure \ref{fig:sim_cmb_rms}. We can see that, except for the larger order of magnitude, the shape of this curve of the rms looks similar to that of the residual map. Therefore, by comparing these two panels, one can easily conclude that the sudden increase in rms close to the galactic plane in the residual map is caused by the galactic foregrounds.

We then compute the corresponding power spectrum of the recovered CMB map, shown in the upper panel of Figure \ref{fig:sim_cmb_spectra}. The orange line is the power spectrum of the recovered CMB map, while the blue line is that of the simulated CMB map. In the lower panel, we show the difference between the recovered spectrum and that of the simulated CMB map. We can see that the recovered CMB signal is quite consistent with the true spectrum at both small scales and large scales, which is different from the fact in \citet{Petroff:2020}, where the recovered spectrum diverges from the true spectrum at $\ell>900$.

We note that, at $\ell>750$, the recovered power spectrum begins to fluctuate, which should be caused by instrument noise and instrument beam effects. To prove this, using the same procedure, we trained another network model with data that have no instrument noise and no instrument beam effects, and test this network using mock data. In Figure \ref{fig:sim_cmb_spectra_NoBeamNoise}, we show the power spectrum of one set of maps from the test set. We can see that, at $\ell>750$, the difference between the recovered spectrum and that of the simulated CMB map is much better than the results in Figure \ref{fig:sim_cmb_spectra}. Therefore, the instrument noise and instrument beam effects have a great influence on the recovery of the CMB signal.

\begin{figure}
	\centering
	\includegraphics[width=0.45\textwidth]{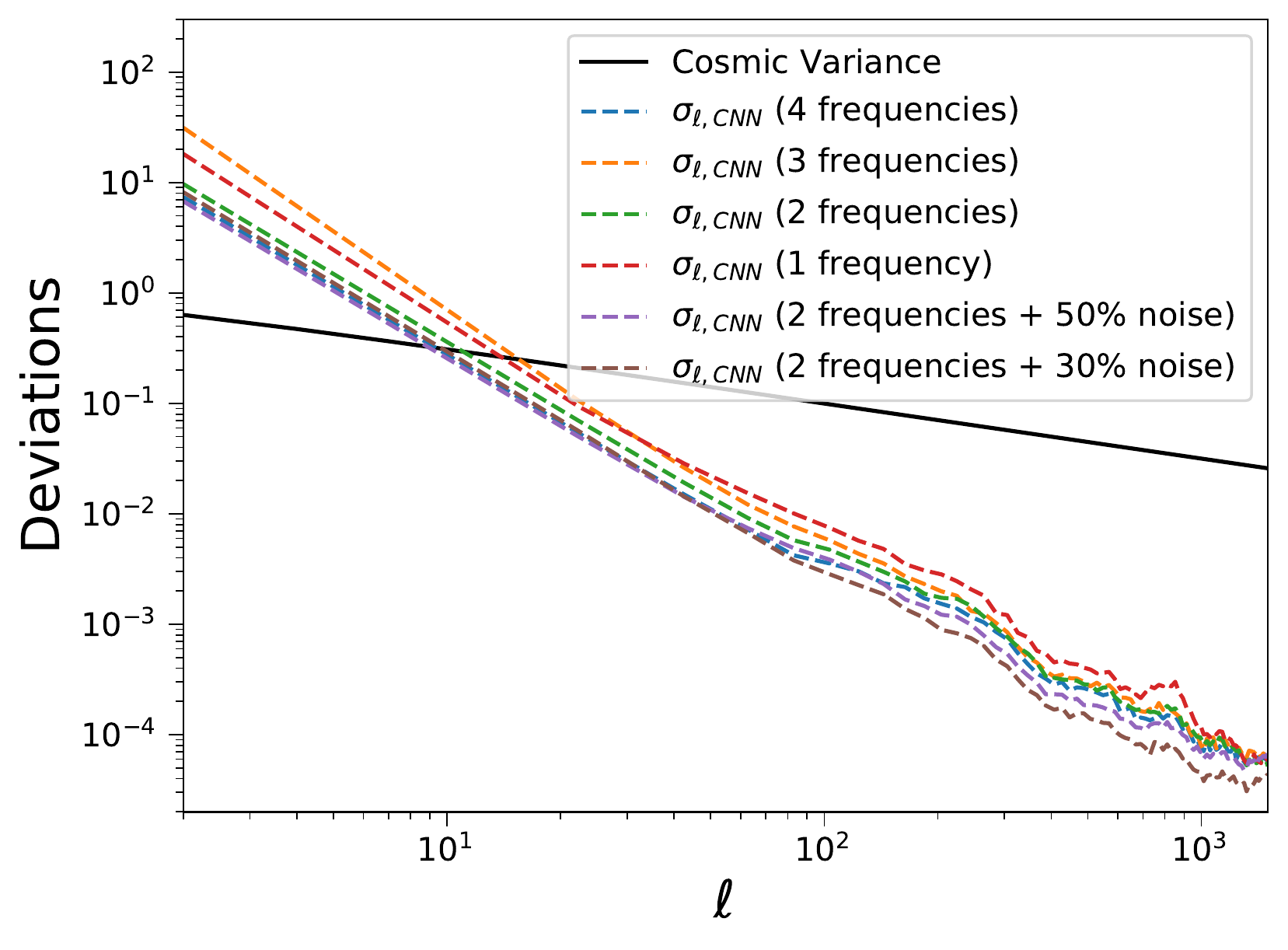}
	\caption{Deviations of the recovered power spectra compared with the cosmic variance.} \label{fig:sim_cmb_spectra_multiFreqs_error_cv}
\end{figure}

\begin{figure}
	\centering
	\includegraphics[width=0.45\textwidth]{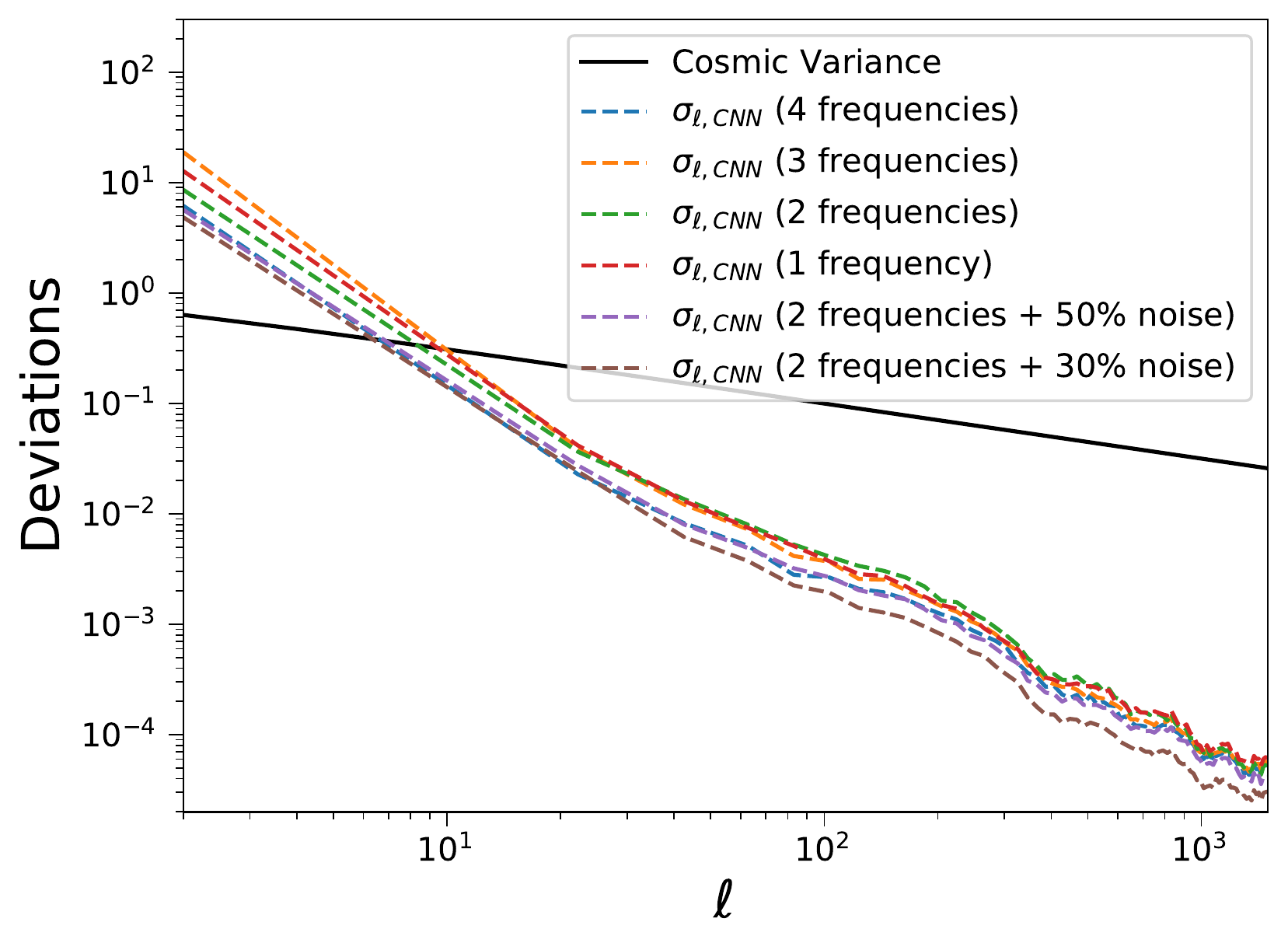}
	\caption{The same as Figure \ref{fig:sim_cmb_spectra_multiFreqs_error_cv}, but now the Planck mask is used when calculating $C_\ell$.} \label{fig:sim_cmb_spectra_multiFreqs_error_cv_plk_mask}
\end{figure}

In addition, we calculate the deviation of the recovered power spectrum, and then compare this deviation with the cosmic variance. First, we calculate the difference between the recovered power spectrum and the simulated ground truth:
\begin{equation}
\Delta C_{\ell,\text{CNN}} = C_{\ell, \text{recover}} - C_{\ell,\text{true}} ,
\end{equation}
where $C_{\ell, \text{recover}}$ is the recovered power spectrum and $C_{\ell,\text{true}}$ is the true power spectrum calculated using the simulated CMB map. Then, we calculate the standard deviation of this difference:
\begin{equation}
\sigma_{\ell,\text{CNN}} = \sqrt{\frac{1}{N}\sum_i^N(\Delta C_{\ell,\text{CNN}} - \overline{\Delta C}_{\ell,\text{CNN}})^2},
\end{equation}
where $N$ is the number of samples in the test set and $\overline{\Delta C}_{\ell,\text{CNN}}$ is the average of $\Delta C_{\ell,\text{CNN}}$. Finally, we compare this deviation with the cosmic variance:
\begin{equation}\label{equ:cosmic_variance}
\sigma_\ell = \left(\frac{\Delta C_\ell}{C_\ell}\right)_\text{cosmic variance} = \sqrt{\frac{2}{2\ell + 1}}.
\end{equation}
Figure \ref{fig:sim_cmb_spectra_multiFreqs_error_cv} shows a comparison of the deviation caused by the CNN and by the cosmic variance, where the black solid line is the cosmic variance and the blue dashed line is the deviation of the recovered power spectrum calculated using the CNN model, corresponding to Figures \ref{fig:sim_cmb_map} and \ref{fig:sim_cmb_spectra}. The result shows that the deviation caused by the CNN is a little larger than the cosmic variance at $\ell<10$. However, with the increase of the multiple $\ell$, the deviation will decreases rapidly, and it will becomes smaller than the cosmic variance by more than two orders of magnitude at $\ell>200$. This deviation is quite small when compared with the cosmic variance, and thus it will have less effect on the recovered CMB power spectrum. To further test the reason why the deviation is larger than the cosmic variance at $\ell<10$, we apply the Planck mask {\it COM\_Mask\_CMB-common-Mask-Int\_2048\_R3.00.fits}\footnote{\url{http://pla.esac.esa.int/pla/\#maps}} when calculating $C_\ell$. The results are shown in Figure \ref{fig:sim_cmb_spectra_multiFreqs_error_cv_plk_mask}. We can see that these results are somewhat better than the previous ones, but the deviations caused by the CNN are still larger than the cosmic variance at $\ell<8$. This means that the deviation is not primarily caused by the foreground contaminations in the galactic plane; instead, it is more likely that the deviation is caused by the CNN itself. Further research is needed to figure out if this issue can be solved fundamentally, which will be shown in our future works.


\begin{table}
	\centering
	\caption{The Number of Frequencies Used to Train the Network.}\label{tab:frequency_number}
	\begin{tabular}{c|c}
		\hline\hline
		Frequency Number & Frequencies (GHz) \\
		\hline
		1 & 143 \\
		2 & 143, 217 \\
		3 & 143, 217, 353 \\
		4 & 100, 143, 217, 353 \\
		\hline\hline
	\end{tabular}
\end{table}

\begin{figure*}
	\centering
	\includegraphics[width=0.23\textwidth]{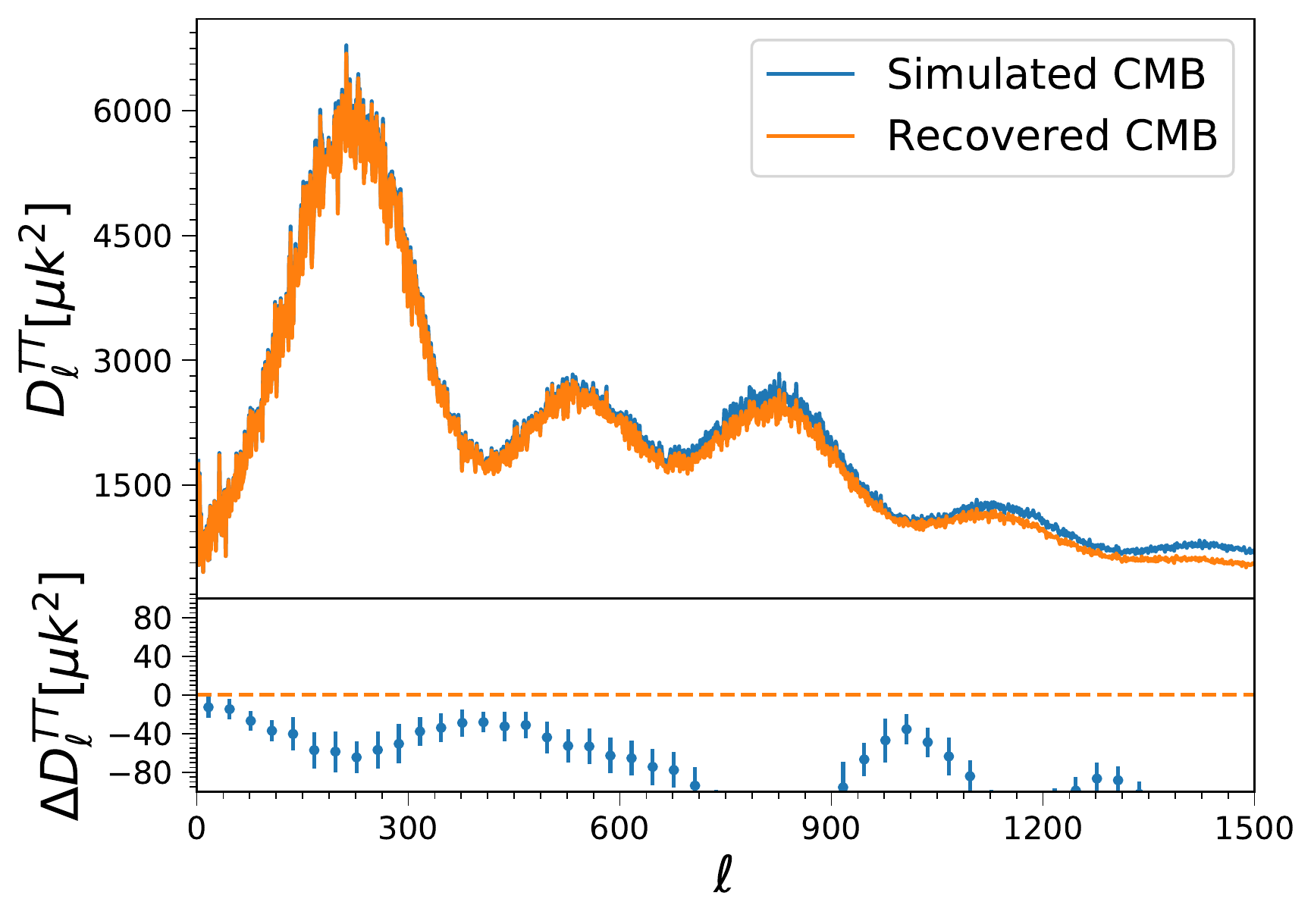}
	\includegraphics[width=0.23\textwidth]{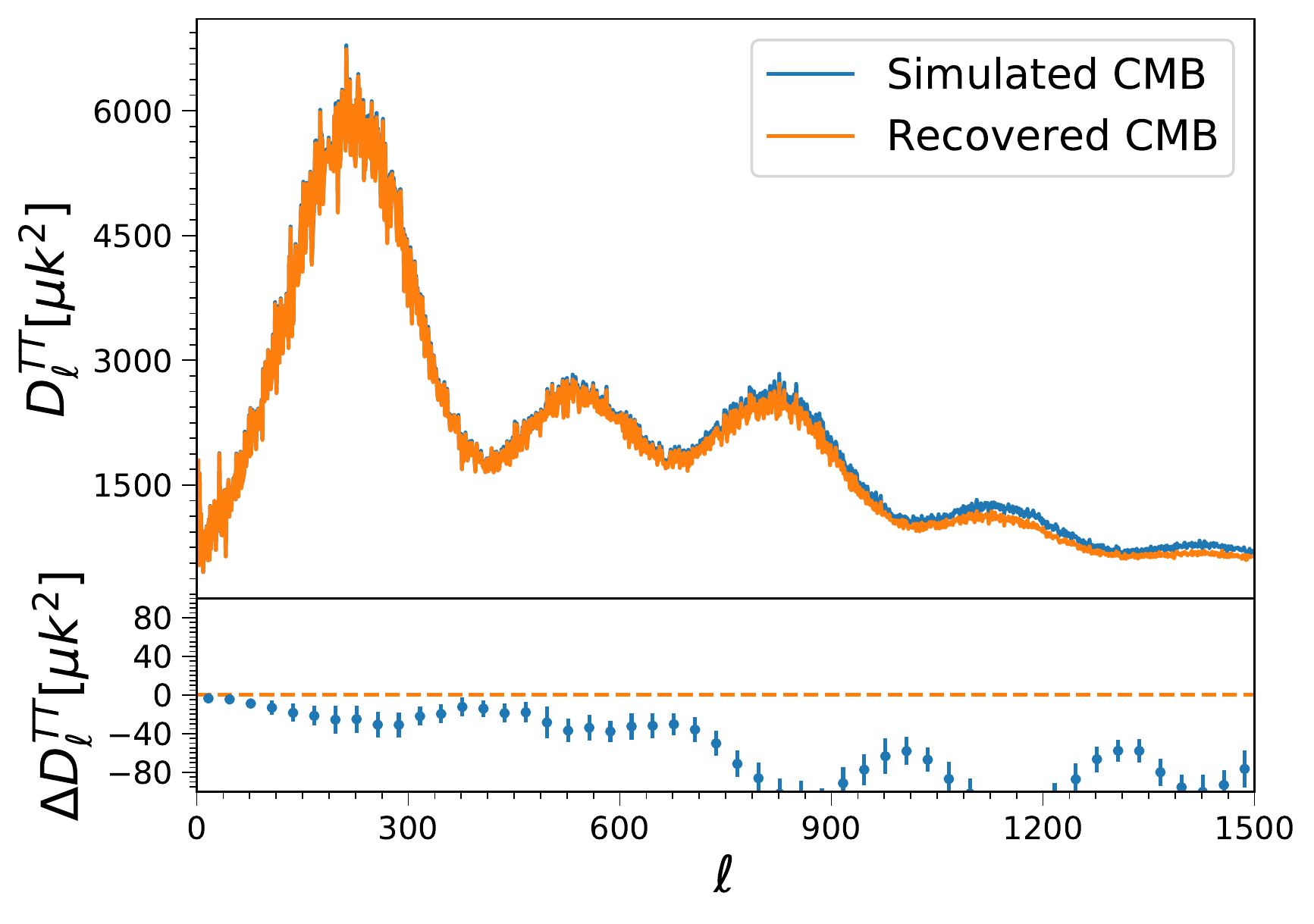}
	\includegraphics[width=0.23\textwidth]{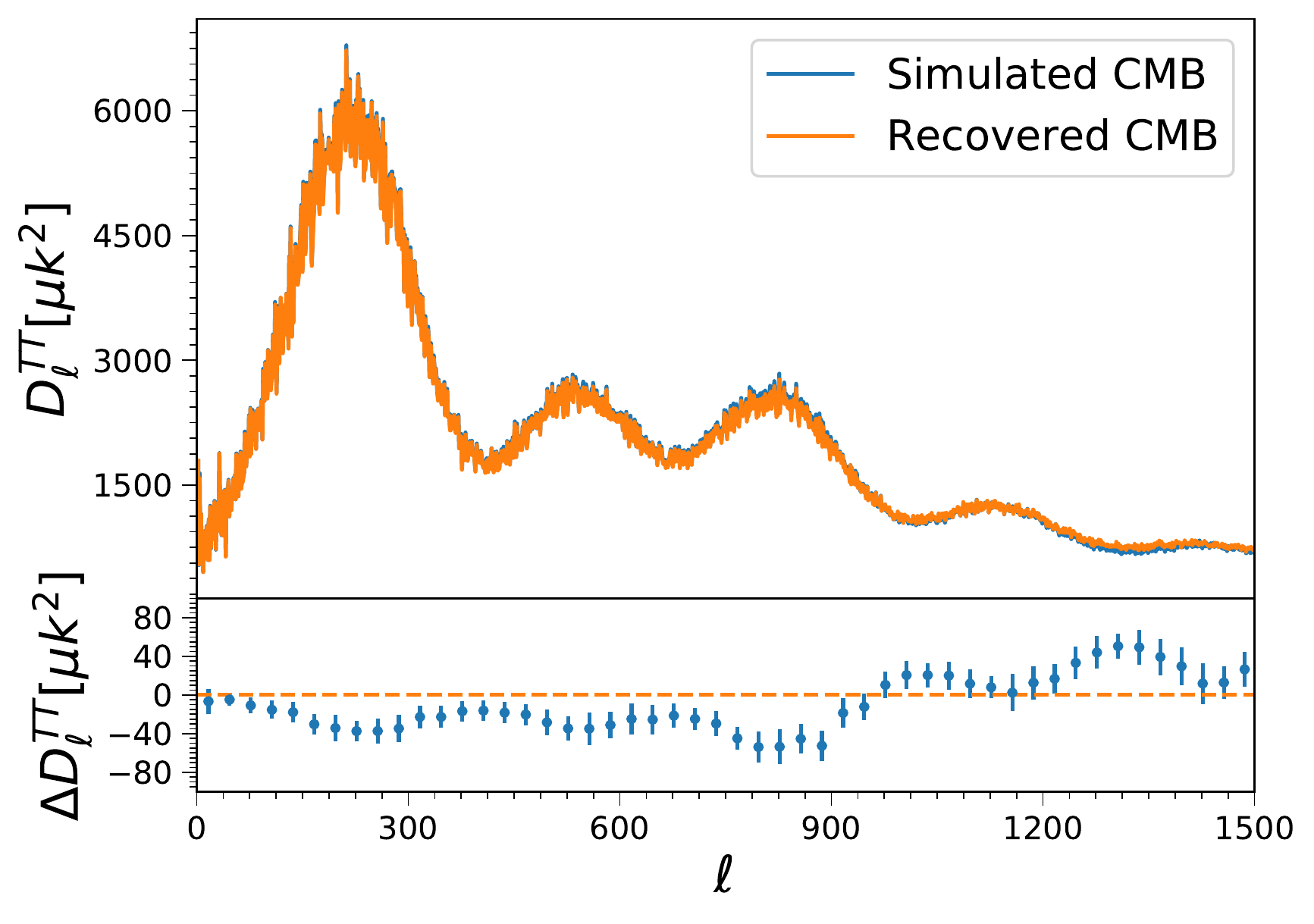}
	\includegraphics[width=0.23\textwidth]{figures/spectra_simcmb_I_10000_038983.pdf}
	\caption{Power spectra of the recovered CMB map (upper panels, using a bin size of 1) and the difference between these spectra and that of the simulated CMB maps (lower panels, using a bin size of 30) when using 1, 2, 3, and 4 frequency bands, respectively.} \label{fig:sim_cmb_spectra_multiFreqs}
\end{figure*}

\begin{figure*}
	\centering	
	\includegraphics[width=0.45\textwidth]{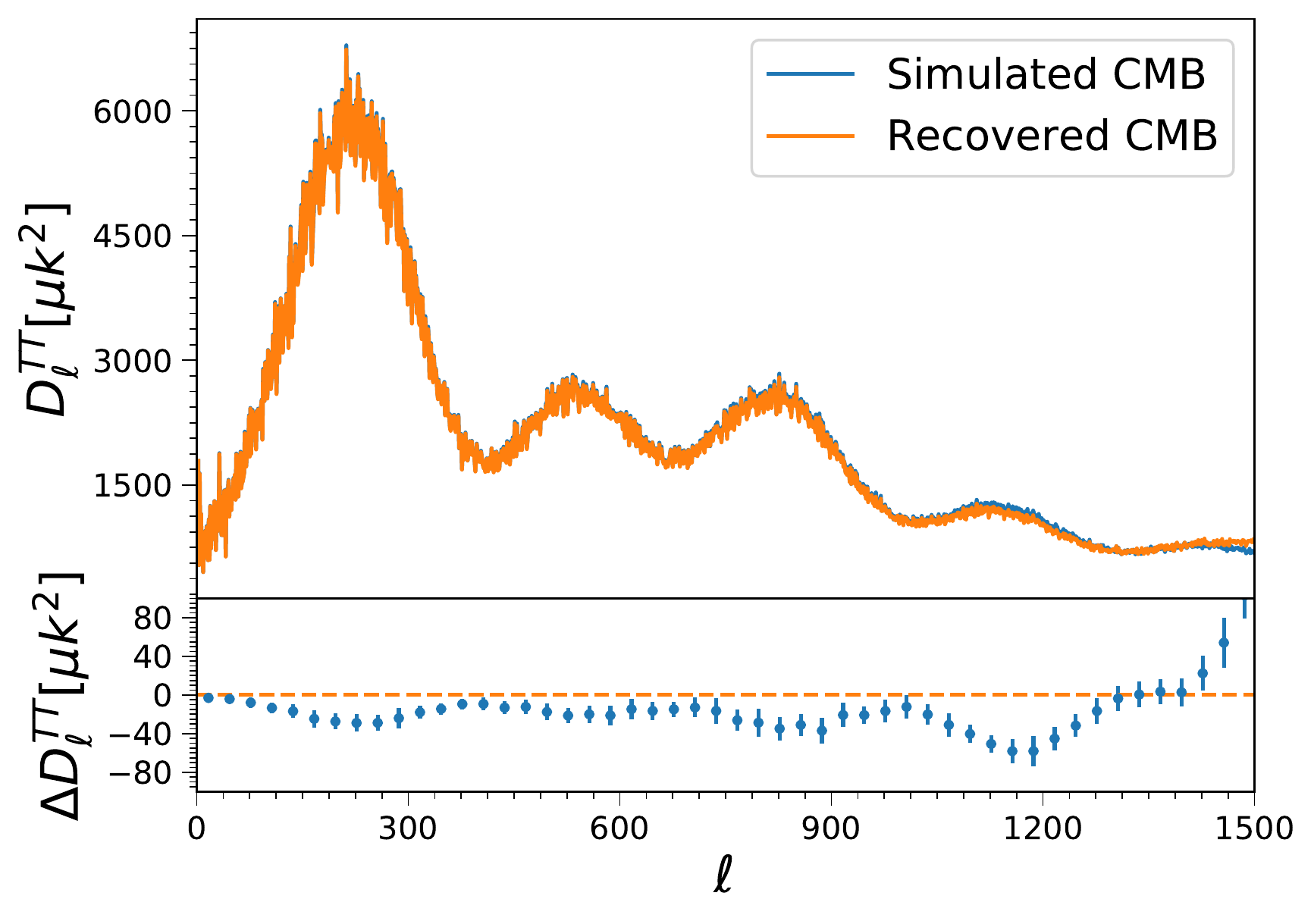}
	\includegraphics[width=0.45\textwidth]{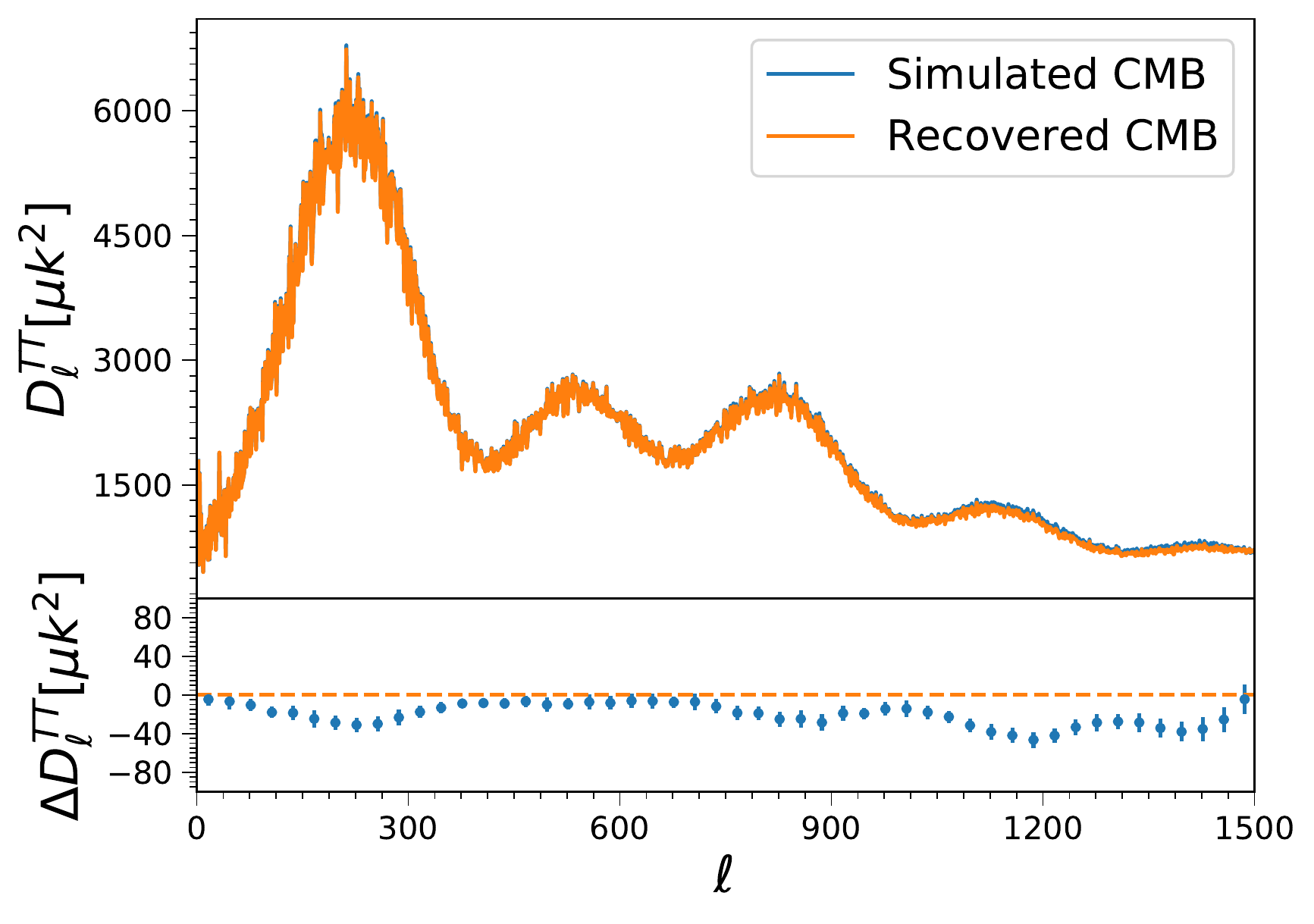}
	\caption{Power spectra of the recovered CMB maps (upper panels, using a bin size of 1), and the difference between these spectra and those of the simulated CMB maps (lower panels, using a bin size of 30), when using two frequency bands. The left panel is for a 50\% noise level and the right panel is for a 30\% noise level.}\label{fig:sim_cmb_spectra_smallNoise}
\end{figure*}

\subsection{Effect of Frequency Bands}\label{sec:effect_of_frequency_bands}

In the above analysis, four frequency bands are used to train the neural network. However, we note that the number of frequency bands may affect the recovered CMB map. Therefore, in this section, we test the effect of frequency bands on the recovery of the CMB signal. Specifically, we select another three cases to train the network, with one, two, and three frequency bands, respectively; thus, there are four cases in total, as shown in Table \ref{tab:frequency_number}. Following the same procedure, we generate the training data using the method of section \ref{sec:simulation}, train the network models for these four respective cases using the method of section \ref{sec:training_process}, and then test them using the corresponding test sets.

In Figure \ref{fig:sim_cmb_spectra_multiFreqs}, we show the recovered power spectrum of one set of maps from the test set for the four respective cases. From left to right, the training set contains one, two, three, and four frequency bands, respectively. Obviously, we can see that as the number of frequency bands increases, the recovered power spectrum gradually matches the true power spectrum. Therefore, the number of frequency bands significantly affects the recovered CMB; and the more frequency bands, the closer the recovered CMB is to the real one. Quantitatively, for the recovered power spectra, the MSEs are 9240.498, 4761.857, 1068.350, and 532.523 $\mu \text{K}^4$, respectively. With the increase in the frequency bands, the MSEs of the power spectra become smaller and smaller, and for the cases that have four frequency bands, the recovered power spectrum can match the true power spectrum very well. This means that for the Planck experiment, the component separation of the CMB can be achieved with four frequency bands.

In addition, we calculate the deviations of power spectrum caused by CNN (see Figure \ref{fig:sim_cmb_spectra_multiFreqs_error_cv}) for all the three cases and compare them with that of case 1. We can see that even all the deviations of these three cases are a little larger than that of case 1, they are all much smaller than the cosmic variance at $\ell>10$. Therefore, this deviation caused by CNN remains a secondary factor when using fewer frequency bands.

\subsection{Effect of Noise}\label{sec:effect_of_noise}

The analysis of section \ref{sec:effect_of_frequency_bands} shows that for the Planck experiment, at least four frequency bands are needed when using the neural network to recover the CMB signal (see Figure \ref{fig:sim_cmb_spectra_multiFreqs}). However, this may not be the case for other experiments because the results can be affected by the noise level of the instrument. To further test the effect of instrument noise on the recovery of the CMB signal, we consider another two different noise levels of the instrument: 50\% and 30\% of the Planck noise level. To more clearly illustrate the effect of noise on the recovery of the CMB signal, we train the network model using two frequency bands.

We first consider the case of 50\% Planck noise level. The recovered power spectrum of one set of maps in the test set is shown in the left panel of Figure \ref{fig:sim_cmb_spectra_smallNoise}. The recovered power spectrum is a little better than that when using Planck noise (see the second panel of Figure \ref{fig:sim_cmb_spectra_multiFreqs}). Moreover, we can clearly see that the recovered power spectrum is more consistent with the truth than that of section \ref{sec:effect_of_frequency_bands}. Therefore, the noise level will greatly affect the recovery of the CMB signal when using the neural network.

Furthermore, we train the network model using data that have a 30\% Planck noise level, and the recovered power spectrum of one set of maps from the test set is shown in the right panel of Figure \ref{fig:sim_cmb_spectra_smallNoise}. We can see that the recovered power spectrum is quite consistent with the true spectrum at both small scales and large scales, which is similar to those using four frequency bands (see Figure \ref{fig:sim_cmb_spectra}). Furthermore, the recovered power spectrum is much better than that using the Planck noise, and even better than that using a 50\% Planck noise level. This indicates that for CMB experiments that have lower noise levels, the neural network is capable of recovering the CMB signal using only two frequency bands. Therefore, this means that the neural network will play an important role in the component separation of future higher-precision CMB experiments.

Moreover, following the same procedure, we calculate the deviations of the power spectra using the training set, as shown in Figure \ref{fig:sim_cmb_spectra_multiFreqs_error_cv}. The deviations are much smaller than the cosmic variance at $\ell>10$, and also smaller than the cases using different frequency bands, in previous sections. This means that the deviations of the power spectra caused by the CNN will be reduced when there are smaller instrument noise levels.

\begin{figure}
	\centering
	\includegraphics[width=0.45\textwidth]{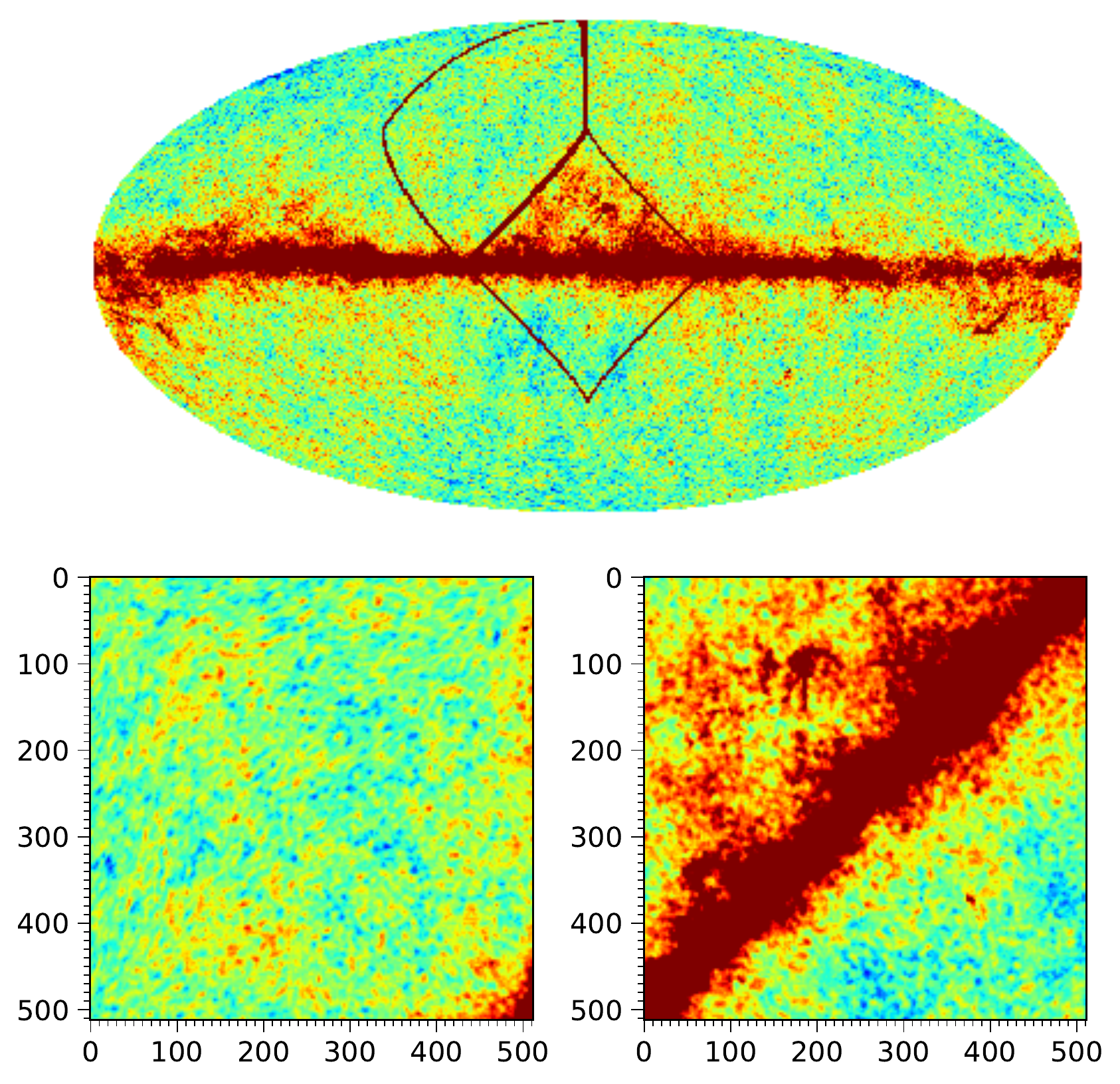}
	\caption{The two blocks (the areas inside the brown lines) of the full-sky contaminated CMB map used to train the neural network.} \label{fig:train_block}
\end{figure}

\subsection{Effect of the Galaxy Foregrounds}\label{sec:effect_of_the_Galaxy}

Besides the effects of instrument noise and frequency bands, the foregrounds from our Galaxy constitute the other main factor. To prove this, we use two parts of the full-sky map to test the CNN method. Specifically, the areas inside the brown lines in Figure \ref{fig:train_block} show two blocks of the full-sky map of the CMB observations, and we train the network model using these two blocks, respectively. The first block is at high galactic latitudes, which are far away from the Galaxy plane and are less contaminated by Galaxy foregrounds. The second block is near the Galaxy plane, and is greatly contaminated. Note that the size of the image is $512\times 512$, which is smaller than the image using the full-sky map. Therefore, the network model here actually contains five convolutional layers and five deconvolutional layers.

First, we train a network model for the first block. After selecting the optimal model with the validation set, we feed maps of the test set to the network model to obtain a series of foreground-cleaned CMB maps. We show one set of maps from the test set in Figure \ref{fig:sim_cmb_map_block0}. The upper left panel is the simulated CMB map, and the upper right panel is the CMB map recovered by the network. We can see that these two maps are almost identical, which can also be concluded from the residual map in the lower left panel. To further look at the recovery of the CMB map at a smaller scale, we select three patches with $3\times3$ deg$^2$ that have different latitudes, as shown in Figure \ref{fig:sim_cmb_map_block0_miniPatch}. We can see that the recovered CMB maps are almost the same as the simulated ones for all the three patches, and that there is little information left in the residual maps. Furthermore, we calculate the power spectrum of the recovered CMB map, and the difference between this spectrum and that of the simulated CMB map, as shown in the lower right panel of Figure \ref{fig:sim_cmb_map_block0}. We can see that the recovered CMB signal is quite consistent with the true spectrum at both small scales and large scales, which is similar to the results of section \ref{sec:recover_sim_cmb} (see Figure \ref{fig:sim_cmb_spectra}). Therefore, this indicates that the CNN method can also be used for the component separation of partial-sky experiments, thereby increasing the practicability of the method.

Then, using the same procedure, we train another network model for the second block, and test it with the corresponding test set. In Figure \ref{fig:sim_cmb_map_block4}, we show one set of maps from the test set. It can be seen from the residual map that, compared with the first block (Figure \ref{fig:sim_cmb_map_block0}), the recovered CMB map is greatly affected by the Galaxy foregrounds. Figure \ref{fig:sim_cmb_map_block4_miniPatch} shows three patches with $3\times3$ deg$^2$, selected from different latitudes. For the first and third patches, the recovered CMB maps are almost the same as the simulated ones, while for the second patch, the recovered CMB map is slightly different from the simulated one, and there is a lot information in the residual map. The reason is that this patch is located in the galactic plane. From the lower right panel of Figure \ref{fig:sim_cmb_map_block4}, we can see that the recovered power spectrum of the CMB is consistent with the ground truth at $\ell<1000$. However, the recovered power spectrum greatly deviates from the ground truth at $\ell>1000$. This means that the Galaxy foregrounds have less influence on the large-scale CMB information, but have greater influence on the small-scale CMB information.

\setcounter{figure}{15}
\begin{figure}
	\centering
	\includegraphics[width=0.45\textwidth]{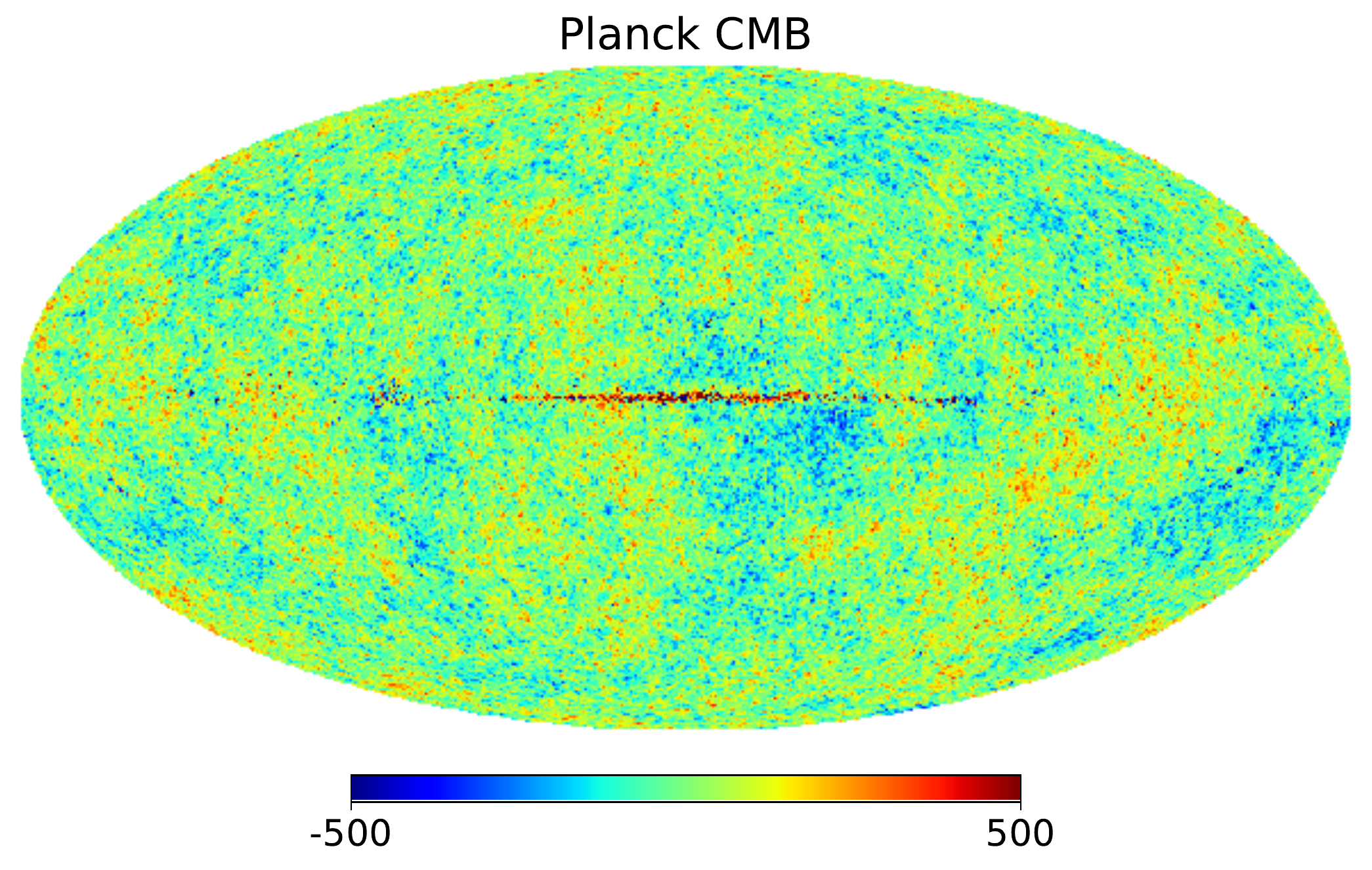}
	\includegraphics[width=0.45\textwidth]{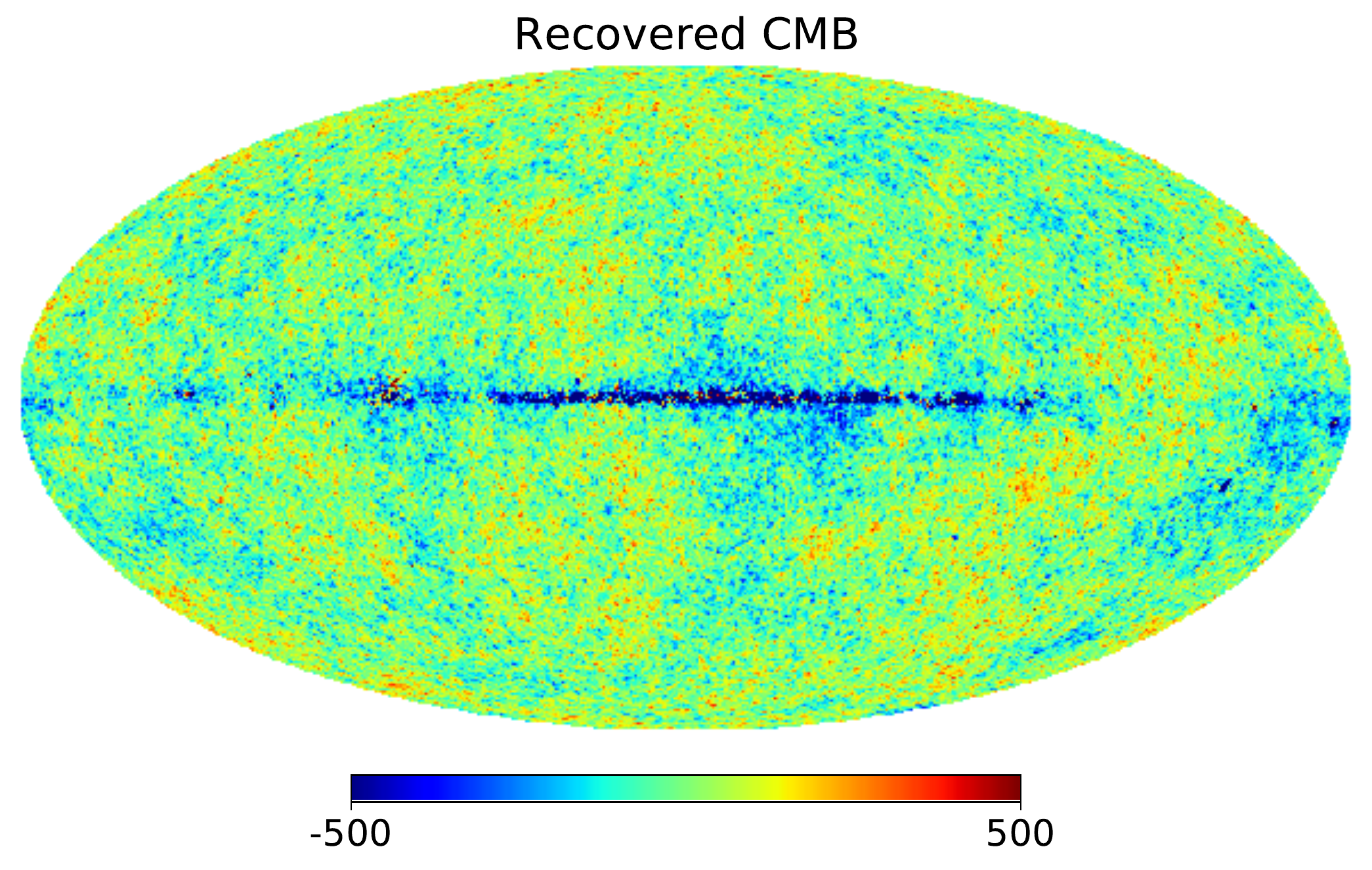}
	\includegraphics[width=0.45\textwidth]{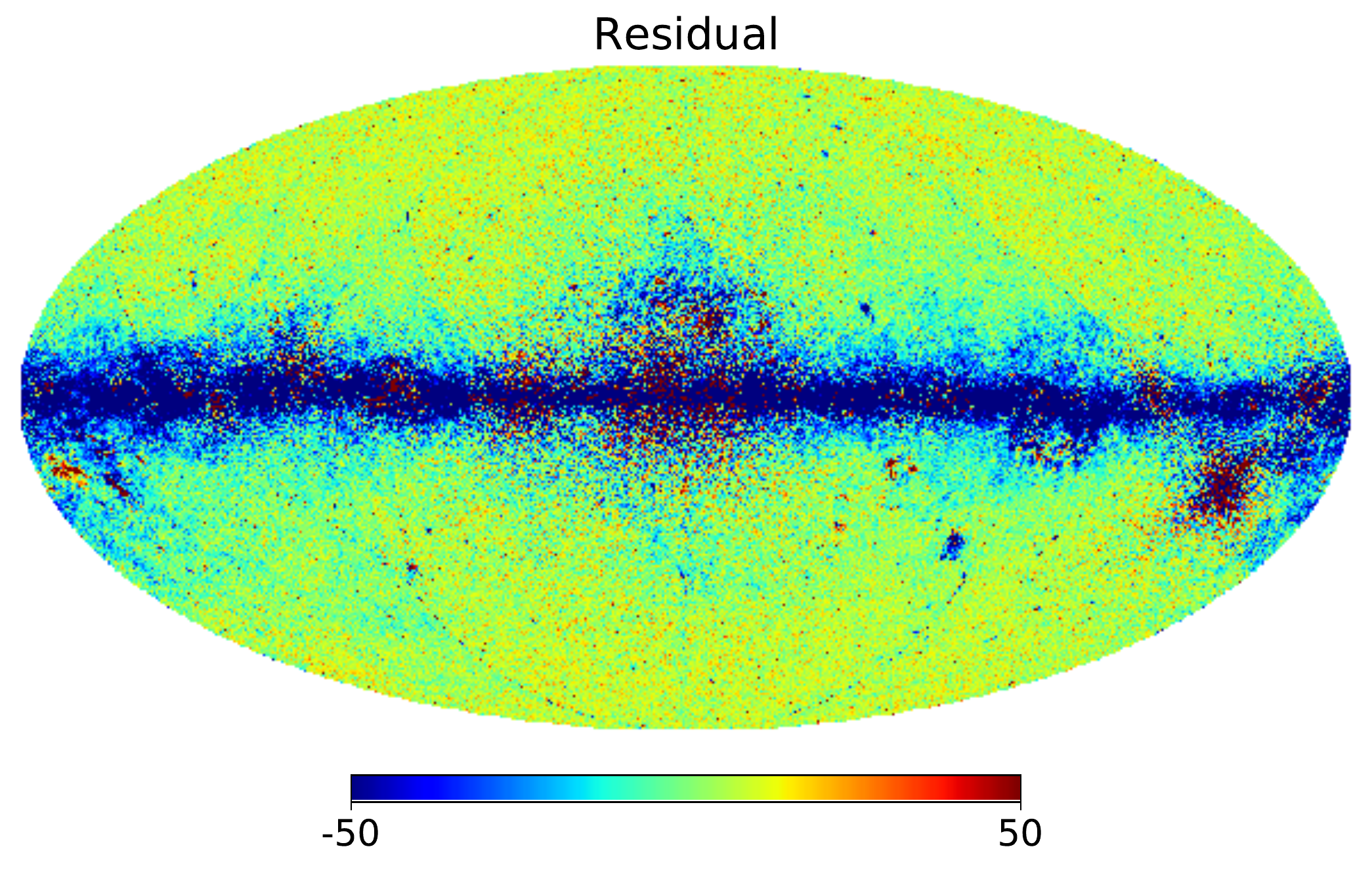}
	\caption{Planck CMB temperature map recovered by the neural network. {\it Upper panel:} the Planck Commander foreground-cleaned CMB map. {\it Middle panel:} the CMB map recovered by the neural network. {\it Lower panel:} the residual map of the recovered CMB map.} \label{fig:planck_cmb_map}
\end{figure}

\begin{figure*}
	\centering
	\includegraphics[width=0.9\textwidth]{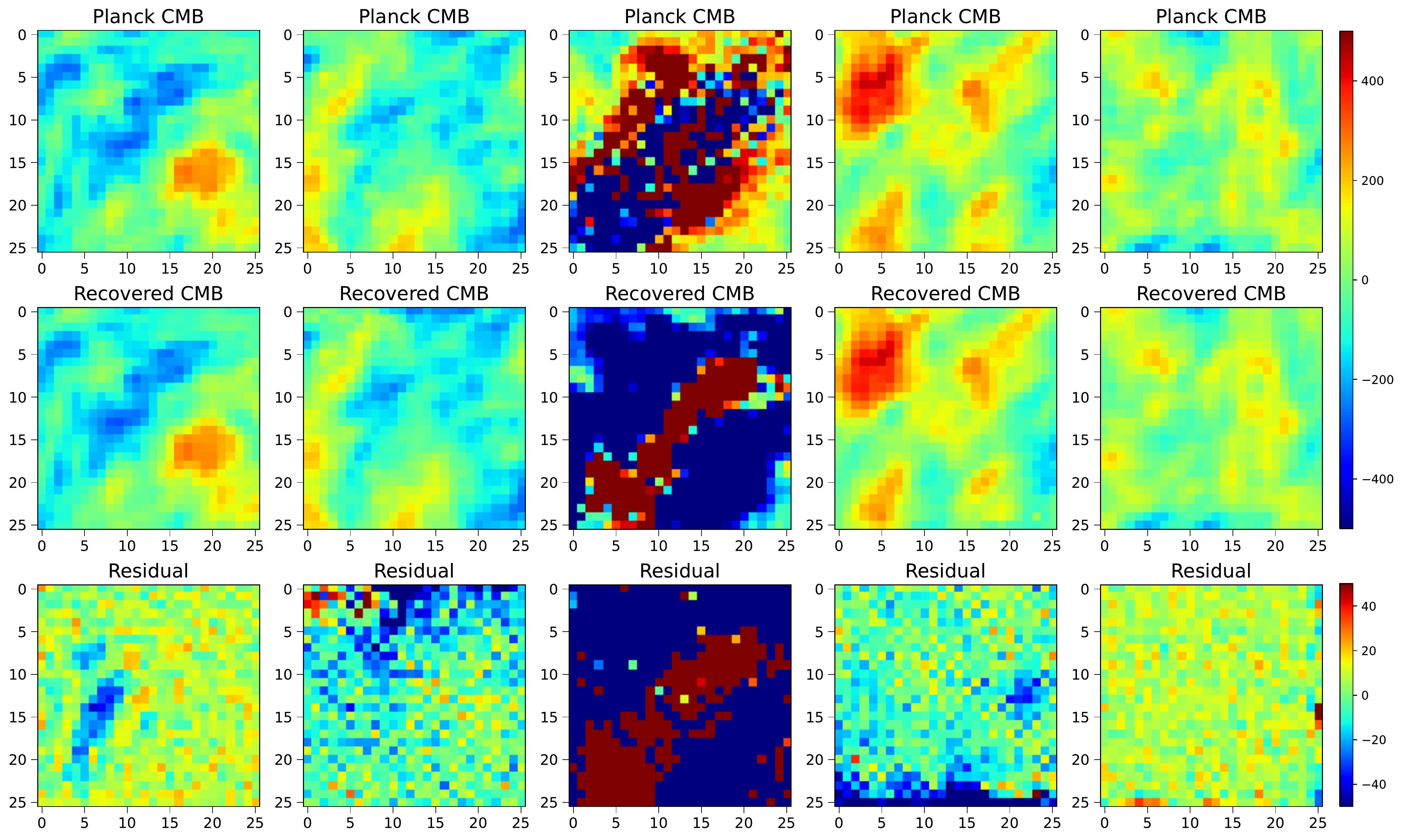}
	\caption{Five small patches with $3\times3$ deg$^2$, selected from Figure \ref{fig:planck_cmb_map}. These patches are selected from the north pole to the south pole, with different latitudes.} \label{fig:planck_cmb_map_miniPatch}
\end{figure*}

\begin{figure}
	\centering
	\includegraphics[width=0.45\textwidth]{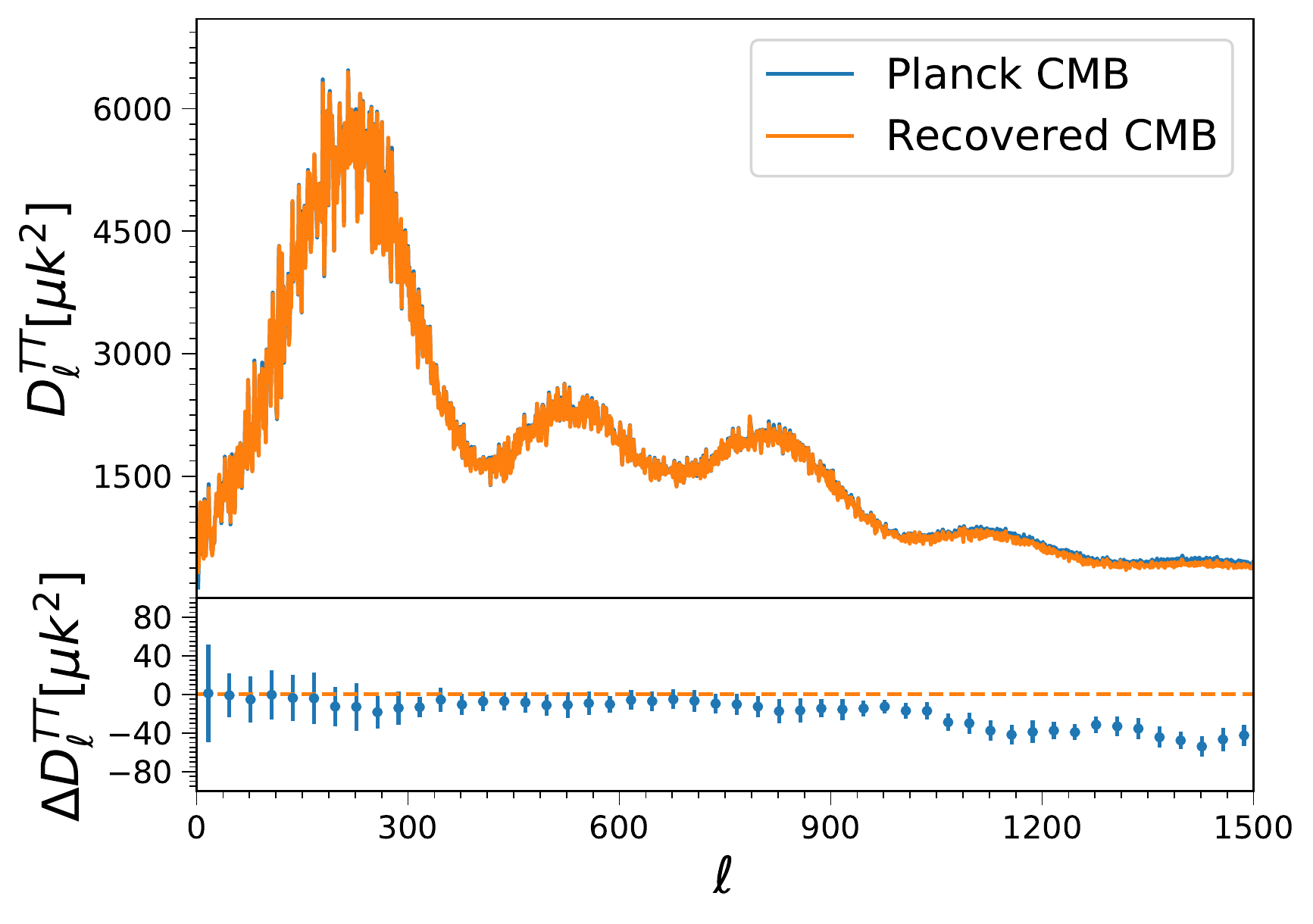}
	\caption{Power spectrum of the recovered Planck CMB map (upper panel, using a bin size of 1), and the difference between this spectrum and that of the Planck Commander foreground-cleaned CMB map (lower panel, using a bin size of 30).} \label{fig:planck_cmb_spectra}
\end{figure}

\begin{figure}
	\centering
	\includegraphics[width=0.45\textwidth]{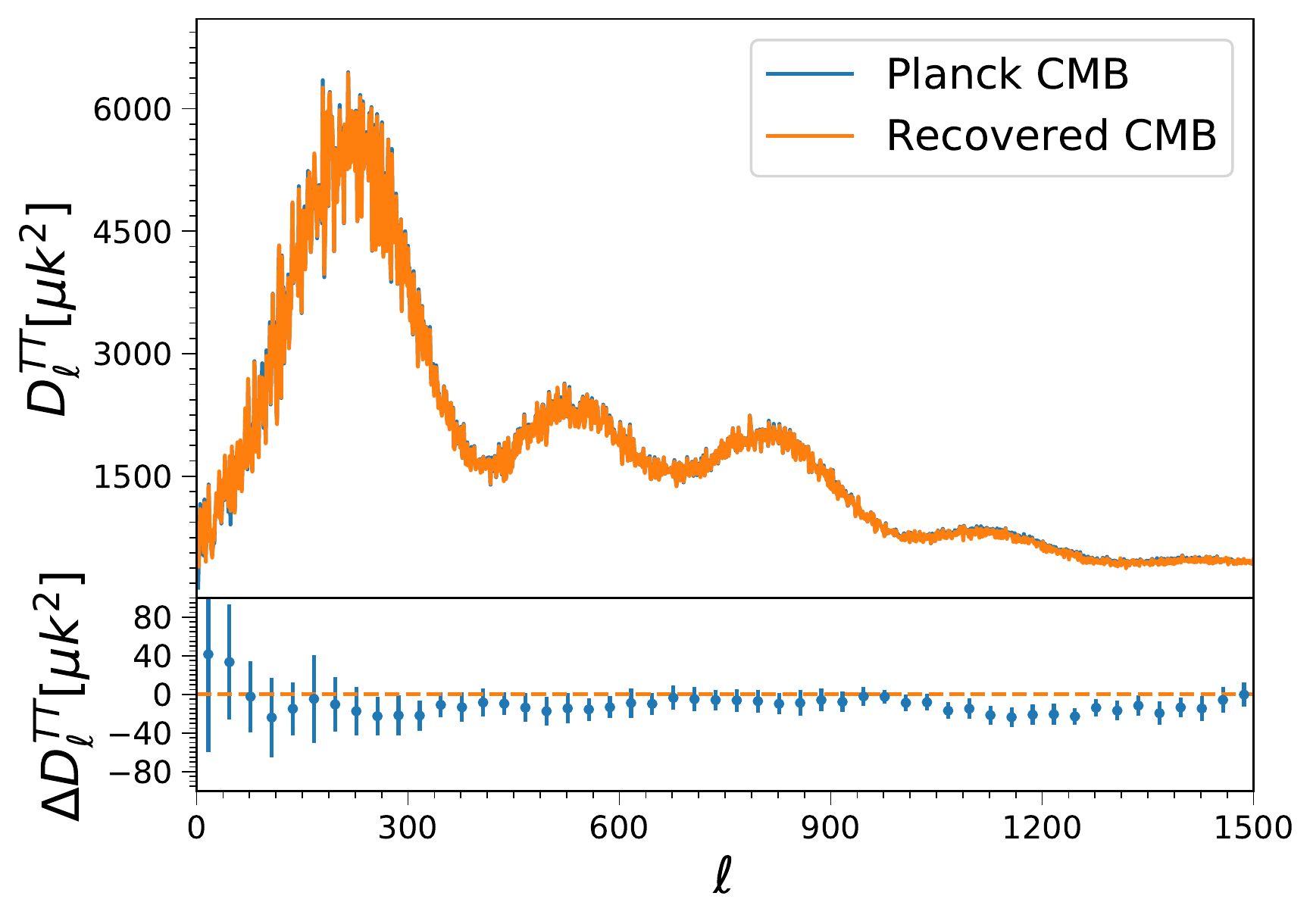}
	\caption{The same as Figure \ref{fig:planck_cmb_spectra}, but now the Planck mask is applied to the training set.} \label{fig:planck_cmb_spectra_trainWithMask}
\end{figure}

\begin{figure*}
	\centering	
	\includegraphics[width=0.9\textwidth]{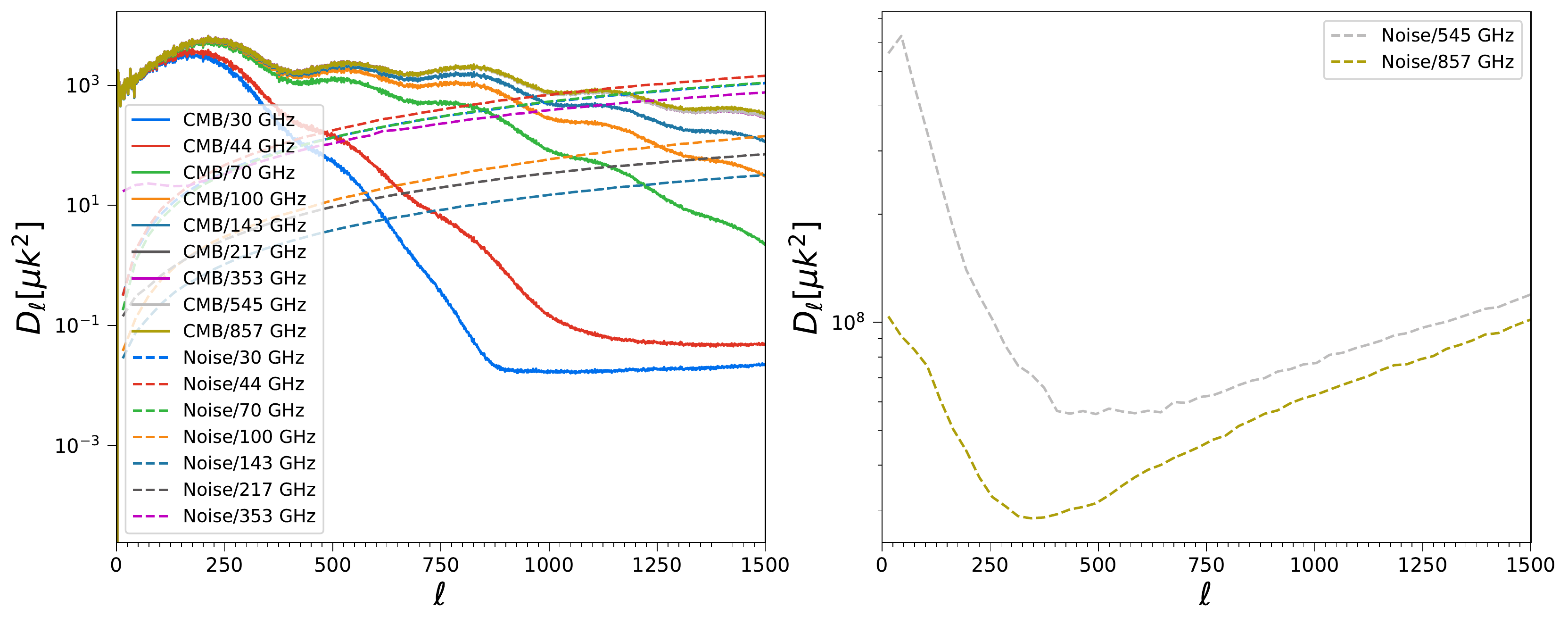}
	\caption{Power spectra of a simulated CMB map, considering Planck beam effects, and power spectra of instrument noise at nine observational frequency bands.}\label{fig:analysis_cmb_noise}
\end{figure*}

\subsection{Test with Planck CMB Maps}\label{sec:recover_planck_cmb}

In view of the success of the CNN in recovering the CMB signal from various huge foreground contaminations, we finally apply this pipeline to the Planck mission. As a preliminary test, we only consider the recovery of CMB temperature fluctuations. Specifically, we apply the CNN model obtained in section \ref{sec:recover_sim_cmb} to the Planck full mission datasets. Therefore, the inputs of the network are four-frequency contaminated CMB maps that contain the beam and noise information, while the target of the network is a foreground-cleaned CMB map with the beam FWHM=7$^\prime$.27. 

The main results are shown in Figure \ref{fig:planck_cmb_map}. The upper panel is the Planck Commander foreground-cleaned CMB map, which is publicly available on the Planck Legacy Archive and is considered the ground truth; the middle panel is the CMB map recovered by the neural network; and the lower panel is the corresponding residual map. We can see that the CMB map recovered by the neural network is quite similar to the Commander-based CMB map at high galactic latitudes, but in the low galactic latitudes, especially close to the galactic plane, there is a little difference, which can also be seen from the residual map. Following the same method in section \ref{sec:recover_sim_cmb}, we select five small patches with $3\times3$ deg$^2$ from the north pole to the south pole, shown in Figure \ref{fig:planck_cmb_map_miniPatch}. We can see that the recovered CMB maps are consistent with the simulated ones for the first, second, fourth, and fifth patches, while the recovered CMB map is different from the simulated one for the third patch, which should be caused by the foreground contaminations in the galactic plane.

For a quantitative comparison, we calculate the power spectra of the recovered CMB and Planck CMB maps, shown in the upper panel of Figure \ref{fig:planck_cmb_spectra}. Note that in the galactic plane, the CMB signal is not recovered very well for either the Commander-based CMB or the CNN-based CMB maps; therefore, the Planck mask is applied to the CMB maps in the calculation of the power spectra. Here, the obtained power spectrum $D_\ell$ is divided by the Gaussian beam window function $B(\ell)$, thus the beam information has been removed from the power spectra. Obviously, these two power spectra are almost identical. Furthermore, we calculate the difference between these two power spectra, shown in the lower panel of Figure \ref{fig:planck_cmb_spectra}. We can see that the difference is quite small for $\ell<1000$, while there is a deviation for $\ell>1000$. This may be caused by the application of the Planck mask to the recovered CMB map, because the Planck mask was not used when training the network. A more reasonable approach might be to apply the Planck mask to the training set, and then train the network using the masked data. To do this, we train a new network model using the masked data, and apply it to the Planck datasets to calculate the power spectrum. The recovered power spectrum is shown in Figure \ref{fig:planck_cmb_spectra_trainWithMask}. Obviously, this result is better than that in Figure \ref{fig:planck_cmb_spectra} for $\ell>1000$, and it is also better than that of \citet{Petroff:2020}. Therefore, this illustrates that the CNN method can successfully recover the CMB signal from the Planck data sets, and it also indicates the feasibility of this method in processing observational data. 

However, it should be noted that only four main foreground components are considered in the training set, and we are here only proving the feasibility of this method in processing observational datasets, thus all these analyses are simple and rough. Therefore, when applying this method to observational datasets, a detailed analysis using more comprehensive foreground models should be elaborated in future works.

\section{Discussions}\label{sec:discussion}

\subsection{Frequency Selection for Planck Mission}

In sections \ref{sec:recover_sim_cmb} and \ref{sec:effect_of_noise}, we have shown that both instrument noise and beam effects have an influence on the recovery of the CMB signal. This means that either strong instrument noise or strong instrument beam effects will cause the CMB signal extracted by the CNN to be biased. This is the main reason why only four frequency bands are considered in our analysis.

In order to show the instrument noise and beam effect of the Planck satellite, we plot in Figure \ref{fig:analysis_cmb_noise} the power spectra of a simulated CMB map, considering Planck beam effects and the power spectra of instrument noise at nine observational frequency bands. For 30 GHz and 44 GHz, a lot of the CMB information at $\ell>600$ is lost due to the beam effect. Therefore, these two frequency bands can almost only provide large-scale information. Furthermore, the noise power spectra are larger than the power spectra of the CMB at $\ell>500$, which makes it difficult for the CNN to extract the CMB signal at $\ell>500$. Therefore, these two frequency bands are not considered in our analysis. For 70 GHz, the CMB information is retained at $\ell<1000$, but the noise power spectrum becomes larger than that of the CMB at $\ell>850$; thus, this frequency band is also not considered in our analysis. For 545 GHz and 857 GHz, we can see that the noise power spectra are much larger than those of the CMB signal in the order of magnitude. Therefore, these two frequency bands should not be used, even though small-scale information is not affected by the beam effects. Therefore, considering the instrument noise and beam effects, only four frequency bands (100 GHz, 143 GHz, 217 GHz, and 353 GHz) are considered in our analysis.

We note that the effective beam effect of the output CMB map should also be selected reasonably. For the four selected frequency bands, the effective Gaussian beams are 9$^\prime$.66, 7$^\prime$.27, 5$^\prime$.01, and 4$^\prime$.86, respectively. Obviously, it is not reasonable to select a beam effect with FWHM<4$^\prime$.86, since the CNN will not generate additional CMB information. On the contrary, it is reasonable to select a larger value of FWHM among these four frequency bands. Therefore, an effective beam effect with FWHM=7$^\prime$.27 is considered in our analysis, which is reasonable for recovering the CMB signal. Therefore, we propose that for other current or future CMB experiments, the training data should also be determined according to the specific instrument noise and beam effect when using this method.

\begin{figure*}
	\centering	
	\includegraphics[width=0.45\textwidth]{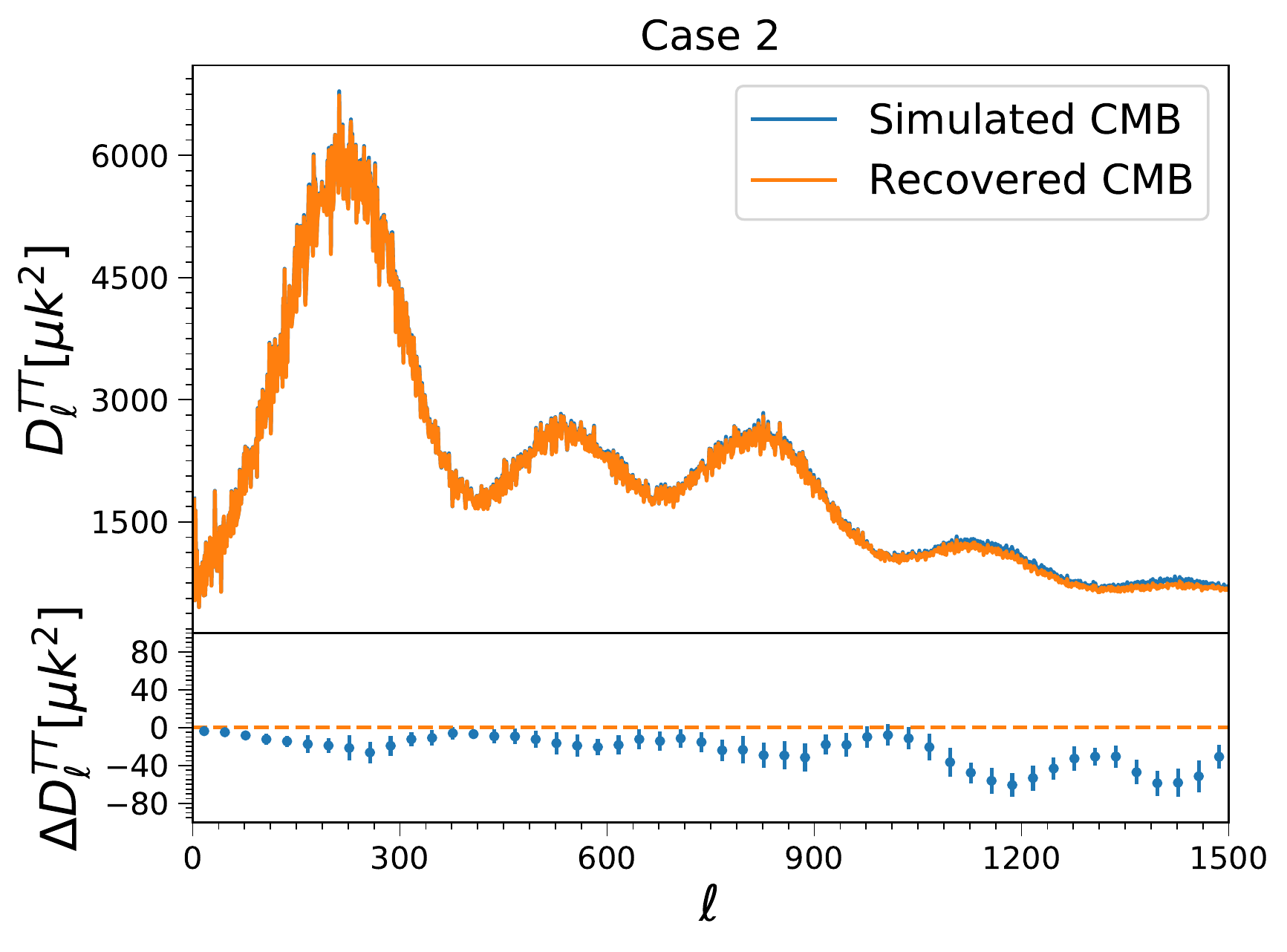}
	\includegraphics[width=0.45\textwidth]{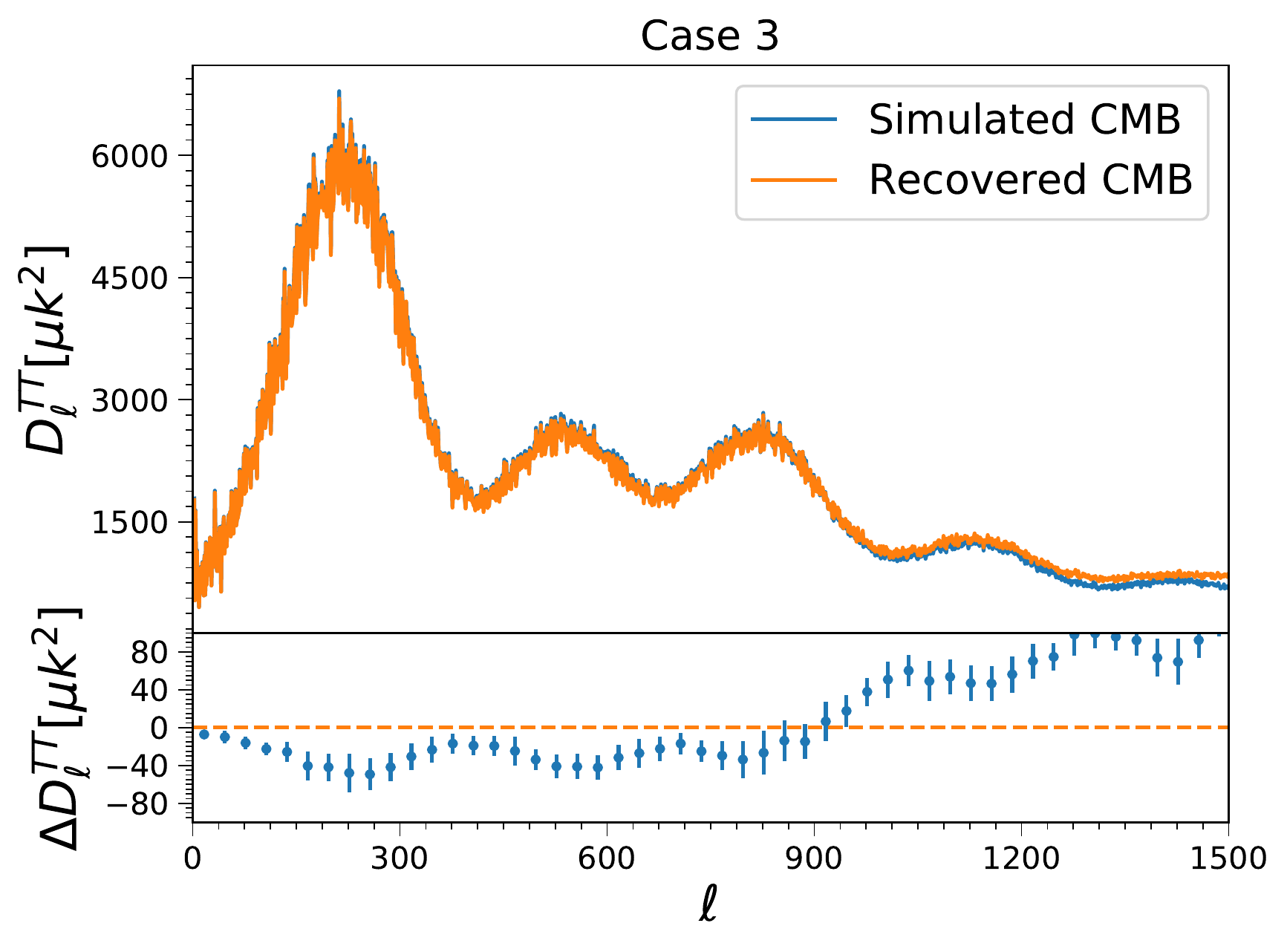}
	\includegraphics[width=0.45\textwidth]{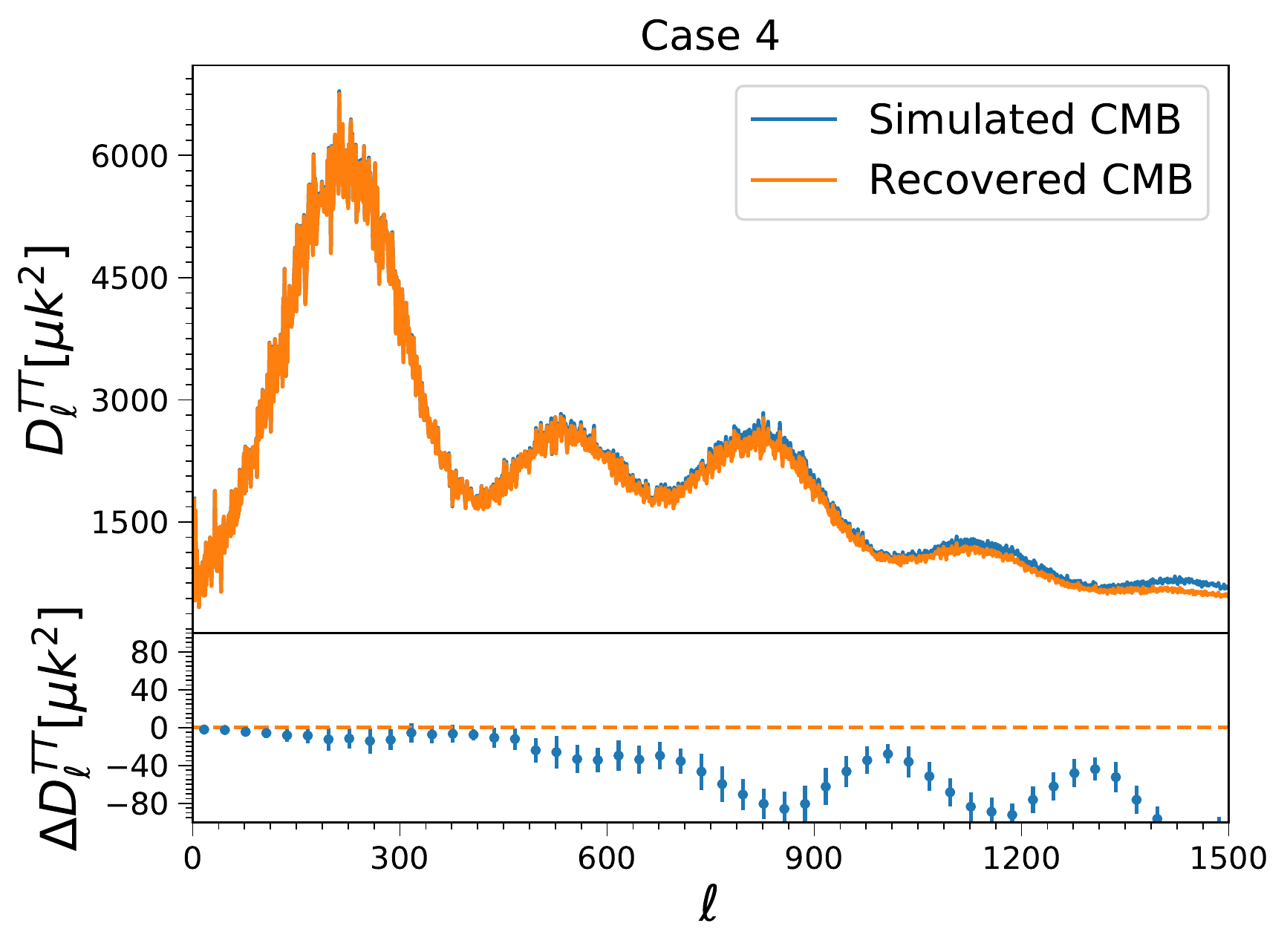}
	\includegraphics[width=0.45\textwidth]{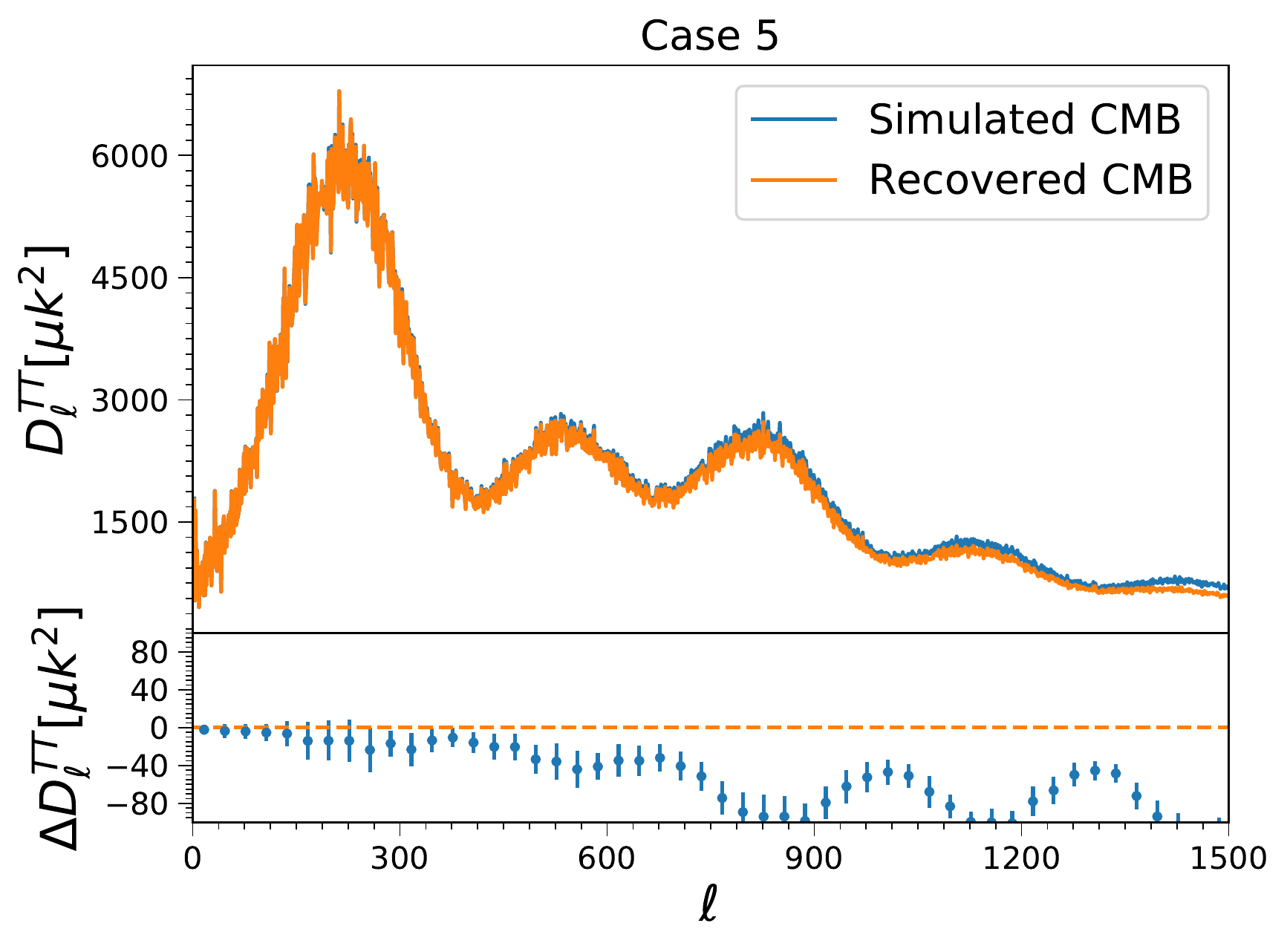}
	\caption{Power spectra of the recovered CMB maps, and the difference between these spectra and those of the simulated CMB maps, for cases 2, 3, 4, and 5, respectively.}\label{fig:vary_fg_params_cmbPowerSpectra}
\end{figure*}

\begin{figure*}
	\centering
	\includegraphics[width=0.45\textwidth]{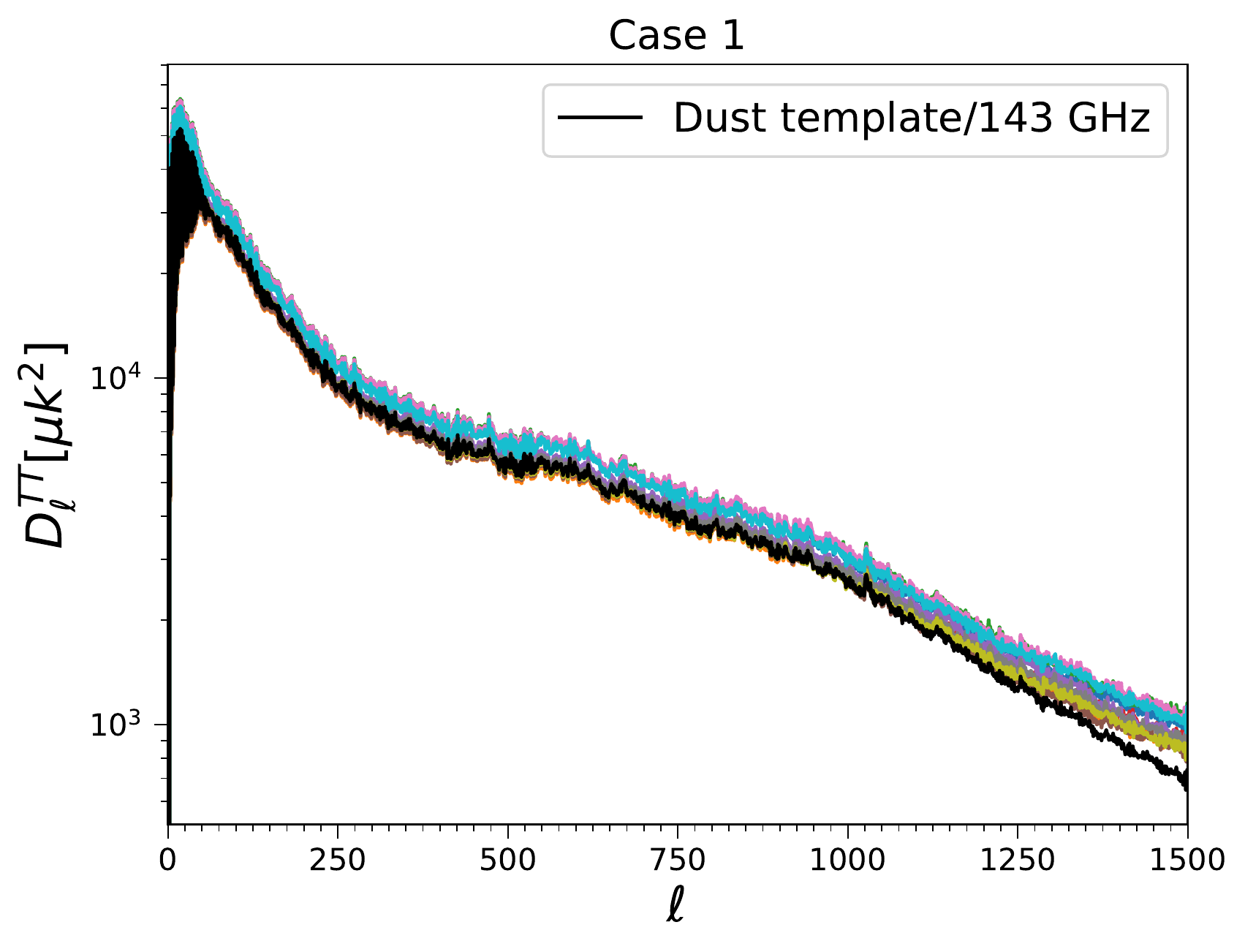}
	\includegraphics[width=0.45\textwidth]{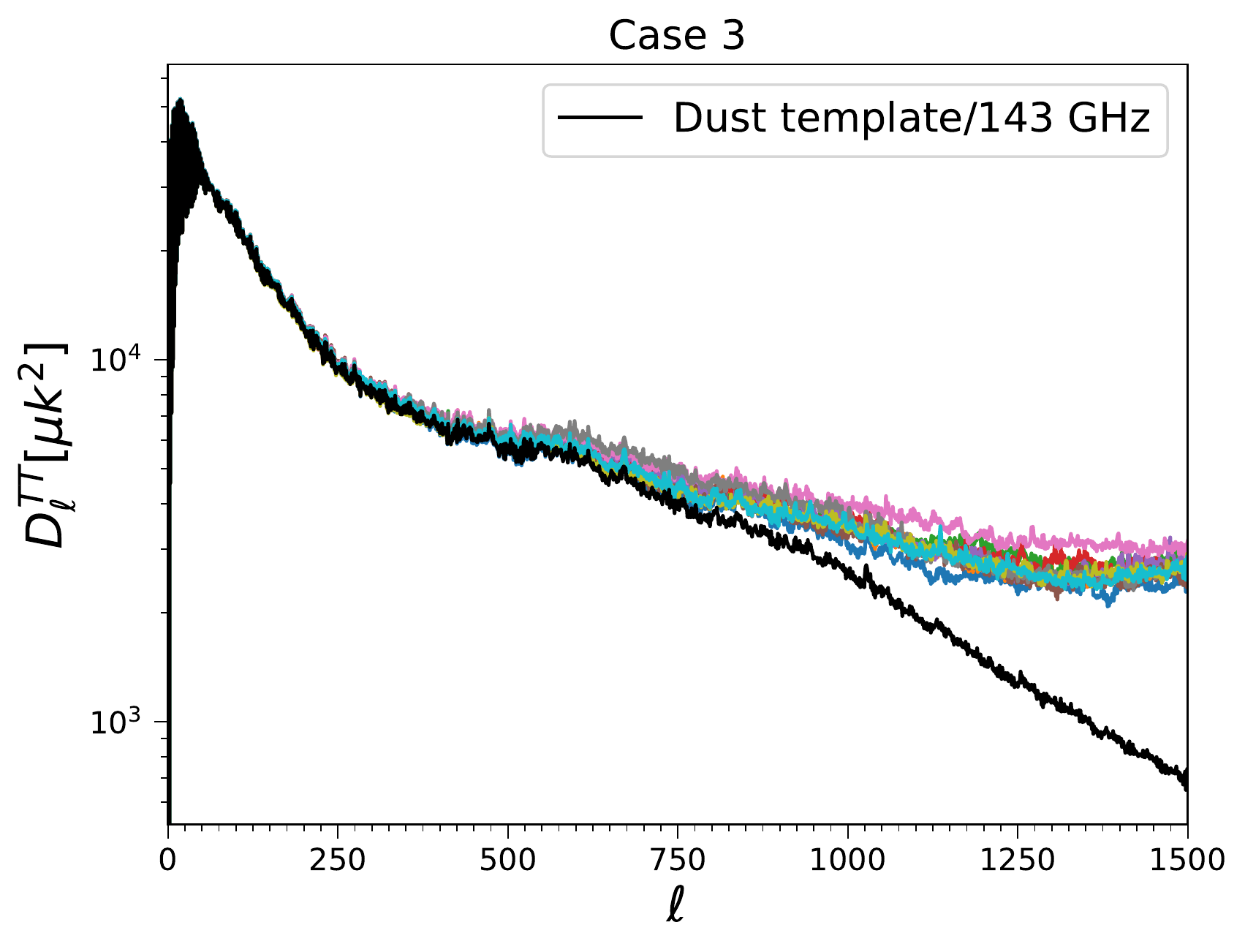}
	\includegraphics[width=0.45\textwidth]{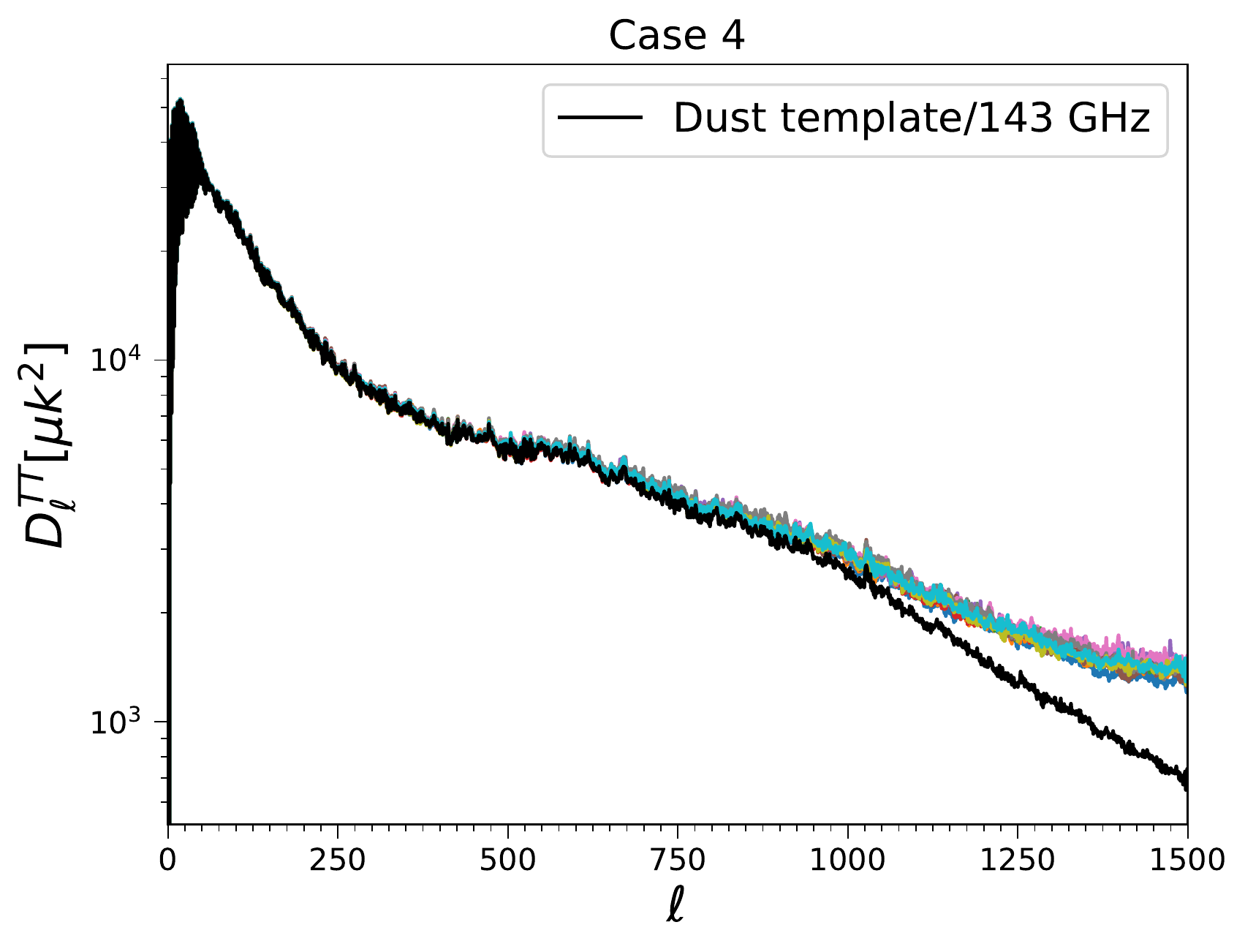}
	\includegraphics[width=0.45\textwidth]{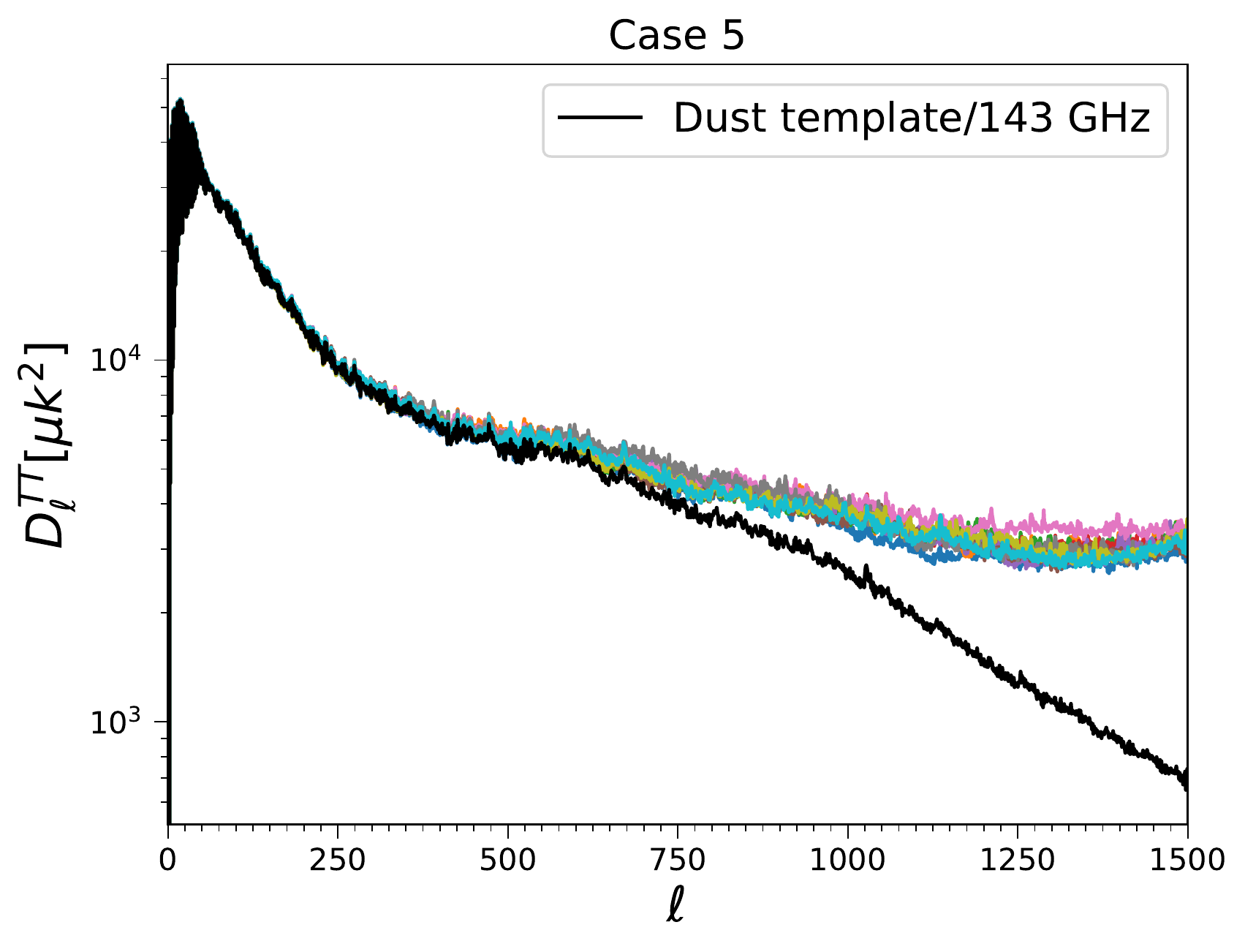}
	\caption{Power spectra of thermal dust maps for cases 1, 3, 4, and 5, respectively. The black lines refer to the power spectrum of the dust template, while the others are the power spectra of ten sets of simulated thermal dust maps.}\label{fig:vary_fg_params_fgPowerSpectra}
\end{figure*}

\subsection{Variability of Foregrounds}\label{sec:variability_of_foregrounds}

\begin{table}
	\centering
	\caption{Cases of Variations of the Spectral Parameters, Amplitude $A$ and Spectral Index $\beta$ (see Equation (\ref{equ:spectral_params})) of the Foregrounds, and the Corresponding Results on the Test Data.}\label{tab:cases_of_spectral_params}
	\begin{tabular}{c|c|c|c}
		\hline\hline
		& $A$ & $\beta$ & Random Pixel \\
		&  &  & Independently \\
		\hline
		Case 1 & ... & 10\% & No \\
		Case 2 & ... & ... & ... \\
		Case 3 & 10\%  & ... & Yes \\
		Case 4 & ... & 10\% & Yes \\
		Case 5 & 10\%  & 10\% & Yes \\
		\hline\hline
	\end{tabular}\\
	Note: $10\%$ refers to the standard deviation of the spectral parameters relative to the template parameter value, and ``random pixel independently'' means whether a pixel is randomized independently or not.
\end{table}

Since the foregrounds are much larger than the CMB signal in the order of magnitude, foregrounds are very important elements in the component separation of CMB. In addition, the CNN method is a data-driven method that strongly relies on simulated data. Therefore, when generating training data, the foregrounds should be simulated reasonably and correctly. Considering the fact that the spectral parameters of the foregrounds will vary across the sky, the spectral parameters should be generated randomly when simulating the foregrounds. For the simulation of the foregrounds in section \ref{sec:simulation}, we only considered the variation of the spectral index $\beta$, with a $10\%$ standard deviation. In this section, we will consider more cases of the variation of the spectral parameters to test the impact of this on the recovery of the CMB signal.

In table \ref{tab:cases_of_spectral_params}, we show five cases of variations of the spectral parameters. Case 1 refers to the method illustrated in section \ref{sec:simulation}, which only considers the variation of the spectral index $\beta$, with a $10\%$ standard deviation. For case 2, there are no variations in either the amplitude $A$ or the spectral index $\beta$, which means that the template of the foregrounds in PySM is used directly for the training data. For cases 3 and 4, we only consider the variation of the amplitude or the spectral index, respectively. For case 5, we simulate data by varying both the amplitude and the spectral index. It should be noted that for case 1, all pixels are varied together, which means that all the pixels of a map share the same random number. On the contrary, for cases 2, 3, 4, and 5, the pixels are varied independently, which means that each pixel of a map has an independent random number.

Following the same procedure as the analysis above, for case 2, we generate training data and train and test the network. In the upper left panel of Figure \ref{fig:vary_fg_params_cmbPowerSpectra}, we show the recovered CMB power spectrum, and the difference between this power spectrum and that of the simulated one. We can see that the recovered power spectrum is a little lower than the ground truth, especially for $\ell>1000$, when compared with case 1 (see Figure \ref{fig:sim_cmb_spectra}). We note that only one set of foregrounds (the template) is used in the training data. The results may indicate that, compared with case 1, the variation of the spectral parameters will improve the robustness of the network and increase the accuracy of the recovered power spectrum.

For case 3, we generate new data and train another network model to recover the CMB signal. The results are shown in the upper right panel of Figure \ref{fig:vary_fg_params_cmbPowerSpectra}. We can see that the recovered power spectrum becomes larger than the ground truth at $\ell>900$, and gradually deviates with the increase in multipoles, which is unacceptable.

The main difference between case 3 and case 1 is the variation of the spectral parameters. To illustrate this difference, take the thermal dust at 143 GHz as an example---we show the power spectra of ten sets of samples in Figure \ref{fig:vary_fg_params_fgPowerSpectra}. The black lines refer to the power spectrum of the dust template, while the other colored lines are the power spectra of the simulated maps. For case 1, the simulated power spectra are almost coincident with the template, at both large scale and small scale. However, for case 3, the simulated power spectra are coincident with the template at large scale, but gradually deviate from the template at $\ell>900$. It looks like the simulated thermal dust is obtained by adding a noise map to the template. As we have shown in section \ref{sec:effect_of_noise}, noise will greatly affect the recovery of the CMB signal, thus the deviation of the recovered power spectrum at $\ell>900$ should be caused by the variation of the spectral parameters for the foreground components. 

Finally, following the same procedure, we also test cases 4 and 5. Examples of the recovered CMB power spectra are shown in the lower left panel and lower right panel of Figure \ref{fig:vary_fg_params_cmbPowerSpectra}, respectively. For both cases, we can see that the recovered power spectra become lower than the ground truth at $\ell>450$, and gradually deviate with the increase in multipoles. Furthermore, we can see that the power spectra of the simulated dust maps (the lower left panel and the lower right panel of Figure \ref{fig:vary_fg_params_fgPowerSpectra}) deviate from the template at small scale, which is similar to case 3. Therefore, the deviations of the power spectra in Figure \ref{fig:vary_fg_params_cmbPowerSpectra} are caused by the randomness of the foreground spectral parameters. Comparing the results of case 4 and case 1, we may conclude that the power spectra of the foreground components will be destroyed (especially at small scale) if we randomize the spectral index $\beta$ for each pixel independently, and more additional noise will be introduced.

The results of these five cases show that the random form of the spectral parameters will have a great impact on the recovery of the CMB signal. Therefore, the spectral parameters of the foregrounds should be randomized reasonably. We propose that the power spectrum of the simulated foregrounds should be statistically coincident with that of the template at both large scale and small scale. We note that the method of varying the spectral parameters used in this paper is very simple, and that further efforts are needed to study how to simulate the foregrounds more reasonably. We will investigate this issue in our future works.

\begin{figure*}
	\centering	
	\includegraphics[width=0.32\textwidth]{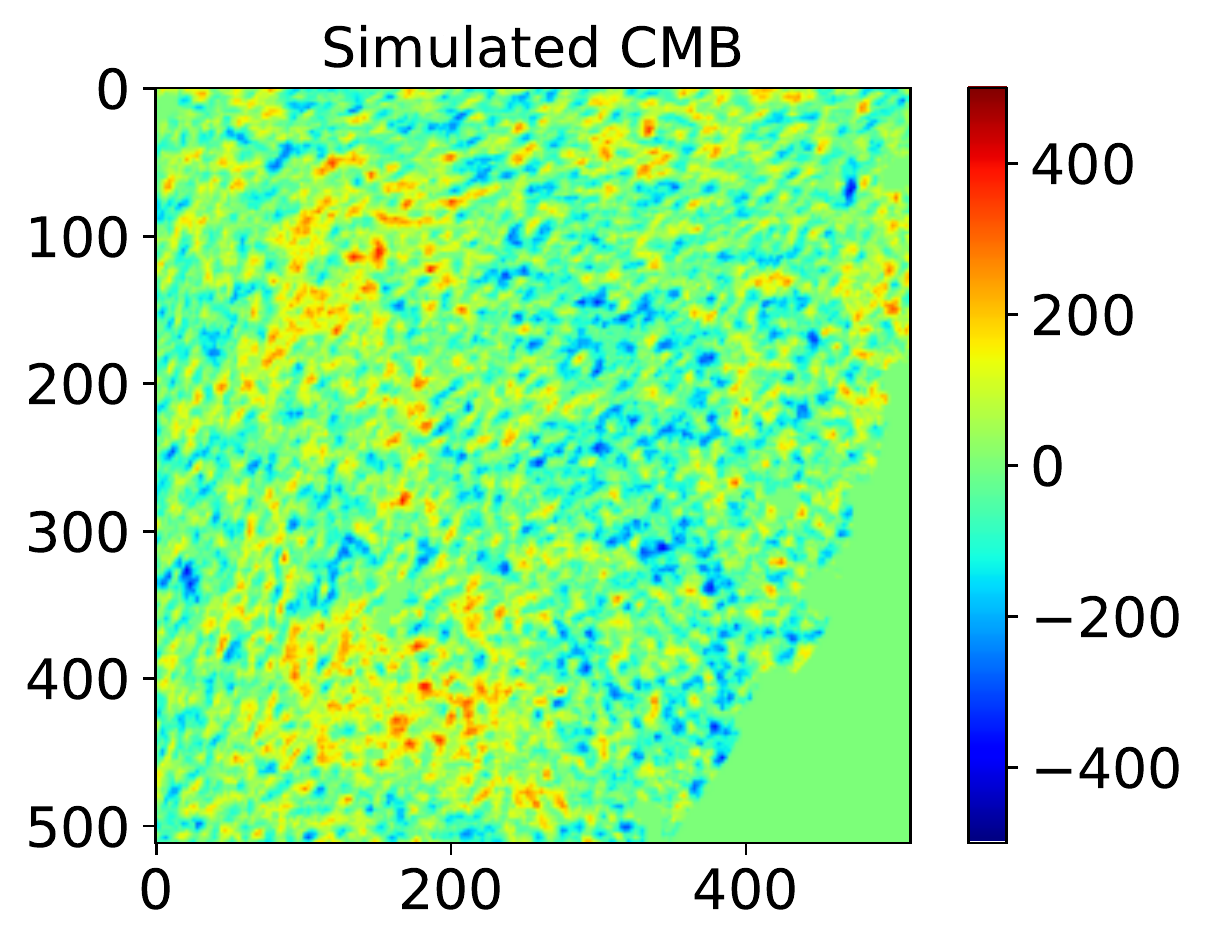}
	\includegraphics[width=0.32\textwidth]{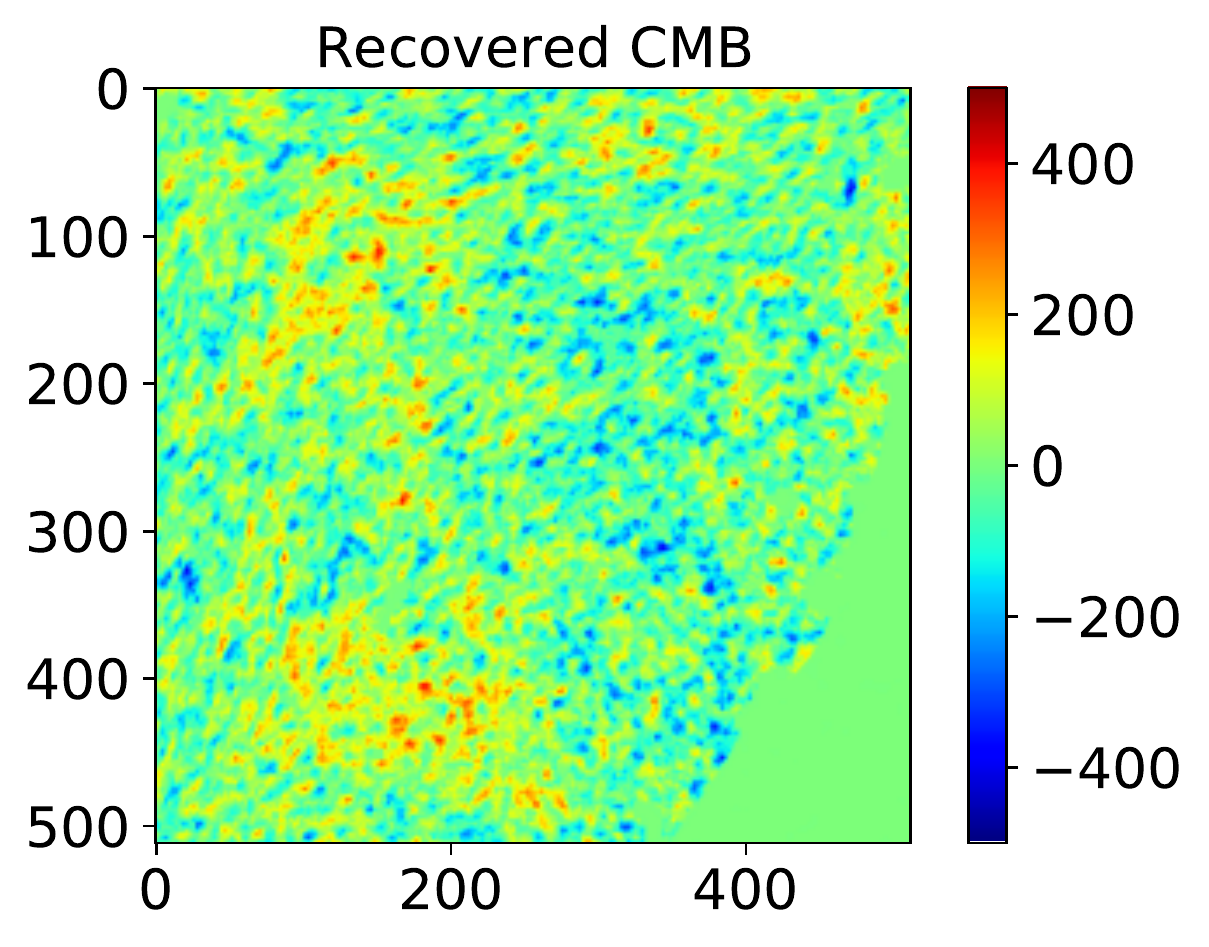}\\
	\includegraphics[width=0.32\textwidth]{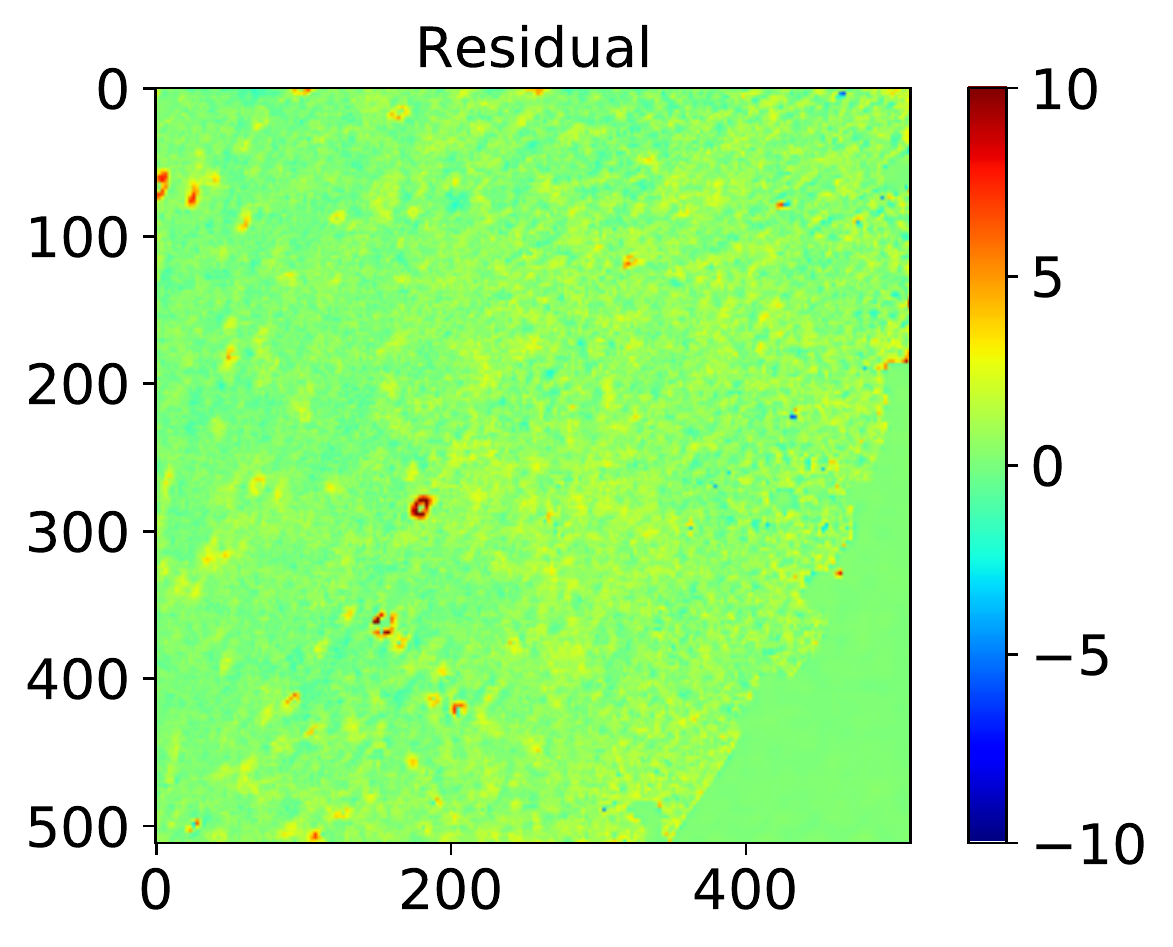}
	\includegraphics[width=0.32\textwidth]{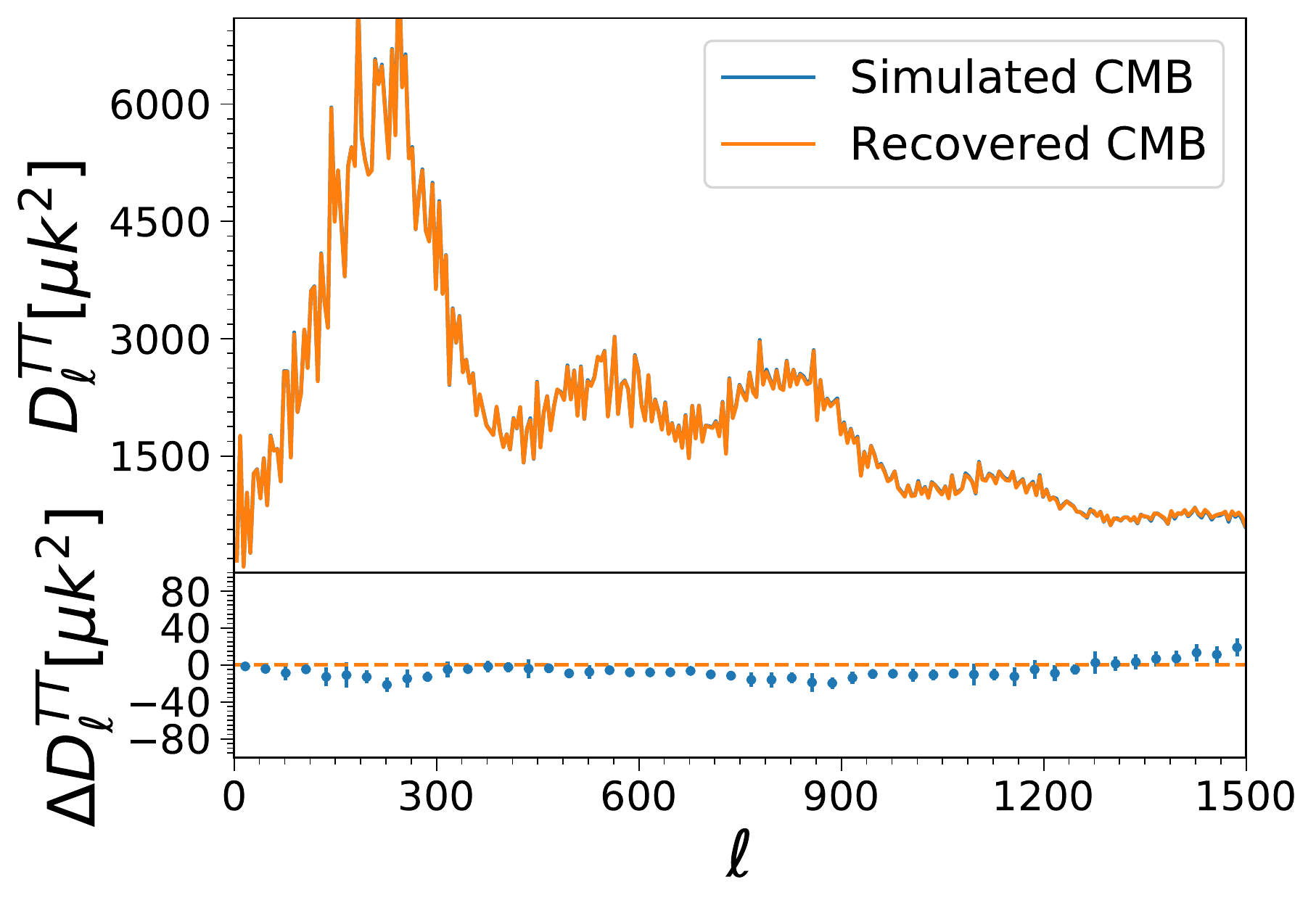}
	\caption{The same as the Figure \ref{fig:sim_cmb_map_block0}, but now the Planck mask is adopted in the training data.}\label{fig:sim_cmb_map_block0_mask}
\end{figure*}

\subsection{Partial-sky Experiments}\label{sec:discussion_partial-sky_experiments}

The analysis in section \ref{sec:effect_of_the_Galaxy} shows that the CNN method can also be used for partial-sky experiments. However, the block maps used in section \ref{sec:effect_of_the_Galaxy} are regularly shaped sky areas, which may be difficult to achieve in actual observations. For a specific partial-sky experiment, the observed sky area may have an irregular shape. In this section, we provide a method for training the network using irregular maps. For an irregular sky area, we can always transform it and fill it into a rectangle. Taking the simulated data as an example, we apply the Planck mask to the first block of Figure \ref{fig:train_block} and use it to train a network model. This means that the number of pixels of the unobserved area is set to zero. Using the same procedure as section \ref{sec:effect_of_the_Galaxy}, we train the network on the training set and test it on the test set. 

In Figure \ref{fig:sim_cmb_map_block0_mask}, we show one set of maps from the test set. From the simulated CMB map (the upper left panel of this figure), we can see that, since there is no observation, the pixel value of the area in the lower right corner is set to zero. The recovered CMB map is shown in the upper right panel, and it looks similar to the simulated one. From the lower left panel, we can see that there is very little information is left in the residual map. Furthermore, we calculate the power spectra of both the recovered CMB map and the simulated CMB map.
Specifically, we first fill the partial-sky map into a full-sky map, and make sure the other pixels have zero values, then calculate the power spectrum using this full-sky map. The results (see the lower right panel) show that the recovered CMB signal is quite consistent with the true spectrum at both small scales and large scales. Therefore, this indicates that the CNN method can also be used for the component separation of partial-sky experiments with irregular sky areas.

We note that the sky area used here is generated manually, by applying Planck mask, which may be more ideal than the actual sky area. Therefore, further research is needed when applying this method to actual observational data. We will investigate this interesting issue in future works.

\begin{table}
	\centering
	\caption{Instrumental Specifications of the CMB-S4 Experiment for the Six Selected Frequency Bands \citep{Campeti:2019}.}\label{tab:CMB_S4_Instrumental_specifications}
	\begin{tabular}{c|c|c}
		\hline\hline
		Frequency & Sensitivity & FWHM \\
		(GHz) & ($\mu$K-arcmin) & (arcmin) \\
		\hline
		85  & 1.6 & 27.0 \\
		95  & 1.3 & 24.2 \\
		145 & 2.0 & 15.9 \\
		155 & 2.0 & 14.8 \\
		220 & 5.2 & 10.7 \\
		270 & 7.1 & 8.5  \\
		\hline\hline
	\end{tabular}
\end{table}

\begin{figure*}
	\centering	
	\includegraphics[width=0.3\textwidth]{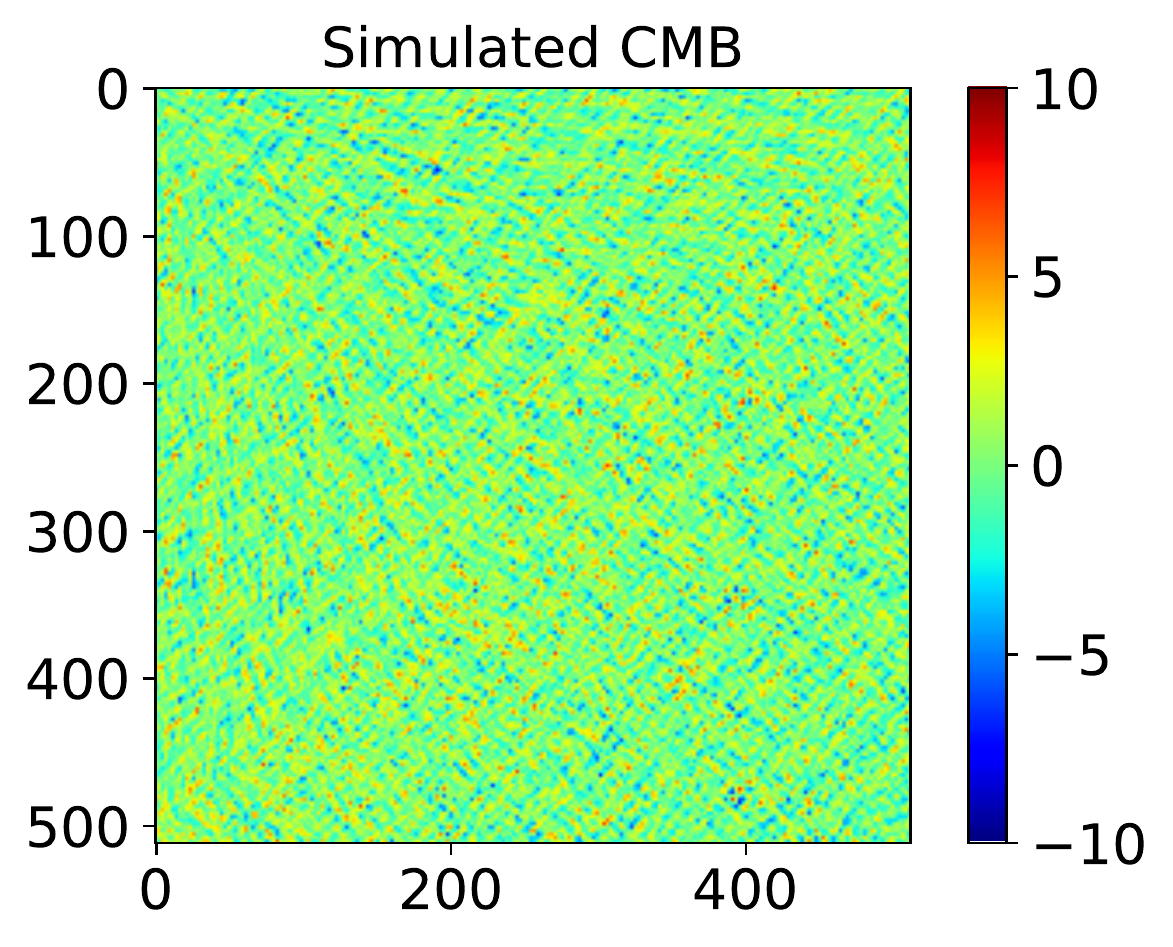}
	\includegraphics[width=0.3\textwidth]{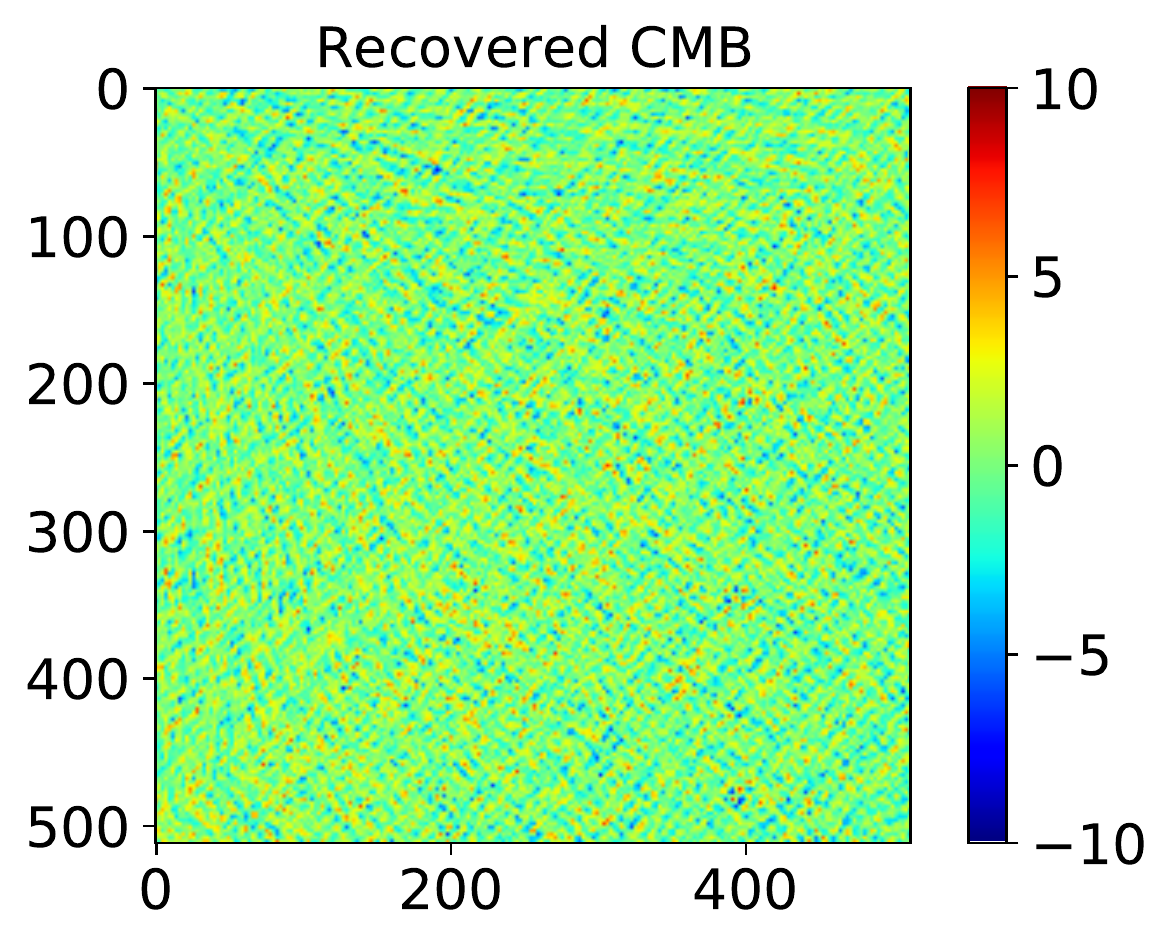}
	\includegraphics[width=0.3\textwidth]{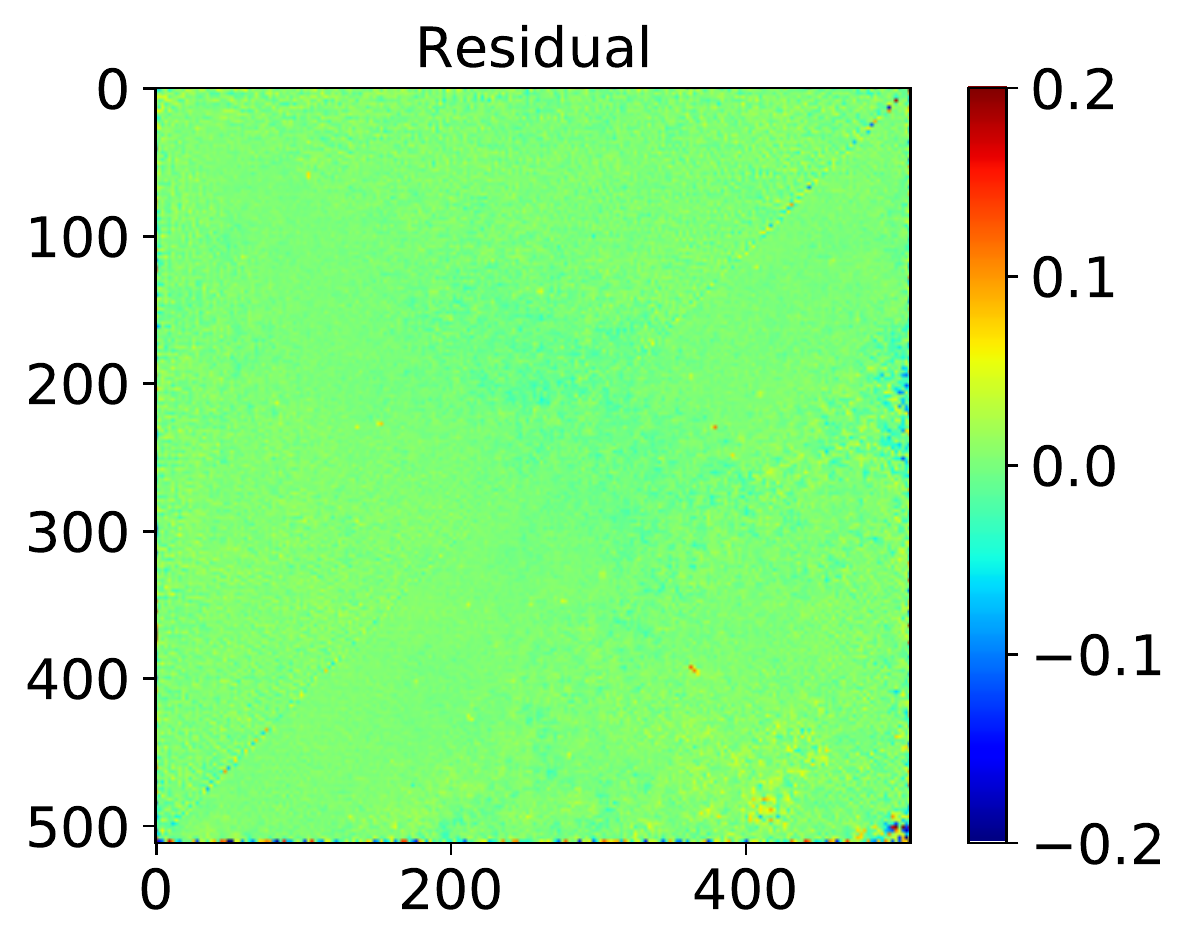}
	\includegraphics[width=0.3\textwidth]{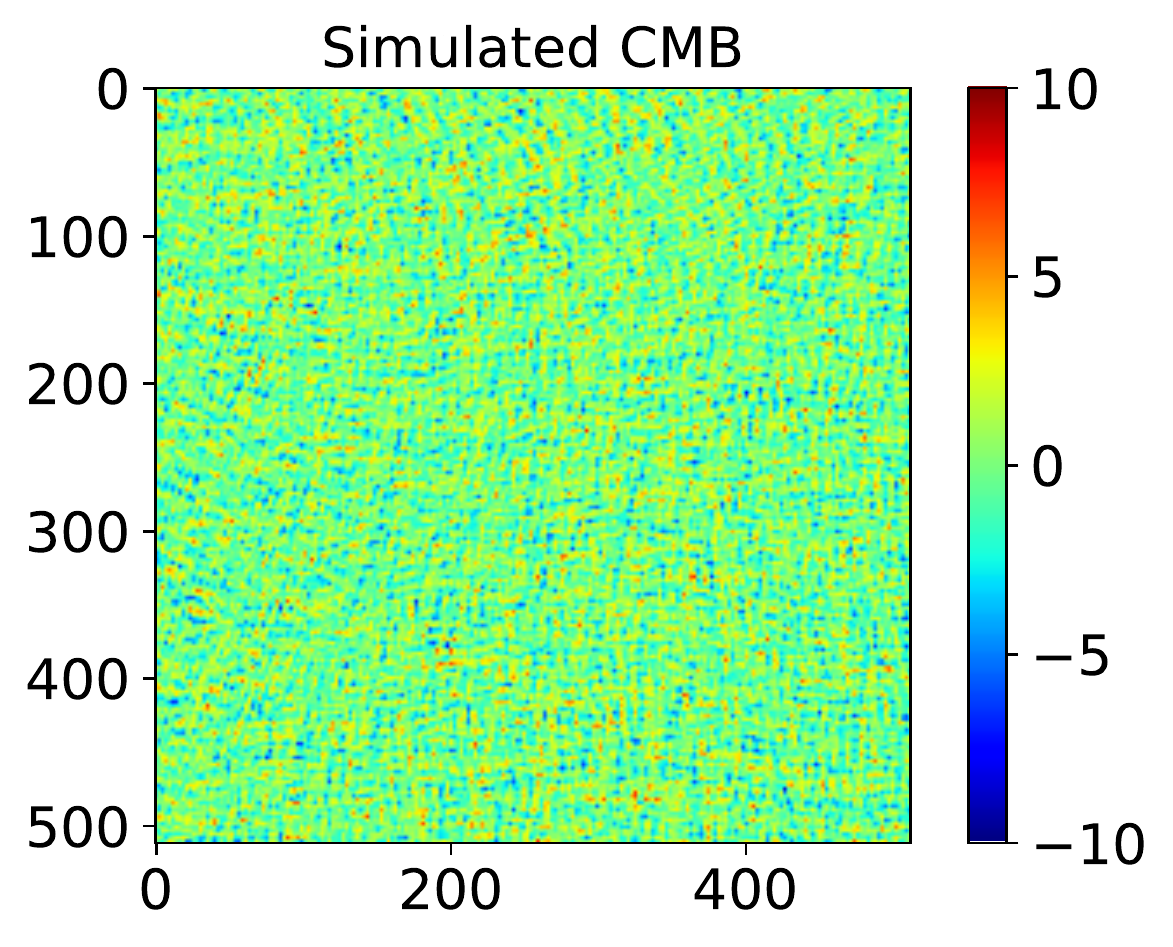}
	\includegraphics[width=0.3\textwidth]{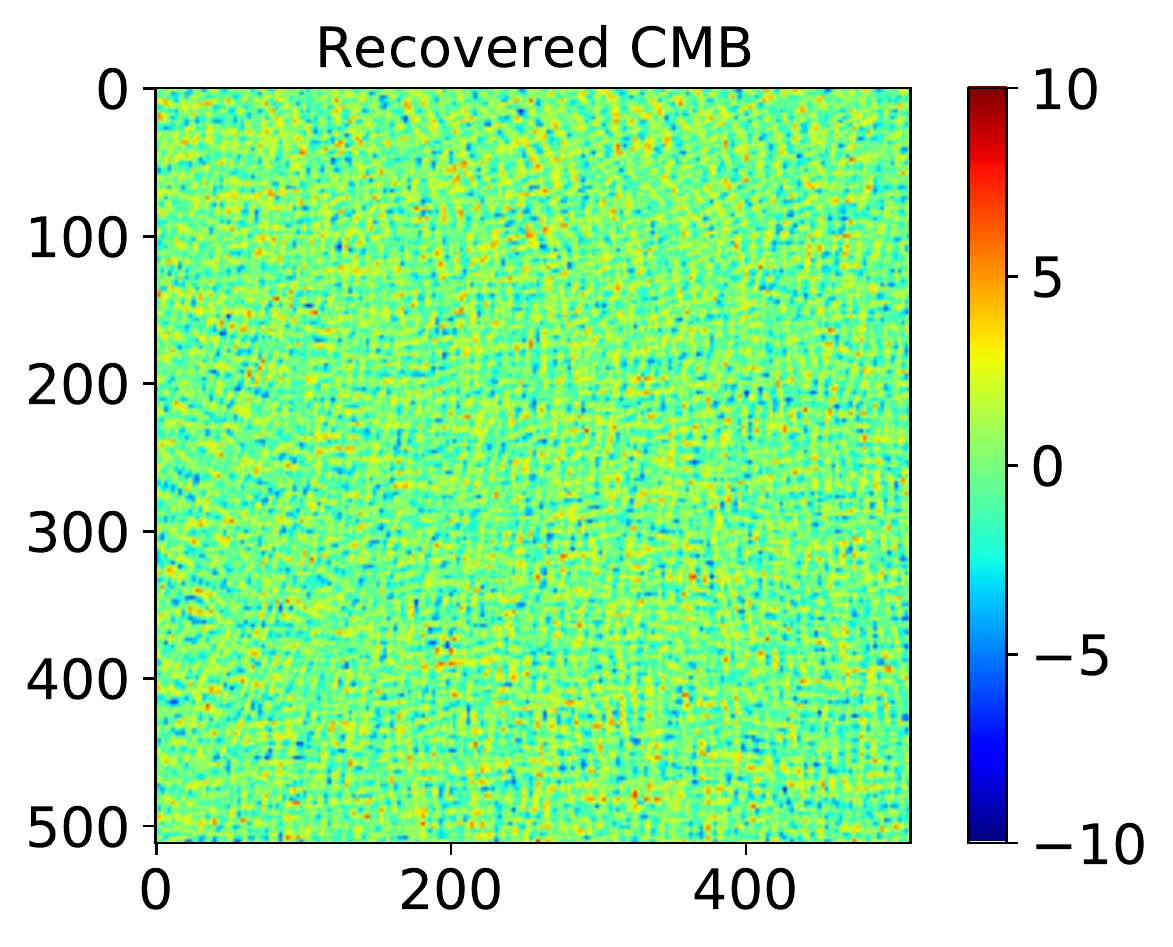}
	\includegraphics[width=0.3\textwidth]{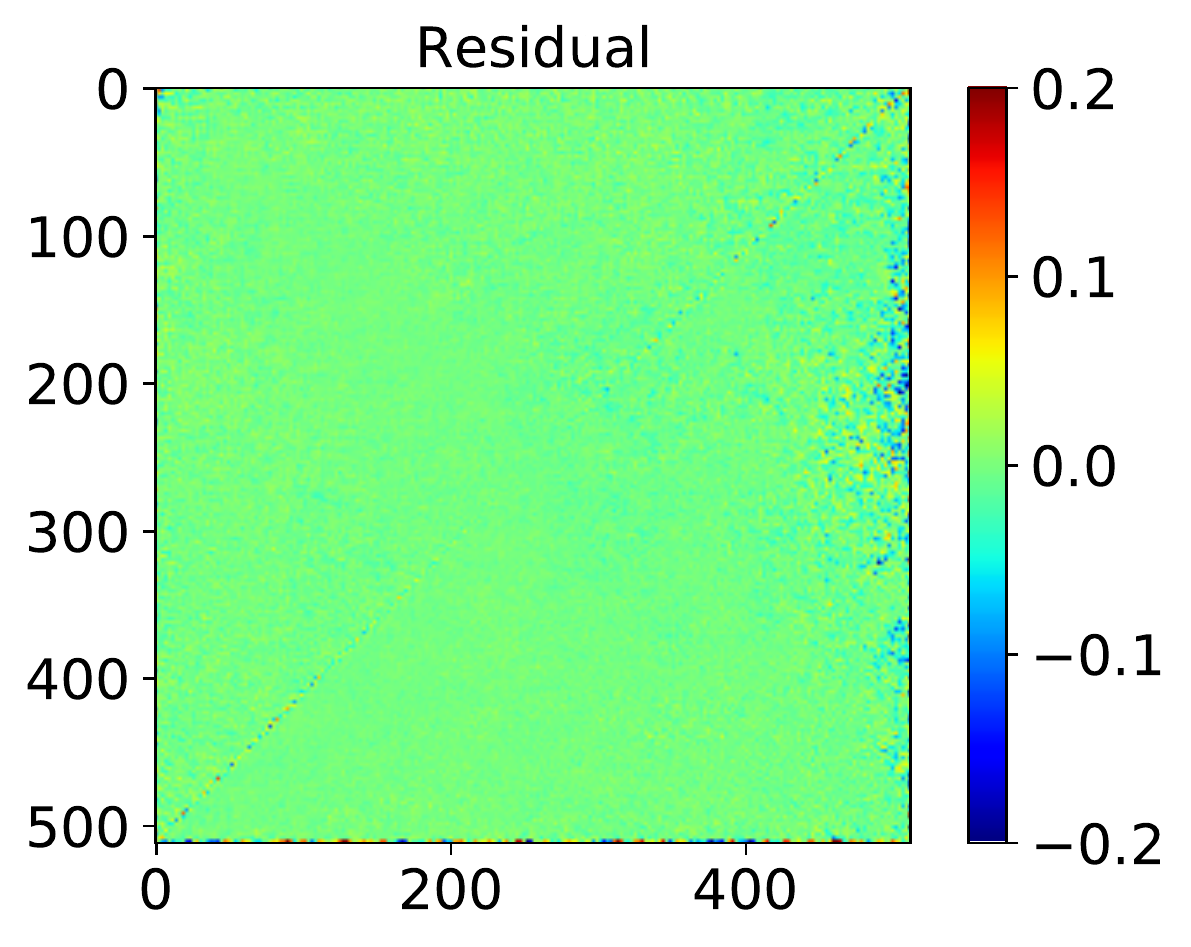}
	\caption{Recovered CMB Q (upper panels) and U (lower panels) maps for the CMB-S4 experiment.}\label{fig:sim_cmb_map_block_QU}
\end{figure*}
\begin{figure*}
	\centering	
	\includegraphics[width=0.9\textwidth]{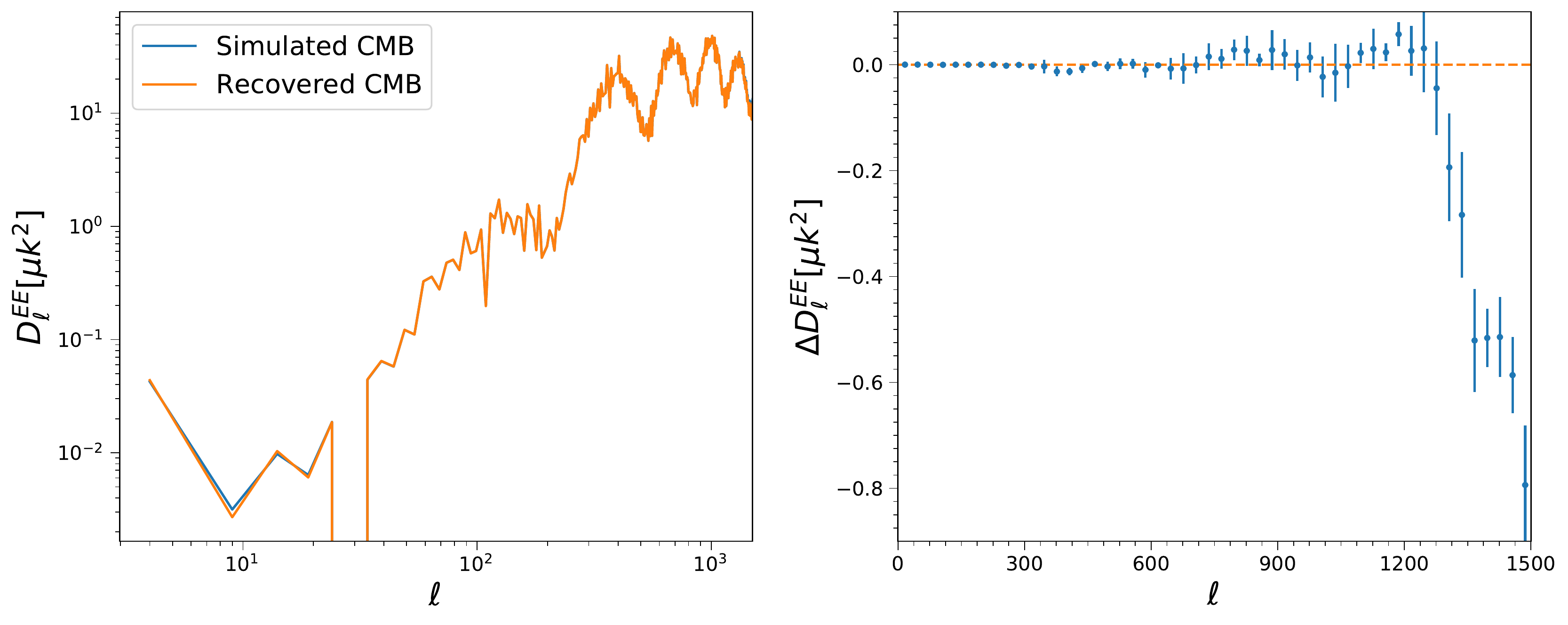}
	\includegraphics[width=0.9\textwidth]{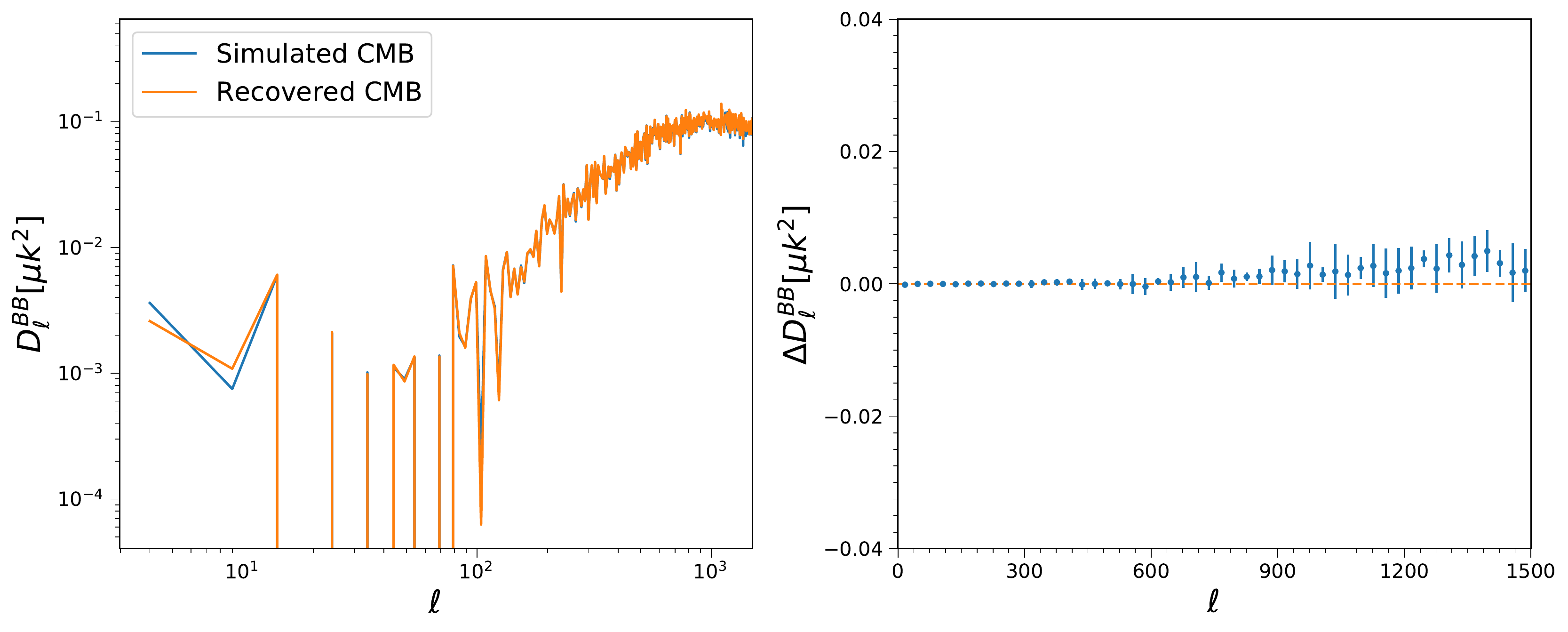}
	\caption{EE and BB power spectra calculated from the recovered CMB Q/U maps in Figure \ref{fig:sim_cmb_map_block_QU} (left panels, using a bin size of 5), and the difference between these spectra and those of the simulated maps (right panels, using a bin size of 30).}\label{fig:sim_cmb_map_block_EE_BB}
\end{figure*}

\subsection{CMB Polarization Maps}\label{sec:discussion_polarization_maps}

The analysis above is mainly focused on the recovery of the CMB temperature signal. However, this method is a common method that can also be used for the recovery of the CMB polarization signal. In this section, taking the CMB-S4 experiment as an example, we test the reliability of the CNN in recovering the CMB Q/U maps.

Specifically, we select six frequency bands of the CMB-S4 experiment, from 85 GHz to 270 GHz. The details of the experimental specifications are shown in Table \ref{tab:CMB_S4_Instrumental_specifications}. For the sake of simplicity, we only consider a block of sky like the first block in Figure \ref{fig:train_block}. Therefore, the sky fraction $f_{\rm sky} = 0.083$ for all frequency bands. The network used here is the same as the one used in section \ref{sec:effect_of_the_Galaxy}, which contains five convolutional layers and five deconvolutional layers. Following the same method as sections \ref{sec:simulation}, \ref{sec:training_set}, \ref{sec:data_preprocessing}, and \ref{sec:training_process}, we simulate the CMB, foregrounds, and noise components for the CMB-S4 experiment, generate the training set, preprocess the data, and finally train the network to recover the CMB Q and U maps, respectively.

After training the network, we test it using the test set. In Figure \ref{fig:sim_cmb_map_block_QU}, we show one set of maps from the test set. It can be seen from the residual maps that very little information is left, which means that the recovered CMB Q/U maps are very similar to the simulated ones. Furthermore, we calculate the EE and BB power spectra from the recovered CMB Q/U maps, shown in Figure \ref{fig:sim_cmb_map_block_EE_BB}. From the residual of the EE power spectrum (the upper right panel of Figure \ref{fig:sim_cmb_map_block_EE_BB}), we can see that the recovered power spectrum is quite consistent with the simulated one at $\ell<1250$, and that it gradually deviates with the increase in multipoles at $\ell>1250$, with the maximum deviation being $\sim 6.5\%$ at $\ell=1500$. Unlike the EE power spectrum, the residual of the BB power spectrum, in the lower right panel of Figure \ref{fig:sim_cmb_map_block_EE_BB}, shows that the BB power spectrum is consistent with the simulated one at both small scales and large scales, which is quite helpful for the detection of the primordial gravitational waves.

\subsection{Comparing with Other Methods}\label{sec:comparing_with_other_method}

In this section, we compare our analysis with that illustrated in \citet{Petroff:2020}. There are two main differences between our analysis and that of \citet{Petroff:2020}: the network model and the data used to train the network.

In our analysis, a general two-dimensional CNN with U-Net architecture is adopted, while in the \citet{Petroff:2020} analysis, a spherical CNN with U-Net architecture based on {\it DeepSphere} is used. The original intention of using a general two-dimensional CNN in our analysis was to enable the method to deal with both the full-sky maps and partial-sky maps. Fortunately, the analysis above shows that the network model based on a general two-dimensional CNN is capable of dealing with both full-sky maps and partial-sky maps. This is useful for experiments that focus on measuring the CMB polarization signal in a part of the sky.

We both tested the CNN method by using the simulated data and by applying the method to the Planck temperature measurement to obtain foreground-cleaned CMB maps. But the difference is that we only considered four frequency bands (100 GHz, 143 GHz, 217 GHz, and 353 GHz) in our analysis, while \citet{Petroff:2020} used seven frequency bands, from 70 GHz to 857 GHz, based on the Planck experiment. We used four frequency bands because we found that both the noise level and the beam effect (see sections \ref{sec:recover_sim_cmb} and \ref{sec:effect_of_noise}) affect the recovery of the CMB signal. 

In addition, according to our understanding, \citet{Petroff:2020} smoothed the maps in all frequencies, using a Gaussian beam with FWHM=13$^\prime$.1 (corresponding to the 70 GHz channel), and the target of the training set contained this beam effect. By contrast, we smoothed the maps in different frequencies with different beams, based on the Planck experiment, and the target of the training set contained a Gaussian beam effect with FWHM=7$^\prime$.27 (corresponding to the 143 GHz channel). From Figure \ref{fig:analysis_cmb_noise}, we can see that the power spectrum of the CMB at 70 GHz will loses more information than that at 143 GHz due to beam effects. Also, the power spectrum of the noise will be larger than that of the CMB at $\ell>900$. Therefore, worse results will be obtained when smoothing the target of the training set using a Gaussian beam with FWHM=13$^\prime$.1.

\subsection{The CNN Method}\label{sec:discussion_the_CNN_method}

The CNN is capable of extracting the features of CMB signals from images, and passing these features to later layers for the further filtering out of interfering signals. The CNN-based component separation method is a new approach to extracting CMB signals from the perspectives of images. The analysis above shows that the CNN method is capable of recovering the tiny CMB temperature and polarization signals from contaminated CMB observations. Therefore, the CNN method is an independent method, and it can be used as an alternative to various other component separation methods. This should be helpful for the data analysis of current and future CMB polarization experiments.

In the pipeline of recovering the CMB signal using the CNN, the neural network should be trained before it is applied to the observational data sets; thus, the CMB and foreground models used to generate the training set are input knowledge. This means that the CNN method is a model-dependent method that relays on the input information. Therefore, the simulation of the training set is important for this task. The training set should contain all possible components that the detector may receive, such as the CMB signal, all possible foreground components, the instrument noise, and the instrument beam effect.

The analysis in this work has mainly focused on the recovery of the CMB signal. However, similar to the CMB signal, various foreground components can also be regarded as signals, therefore the CNN method can also be used to separate these foreground components from observation data sets, which will allow us to have a better understanding of the physical properties of the foreground components. In addition, the CNN method is a general method for processing two-dimensional images; thus, it can be used for the component separation of other sky survey experiments, such as the radio survey experiments. We will investigate these interesting issues in future works.

\section{Conclusions}\label{sec:conclusions}

In this work, we show that a CNN can be used to recover the tiny CMB signals from various huge foreground contaminations. Focusing on CMB temperature fluctuations, we train the CNN model using simulated data, and find that the CMB temperature maps can be recovered with high accuracy, and that the deviation of the power spectrum $C_\ell$ is smaller than the cosmic variance at $\ell>10$. Furthermore, we apply this method to the Planck full mission datasets and find that the recovered CMB map is quite consistent with that disclosed by the Planck collaboration. Therefore, this indicates that the CNN method is capable of recovering the CMB signal from observational datasets, making it an alternative to current methods for the component separation of CMB raw maps, which can be used to analyze future CMB observational datasets, such as the CMB-S4 experiment.

Except for the full-sky maps, the CNN method is also capable of dealing with partial-sky maps. Specifically, we train the CNN model using part of the full-sky CMB map, and find that the recovered CMB signal is consistent with the simulated one. This means that this method can also be used in partial-sky experiments, allowing the method to be more widely used in future observational experiments. 

Moreover, taking the CMB-S4 experiment as an example, we simulate CMB Q/U maps to test the potential of the CNN in recovering the CMB polarization signal. The results show that, for the CMB-S4 experiment, both Q and U maps can be recovered with high accuracy. Furthermore, we calculate the EE and BB power spectra, and find that both of them can be recovered with high accuracy. Therefore, the CNN method may play an important role in experiments that measure the CMB polarization signal.

More interestingly, the CNN method is aimed at analyzing two-dimensional images; therefore, this means that it can become a common method for other similar tasks, such as the component separation of future radio surveys, like the Square Kilometer Array project \citep{SKA}. We will investigate these interesting issues in future works.

\section{Acknowledgement}

J.-Q. Xia is supported by the National Science Foundation of China, under grant Nos. U1931202 and 12021003, and the National
Key Research and Development Program of China, No. 2017YFA0402600. J.-F. Li. is supported by the National Natural Science Foundation of China, under grant No.11722437.

\appendix

Figures here show the effect of the galaxy foregrounds on the recovery of the CMB signal. Corresponding descriptions of these figures can be found in Section \ref{sec:effect_of_the_Galaxy}.

\setcounter{figure}{11}
\begin{figure*}
	\centering	
	\includegraphics[width=0.32\textwidth]{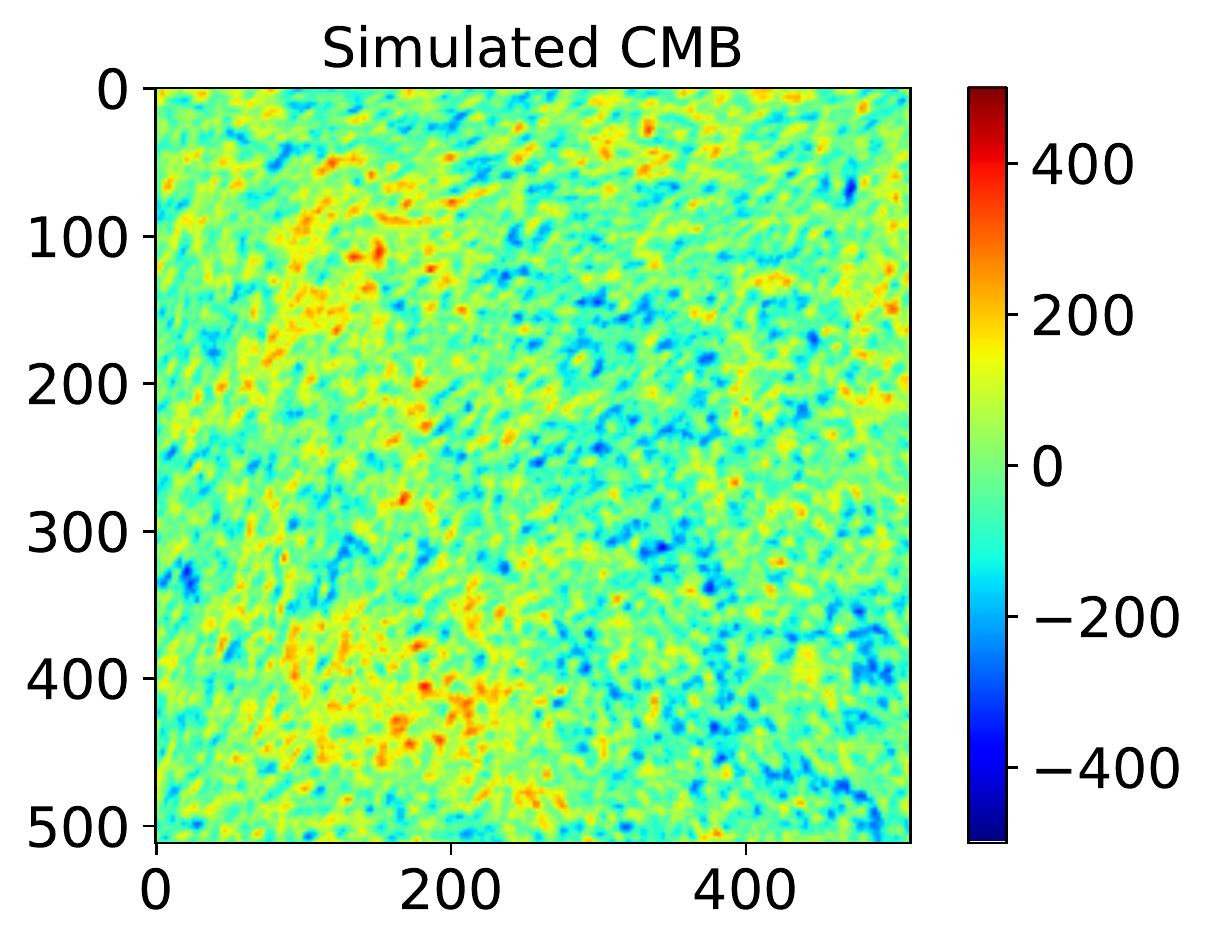}
	\includegraphics[width=0.32\textwidth]{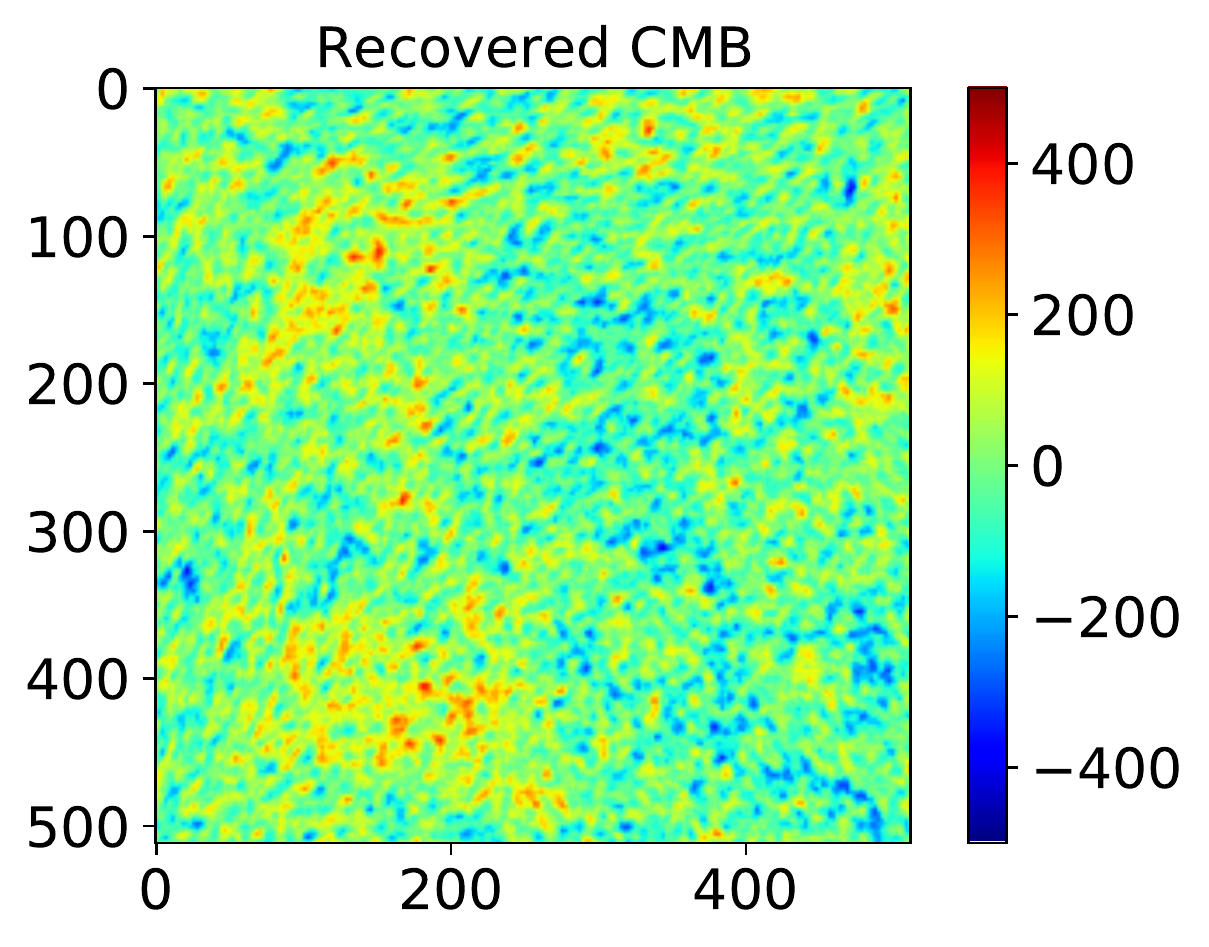}\\
	\includegraphics[width=0.32\textwidth]{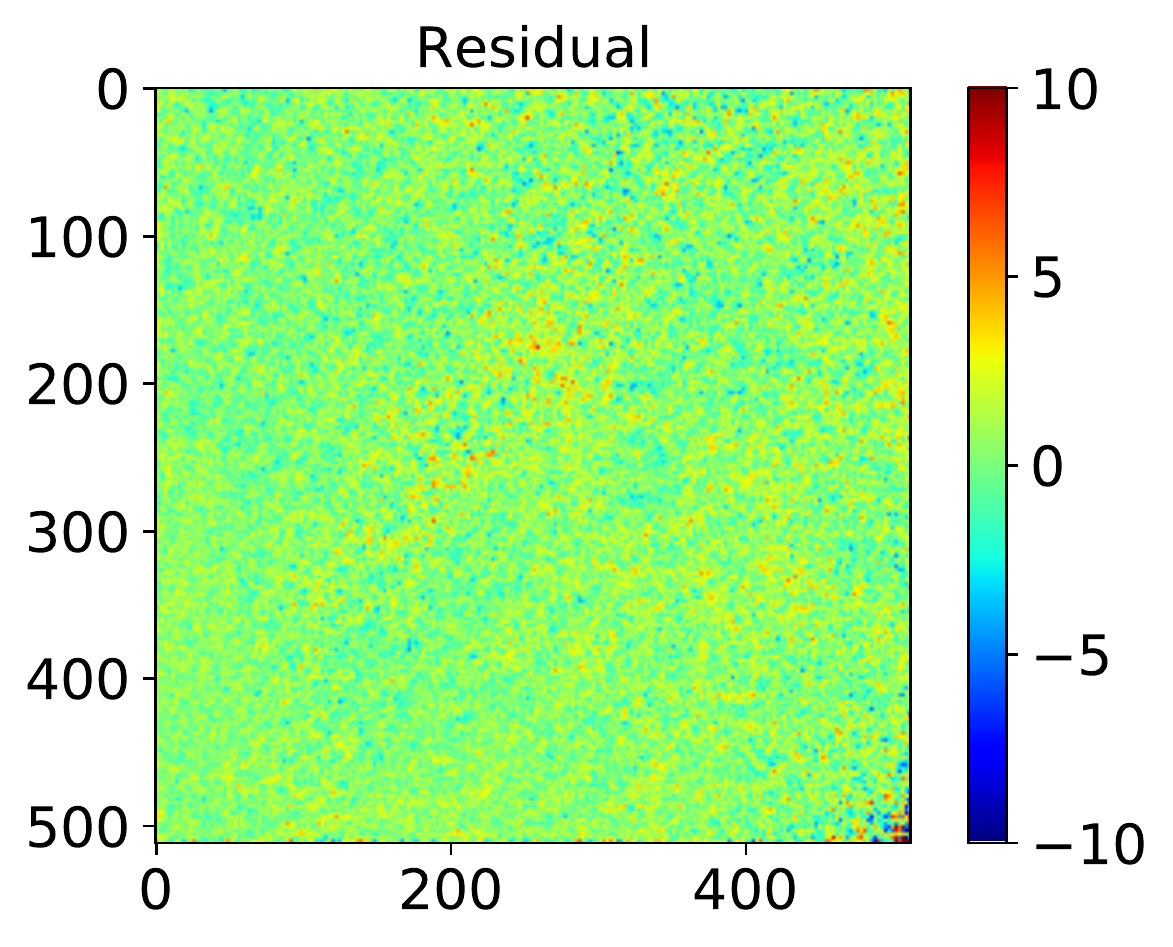}
	\includegraphics[width=0.32\textwidth]{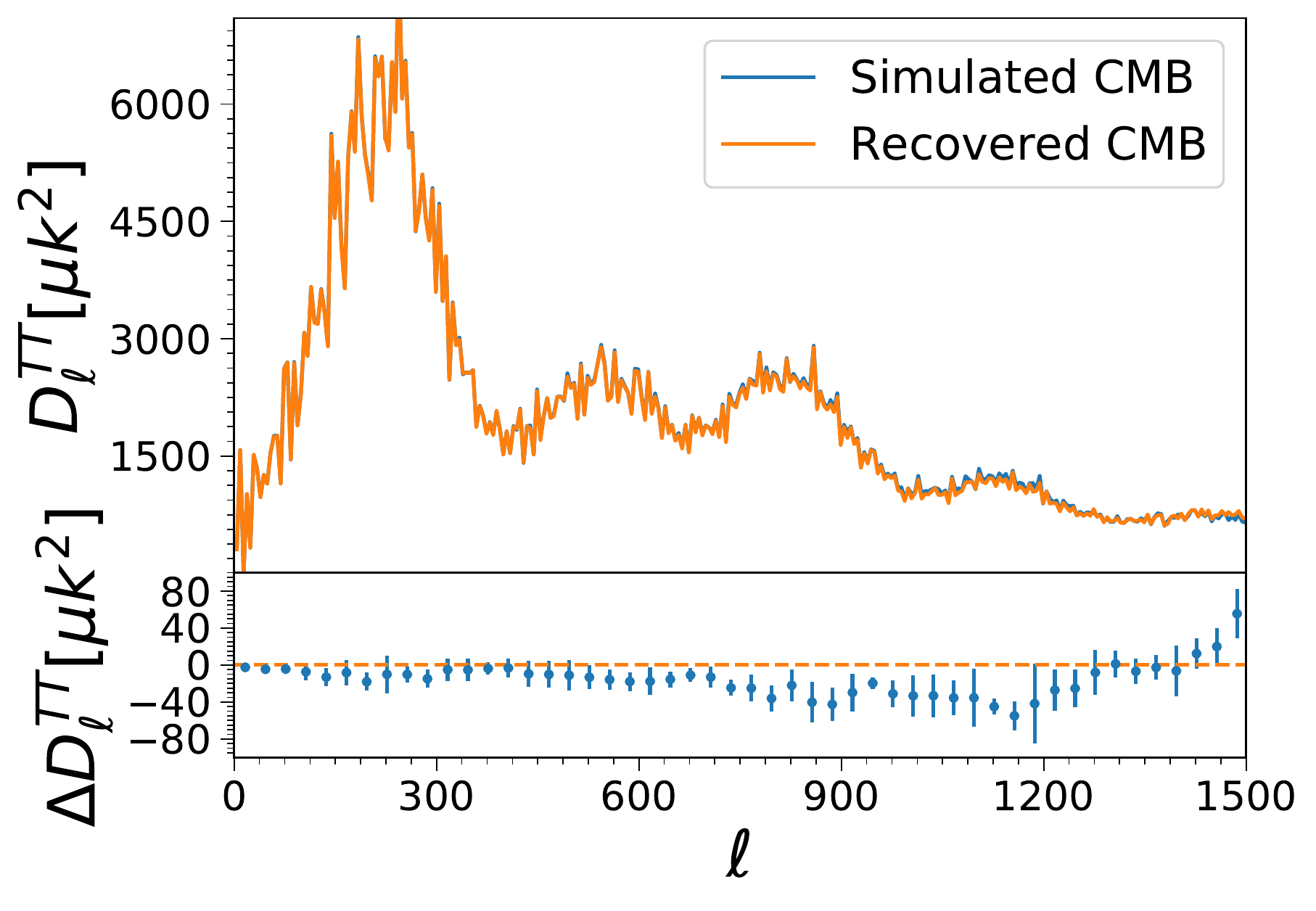}
	\caption{Recovered CMB temperature map and power spectrum when using the first block of Figure \ref{fig:train_block} to train the network. {\it Upper left panel:} the simulated pure CMB map. {\it Upper right panel:} the CMB map recovered by the neural network. {\it Lower left panel:} the residual map of the recovered CMB map. {\it Lower right panel:} power spectrum of the recovered CMB map (with a bin size of 5) and the difference between this spectrum and that of the simulated CMB map (with a bin size of 30).}\label{fig:sim_cmb_map_block0}
\end{figure*}

\begin{figure*}
	\centering	
	\includegraphics[width=0.65\textwidth]{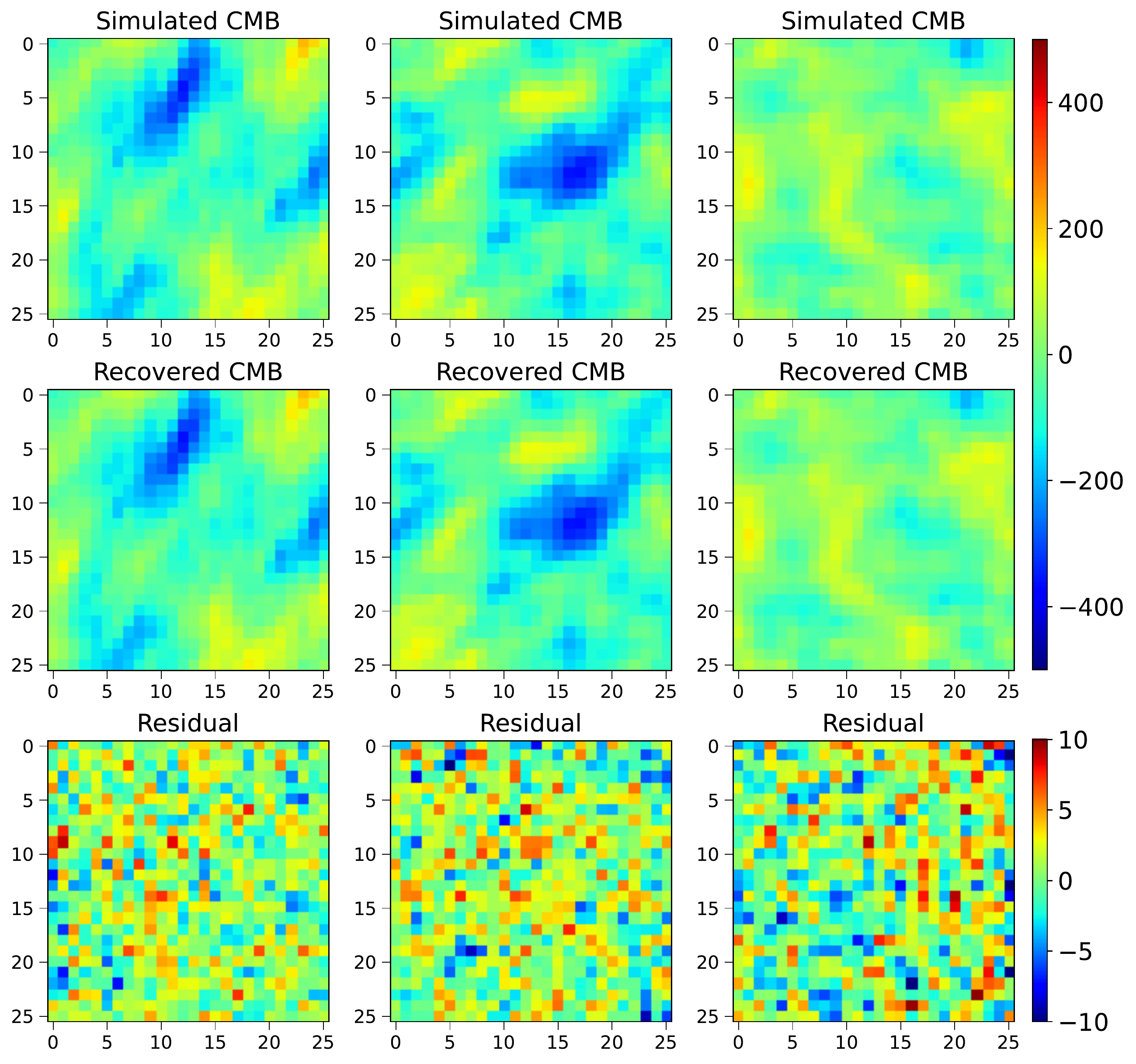}
	\caption{Three small patches with $3\times3$ deg$^2$ selected from Figure \ref{fig:sim_cmb_map_block0}. These patches are selected with different latitude.}\label{fig:sim_cmb_map_block0_miniPatch}
\end{figure*}

\begin{figure*}
	\centering
	\includegraphics[width=0.32\textwidth]{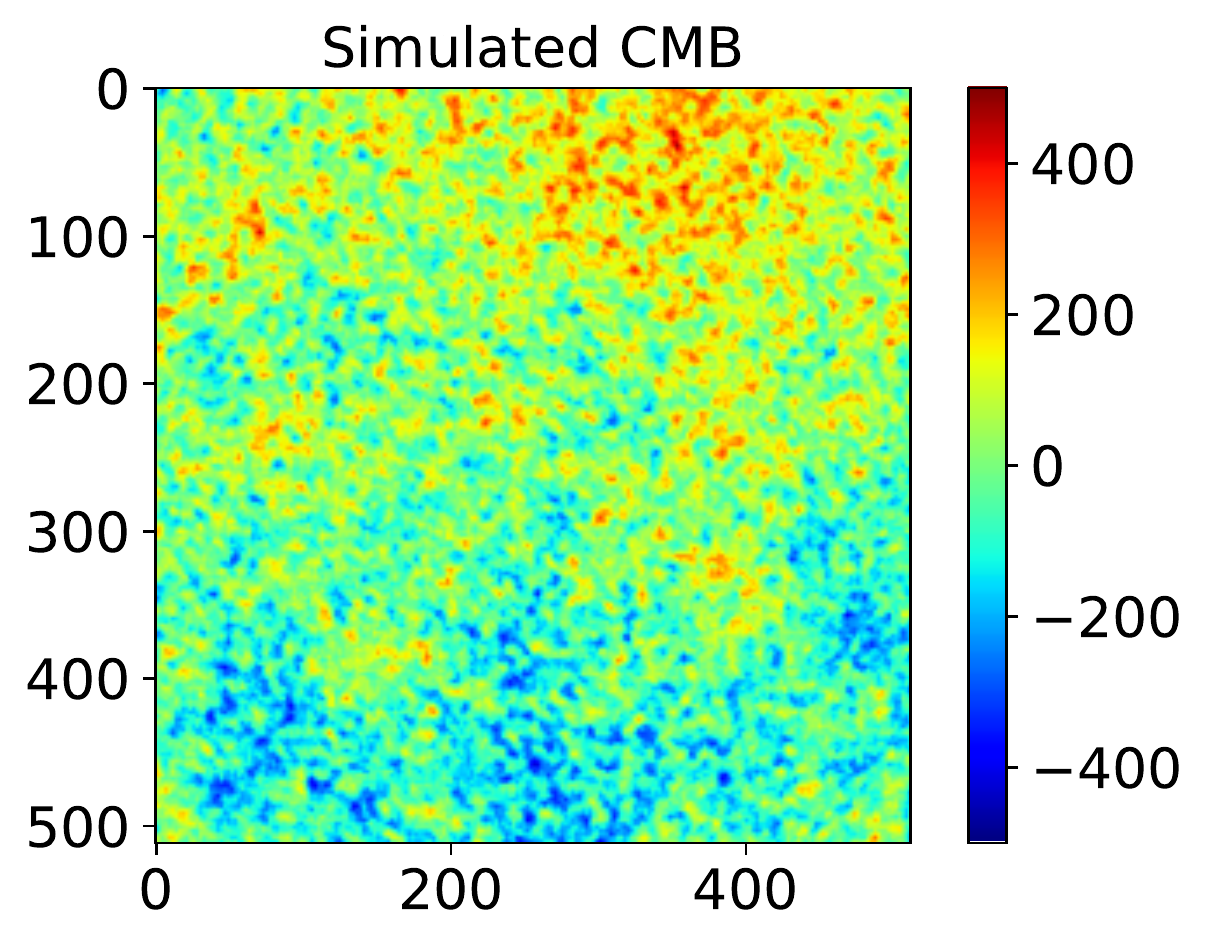}
	\includegraphics[width=0.32\textwidth]{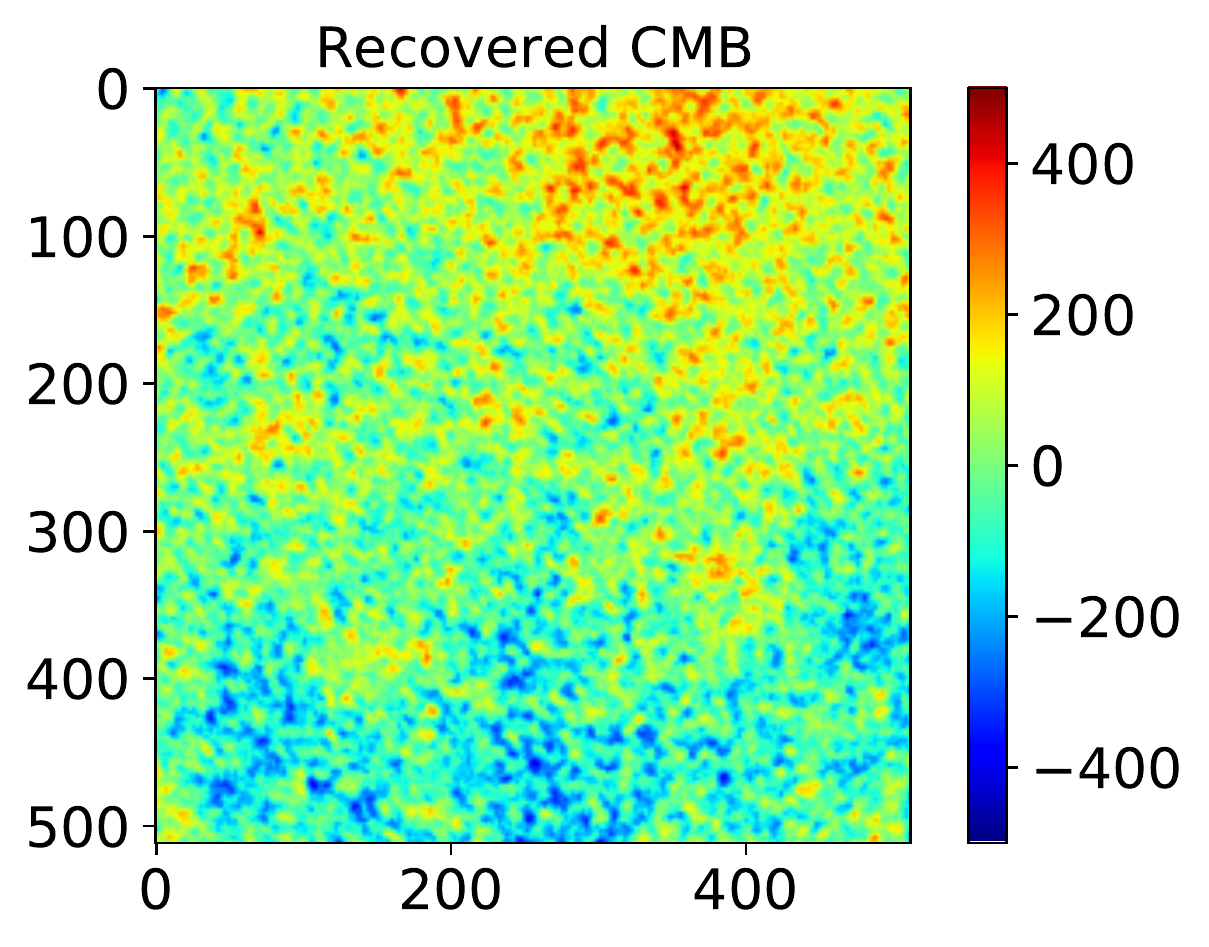}\\
	\includegraphics[width=0.32\textwidth]{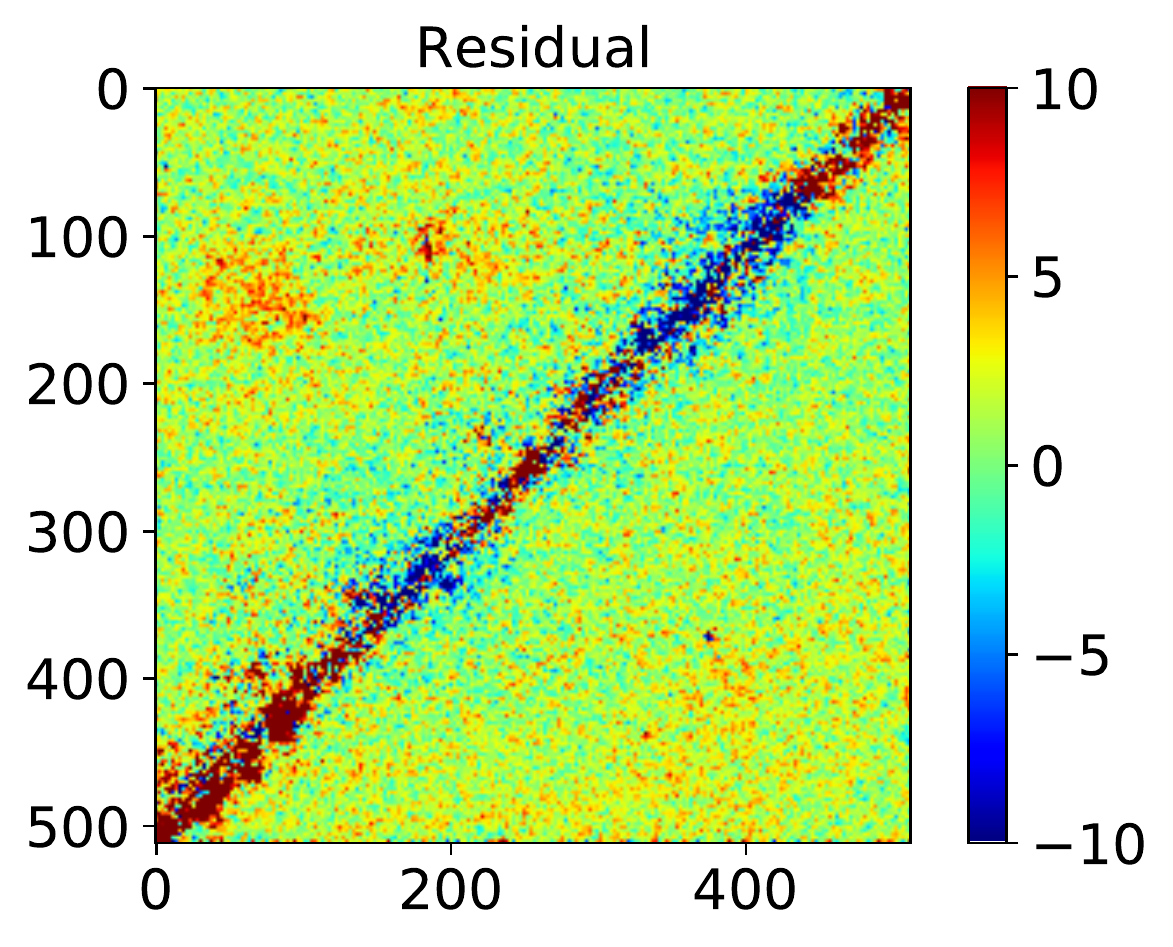}
	\includegraphics[width=0.32\textwidth]{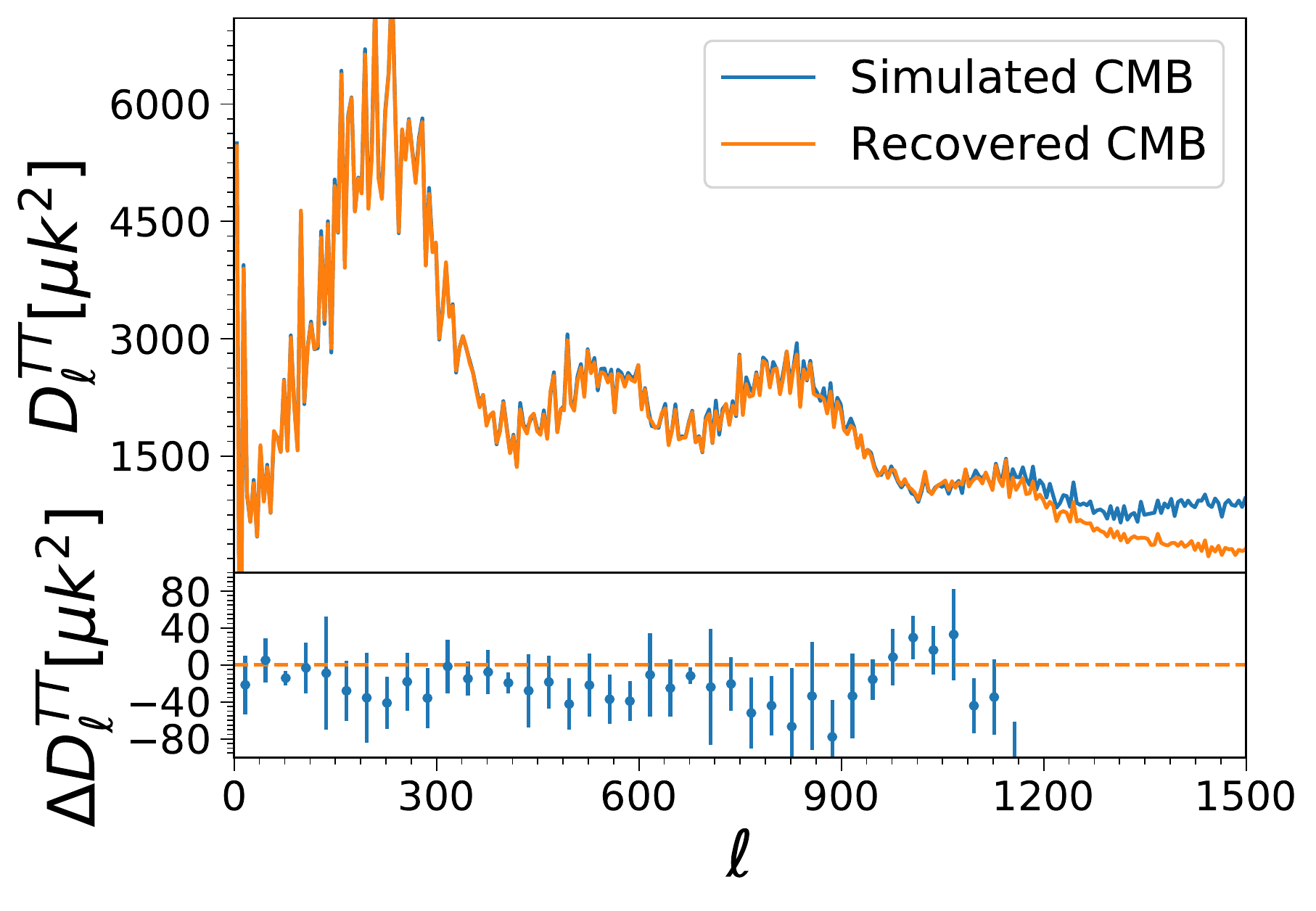}
	\caption{The same as Figure \ref{fig:sim_cmb_map_block0}, but now using the second block of Figure \ref{fig:train_block} to train the network.}\label{fig:sim_cmb_map_block4}
\end{figure*}

\begin{figure*}
	\centering
	\includegraphics[width=0.65\textwidth]{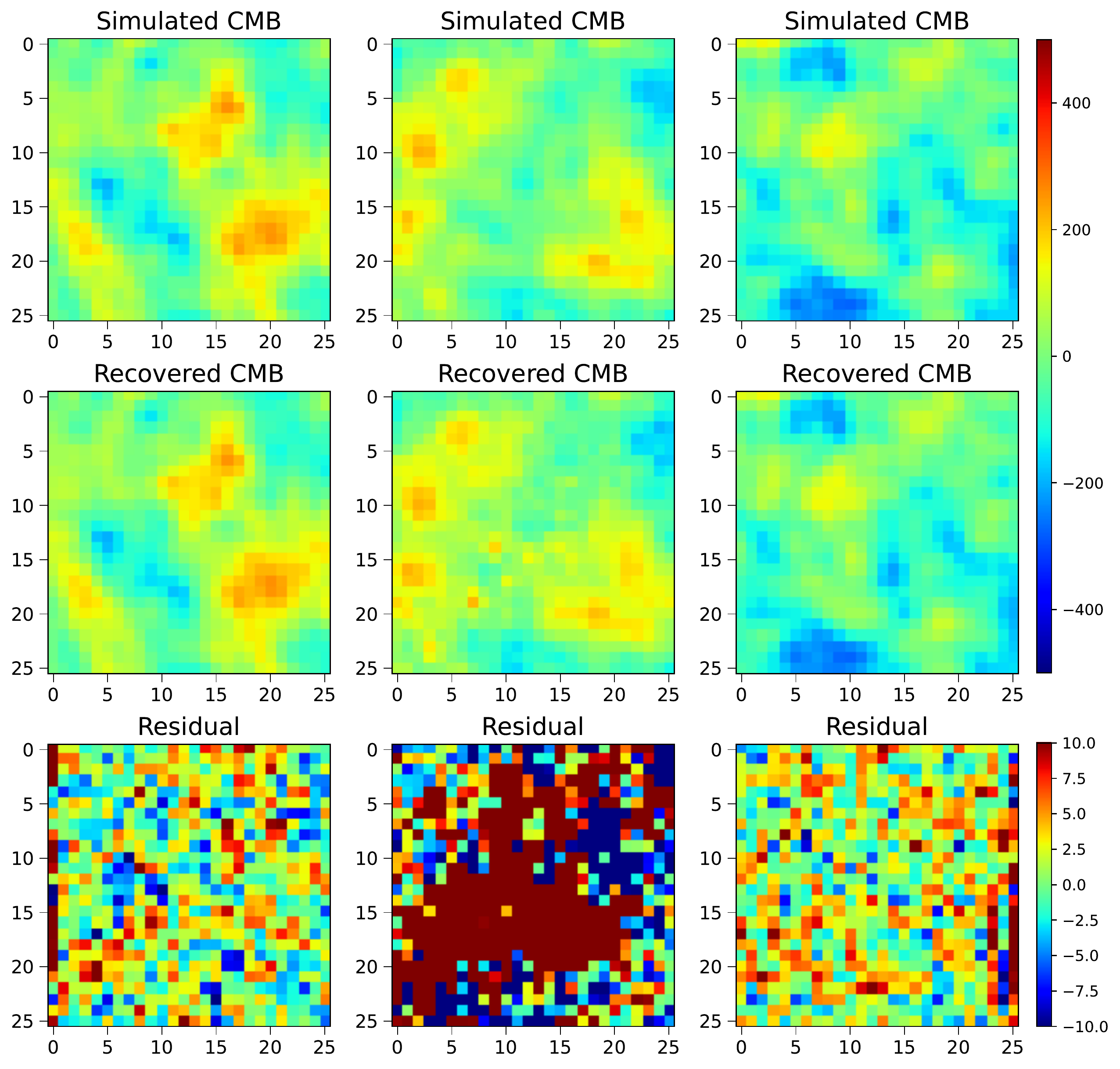}
	\caption{Three small patches with $3\times3$ deg$^2$ selected from Figure \ref{fig:sim_cmb_map_block4}. These patches are selected with different latitude.}\label{fig:sim_cmb_map_block4_miniPatch}
\end{figure*}

\end{document}